\documentclass{pasj01}
\usepackage{bm}

\Received{}
\Accepted{}

\begin{document} 

\newcommand{\simgt}{\lower.5ex\hbox{$\; \buildrel > \over \sim \;$}}
\newcommand{\simlt}{\lower.5ex\hbox{$\; \buildrel < \over \sim \;$}}
\newcommand{\vsp}{\vspace*{-1.5mm}}
\newcommand{\vs}{\vspace*{-3.5mm}}
\newcommand{\hsp}{\hspace*{-1.8mm}}
\newcommand{\targeta}{RCS\,J2319$+$0038}
\newcommand{\targetb}{HSC\,J0947$-$0119}
\newcommand{\Chandra}{\textit{Chandra}}
\newcommand{\XMM}{\textit{XMM-Newton}}
\newcommand{\Spitzer}{\textit{Spizter}}
\newcommand{\Hubble}{\textit{Hubble}}
\newcommand{\Planck}{\textit{Planck}}

\title{Galaxy clusters at ${\bm z \sim 1}$ imaged by ALMA 
with the Sunyaev-Zel'dovich effect} 


\author{Tetsu \textsc{Kitayama}\altaffilmark{1},
 Shutaro \textsc{Ueda}\altaffilmark{2,3},
 Nobuhiro \textsc{Okabe}\altaffilmark{4,5,6,7},
 Takuya \textsc{Akahori}\altaffilmark{8,9},
 Matt \textsc{Hilton}\altaffilmark{10,11,12},
 John P. \textsc{Hughes}\altaffilmark{13}, 
 Yuto \textsc{Ichinohe}\altaffilmark{14},
 Kotaro \textsc{Kohno}\altaffilmark{15,16},
 Eiichiro \textsc{Komatsu}\altaffilmark{17,18},    
 Yen-Ting \textsc{Lin}\altaffilmark{2}, 
 Hironao \textsc{Miyatake}\altaffilmark{18,19,20,21,22},
 Masamune \textsc{Oguri}\altaffilmark{23,24}, 
 Crist\'obal \textsc{Sif\'on}\altaffilmark{25},
 Shigehisa \textsc{Takakuwa}\altaffilmark{2,26},
 Motokazu \textsc{Takizawa}\altaffilmark{27},
 Takahiro \textsc{Tsutsumi}\altaffilmark{28},
 Joshiwa \textsc{van Marrewijk}\altaffilmark{29},
 and Edward J. \textsc{Wollack}\altaffilmark{30}
 }

\altaffiltext{1}{Department of
  Physics, Toho University, Funabashi, Chiba 274-8510, Japan}
\altaffiltext{2}{Academia Sinica Institute of Astronomy and Astrophysics
   (ASIAA), No. 1, Section 4, Roosevelt Road, Taipei 10617, Taiwan}
   \altaffiltext{3}{Institute of Astronomy, National Tsing Hua
  University, Hsinchu 30013, Taiwan}
\altaffiltext{4}{Department of Physics, Hiroshima University, 1-3-1 Kagamiyama,
Higashi-Hiroshima, Hiroshima 739-8526, Japan}
\altaffiltext{5}{Astrophysical Science Center, Hiroshima University, 1-3-1
Kagamiyama, Higashi-Hiroshima, Hiroshima 739-8526, Japan}
\altaffiltext{6}{Core Research for Energetic Universe, Hiroshima
  University, 1-3-1, Department of Physics, Hiroshima University, 1-3-1
  Kagamiyama, Higashi-Hiroshima, Hiroshima 739-8526, Japan}
\altaffiltext{7}{Physics Program, Graduate School of Advanced Science
  and Engineering, Hiroshima University, 1-3-1 Kagamiyama,
  Higashi-Hiroshima, Hiroshima 739-8526, Japan}
 \altaffiltext{8}{Mizusawa VLBI observatory, National Astronomical
 Observatory of Japan, 2-21-1 Osawa, Mitaka, Tokyo 181-8588, Japan}
 \altaffiltext{9}{
 Operation Division, SKA Observatory, Jodrell Bank, Lower Withington,
  Macclesfield, SK11 9DL, UK}
 \altaffiltext{10}{Astrophysics Research Centre, University of KwaZulu-Natal, Westville Campus, Durban 4041, South Africa}
\altaffiltext{11}{School of Mathematics, Statistics \& Computer Science,
  University of KwaZulu-Natal, Westville Campus, Durban 4041, South
  Africa}
\altaffiltext{12}{Wits Centre for Astrophysics, School of Physics,
  University of the Witwatersrand, Private Bag 3, 2050, Johannesburg, South Africa}
\altaffiltext{13}{Department of Physics and Astronomy, Rutgers, The
  State University of New Jersey, Piscataway, NJ 08854-8019, USA}  
 \altaffiltext{14}{Department of Physics, Rikkyo University, 3-34-1
  Nishi-Ikebukuro, Toshima-ku, Tokyo 171-8501, Japan}
\altaffiltext{15}{Institute of Astronomy, School of Science, The University of Tokyo, 2-21-1 Osawa, Mitaka, Tokyo 181-0015, Japan}
\altaffiltext{16}{Research Center for the Early Universe, School of Science, 
The University of Tokyo, 7-3-1 Hongo, Bunkyo, Tokyo 113-0033, Japan}
\altaffiltext{17}{Max-Planck-Institut f\"{u}r Astrophysik, Karl-Schwarzschild Str. 1, D-85741 Garching, Germany}
\altaffiltext{18}{Kavli Institute for the Physics and Mathematics of the
  Universe  (Kavli IPMU, WPI), The University of Tokyo Institutes for
  Advanced Study, The University of Tokyo, 5-1-5 Kashiwanoha, Chiba 277-8583, Japan}
         \altaffiltext{19}{Kobayashi-Maskawa Institute for the Origin of
  Particles and the Universe (KMI), Nagoya University, Nagoya, 464-8602, Japan}
      \altaffiltext{20}{Institute for Advanced Research, Nagoya University, Nagoya 464-8601, Japan}
         \altaffiltext{21}{Division of Particle and Astrophysical Science, Graduate School of Science, Nagoya University, Nagoya 464-8602, Japan}
         \altaffiltext{22}{Jet Propulsion Laboratory, California Institute of Technology, Pasadena, CA 91109, USA}
         \altaffiltext{23}{Center for Frontier Science, Chiba University, 1-33 Yayoi-cho, Inage-ku, Chiba 263-8522, Japan}
         \altaffiltext{24}{Department of Physics, Graduate School of Science, Chiba University, 1-33 Yayoi-Cho, Inage-Ku, Chiba 263-8522, Japan}
 \altaffiltext{25}{Instituto de F\'isica, Pontificia Universidad Cat\'olica de Valpara\'iso, Casilla 4059, Valpara\'iso, Chile}
  \altaffiltext{26}{Department of Physics and Astronomy, Graduate School of Science and Engineering,
Kagoshima University, 1-21-35 Korimoto, Kagoshima, Kagoshima 890-0065, Japan}
\altaffiltext{27}{Department of Physics, Yamagata University, 
1-4-12 Kojirakawa-machi, Yamagata, Yamagata 990-8560, Japan}
\altaffiltext{28}{National Radio Astronomy Observatory, P.O. Box O,
 Socorro, NM, 87801, USA}
 \altaffiltext{29}{European Southern Observatory (ESO), Karl-Schwarzschild-Strasse 2,
Garching 85748, Germany}
 \altaffiltext{30}{NASA/Goddard Space Flight Center, Greenbelt, MD 20771, USA}
  
\email{kitayama@ph.sci.toho-u.ac.jp}

\KeyWords{cosmology: observations
-- galaxies: clusters: intracluster medium --
galaxies: clusters:
individual (\targeta, \targetb) -- radio continuum: galaxies --
techniques: interferometric}

\maketitle

  \begin{abstract}   
  We present high angular-resolution measurements of the thermal
  Sunyaev-Zel'dovich effect (SZE) toward two galaxy clusters, \targeta\
  at $z=0.9$ and \targetb\ at $z=1.1$, by the Atacama Large
  Millimeter/submillimeter Array (ALMA) in Band 3. 
  They are supplemented with available \Chandra\ X-ray data,
  optical data taken by Hyper Suprime-Cam on Subaru, and millimeter-wave SZE
  data from the Atacama Cosmology Telescope. Taking into 
  account departures from spherical symmetry, we have reconstructed
  non-parametrically the inner pressure profile of two clusters as well
  as electron temperature and density profiles for \targeta.   This is
  one of the first such measurements for an individual cluster at $z
  \gtsim 0.9$.  We find that the inner pressure profile of both clusters
  is much shallower than that of local cool-core clusters.  Our results
  consistently suggest that \targeta\ hosts a weak cool core, where
  radiative cooling is less significant than in local cool cores. On the
  other hand, \targetb\ exhibits an even shallower pressure profile than
  \targeta\ and is more likely a non-cool-core cluster. The SZE centroid
  position is offset by more than 140
  $h_{70}^{-1}$kpc from the peaks of galaxy distribution in \targetb,
  suggesting a stronger influence of mergers in this
  cluster. We conclude that these distant clusters
are at a very early stage of developing the cool cores typically found in
clusters at lower redshifts. 
  \end{abstract}


\section{Introduction}

The baryonic content of galaxy clusters is dominated by hot ($\simgt
10^7$ K) and tenuous ($\simlt 0.1$ cm$^{-3}$) plasma, referred to as the
intracluster medium (ICM).  The ICM properties of nearby ($z \simlt
0.2$) clusters are well studied thanks to resolved measurements of X-ray
surface brightness and temperature profiles from a large number of
clusters (see e.g., \cite{Bohringer10,Cavaliere13} for review). Such
profiles encode the thermodynamic evolution of clusters driven by
mergers, accretion, and AGN feedback on one hand and radiative cooling
on the other. It is recognized that there is a wide diversity in the
properties of central cores ($\simlt 100$ kpc), within which cooling and
feedback processes play important roles in sculpting their thermodynamic
properties. Broadly speaking, galaxy clusters are divided into two
groups: ``cool-core'' and ``non-cool-core'' clusters (e.g.,
\cite{Peres98,Bauer05}). The former is also characterized as having
peaked or cuspy gas density profiles, whereas the latter often shows 
clear signs of disturbance.

At $z\simgt 1$, the observational situation is much less certain.  There
are only a handful of galaxy clusters whose thermodynamic structure has
been studied at these redshifts in X-rays
\citep{Santos12,McDonald14b,Tozzi15,Brodwin16,Bartalucci17,Sanders18,Mantz20,Ghirardini21}.
This is mainly because the observed X-ray surface brightness decreases as $\propto
(1+z)^{-4}$ and the observed sizes of clusters become small (e.g., typical
core radius of 100 kpc corresponds to $\sim 12''$ at $z\simgt 1$,
whereas the Half-Energy-Width of \XMM\ 
is $\sim 15''$). As a result, it is very challenging to measure
spatially resolved temperature structures from X-ray spectral
analyses. Efforts to measure thermodynamic profiles at high-$z$ often
employ stacking methods to obtain average profiles for samples of
clusters observed over broad redshift ranges, e.g., $0.6<z<1.2$
\citep{McDonald14b}.

In this regard, the Sunyaev-Zel'dovich effect (SZE; \cite{Sunyaev70,Sunyaev72})
serves as a useful probe of distant galaxy clusters (see
\cite{Mroczkowski19} for a recent review).  For given electron density
$n_{\rm e}$ and temperature $T_{\rm e}$, the surface brightness of the
SZE is proportional to $n_{\rm e}T_{\rm e}$, whereas that of X-rays
varies as $n_{\rm e}^2\Lambda(T_{\rm e})(1+z)^{-4}$, where $\Lambda$ is
typically a weak function of $T_{\rm e}$. The SZE brightness is free
from the aforementioned $(1+z)^{-4}$ dimming and provides a direct
measure of electron thermal pressure. The advent of large-area SZE surveys
by the South Pole Telescope (SPT; e.g.,
\cite{Spt09,Spt10,Spt11,Spt13,Spt15,Huang20}) and the Atacama Cosmology
Telescope (ACT; e.g., \cite{Act10,Act11,Act13a,Hilton18,Hilton21}) has
significantly increased the number of galaxy clusters detected at
$z\simgt 1$. Resolved SZE images of clusters at $z \simgt 1$ with an
angular resolution of $< 20''$ have been obtained by MUSTANG on Green
Bank Telescope \citep{Korngut11,Dicker20,Andreon21}, NIKA on the IRAM
telescope \citep{Adam15,Adam18}, and the Atacama Large
Millimeter/submillimeter Array (ALMA) 
\citep{Basu16,Gobat19,Luca20,Luca21}.

In this paper, we present high angular resolution SZE images of galaxy
clusters, \targeta\ at $z=0.90$ and \targetb\ at $z=1.11$, 
taken by ALMA, and supplement them with available
X-ray, optical, and wider-field SZE data.
Because of their much smaller mass, the SZE signal of \targeta\ and \targetb\ is  weaker, 
by a factor of up to $5$, than that of massive galaxy clusters at lower redshifts studied by
ALMA previously, e.g., RX J1347.5--1145 at $z=0.45$ and SPT-CL
J2334--4243 at $z=0.60$ \citep{Kitayama16, Kitayama20}.  The present
paper therefore makes use of one of the deepest ALMA SZE data obtained
so far. Our goal is to reveal thermodynamic structures of clusters at $z\sim 1$ in conjunction with galaxy distributions, and weak lensing mass maps. We develop a method to reconstruct 
electron pressure and temperature profiles of distant clusters 
non-parametrically, taking into account departures from spherical symmetry. We also include or clarify various systematic effects associated with the analysis.

\targeta\ is the most massive galaxy cluster in a spectroscopically
confirmed supercluster at $z \sim 0.9$ \citep{Gilbank08}, 
discovered by the Red-Sequence Cluster Survey (RCS; \cite{Gladders05}).
We adopt for this cluster the spectroscopic redshift of $z=0.90$ from
\citet{Gilbank08}.  Based on the X-ray observation with \Chandra\
\citep{Hicks08}, the density profile of the ICM is well constrained,
whereas only the average spectroscopic temperature is measured owing to
the large distance to the cluster.  More recently,
\targeta\ was also detected in the ACT Data Release 5 (DR5;
\cite{Naess20}) cluster search\footnote{https://lambda.gsfc.nasa.gov/product/act/actadv\_prod\_table.html} \citep{Hilton21}, with signal-to-noise
ratio (S/N) of 5.2.  We note that the centroid position of
the ACT SZE signal is offset from the X-ray peak by $\sim 30''$. We will explore
the origin of this offset by means of higher angular resolution SZE
data.

\targetb\ was discovered more recently by the Hyper
Suprime-Cam Subaru Strategic Program (HSC-SSP;
\cite{Miyazaki18,Komiyama18,Furusawa18,Bosch18,Huang18,Coupon18,Tanaka18,Aihara18a,Aihara18b,Aihara19})
using the Cluster finding Algorithm based on Multi-band Identification
of Red-sequence gAlaxies (CAMIRA; \cite{Oguri14,Oguri18}). Given the
lack of accurate spectroscopic measurements, we adopt for this cluster
the photometric redshift\footnote{$z=1.11$ is
consistent with spectroscopic redshift obtained for several members from
Magellan/LDSS3; however, our observations also suggest possible galaxy
concentrations along the line-of-sight; we are in the process of
obtaining more spectroscopic redshifts in this field.} of $z=1.11$ from
the CAMIRA catalogue for the third public data release of HSC-SSP
\footnote{https://hsc-release.mtk.nao.ac.jp/doc/}. The accuracy of the
photometric redshift in this catalogue is estimated to be $\sigma_z \sim
0.01 (1+z)$ \citep{Oguri18}. The observed richness of \targetb\ is one
of the highest among the clusters at $z>1$ identified by CAMIRA.
\targetb\ is also detected in ACT DR5 with
S/N$=13.2$ \citep{Hilton21}, which is among the highest significance SZE
detection at $z>1$ by ACT. As of this writing, deep X-ray observations for this cluster are unavailable.

Throughout the paper, we adopt a standard set of cosmological density parameters, $\Omega_{\rm
M}=0.3$ and $\Omega_{\rm \Lambda}=0.7$. We use the
dimensionless Hubble constant $h_{70}\equiv H_0/(70
\mbox{km/s/Mpc})$; given existing tensions in  the value of $H_0$
(e.g., \cite{Verde19}) the parameter is unconstrained unless otherwise indicated.  In
this cosmology, the angular size of 1$''$ corresponds to the physical
sizes of 7.79 $h_{70}^{-1}$kpc and 8.19 $h_{70}^{-1}$kpc at $z=0.90$ and
$z=1.11$, respectively. These physical sizes are
insensitive to the values of density parameters and reduce by 1.1\% and
1.2\% at $z=0.90$ and $z=1.11$, respectively, if $\Omega_{\rm M}=0.32$
and $\Omega_{\rm \Lambda}=0.68$ are adopted instead.  Unless otherwise stated, the errors are
given in 1$\sigma$ and the coordinates are given in J2000.

\section{Data and analysis}

\subsection{Millimeter: ALMA Band 3}
\label{sec-almadata}

\targeta\ and \targetb\ were observed by the 12-m and 7-m arrays of ALMA
in Band 3 (project codes 2019.1.00673.S and 2018.1.00680.S) as
summarized in table~\ref{tab-obs}.  The target fields, centered at
(\timeform{23h19m53.280s},~\timeform{0D38'13.400''}) and
(\timeform{9h47m58.565s},~\timeform{-1D20'05.780''}) for \targeta\ and
\targetb, respectively, have diameters of about $1.5'$ covered with 1
 central and 6 surrounding hexagonal mosaic pointings by both arrays.  An equal
spacing of 34.2$''$ between the pointings was adopted, yielding
approximately a Nyquist sampling for the 12-m array and much denser
sampling for the 7-m array.

The observations were executed over the periods listed in table~\ref{tab-obs}, during which the number of antennas varied slightly.  All
the data were taken at four continuum bands centered at 85, 87, 97, and
99 GHz, yielding the overall central frequency of 92 GHz 
($\lambda=3.3$ mm) with an effective bandwidth of 7.5 GHz. The most
compact configuration for the 12m array, C1, was adopted  to cover the overall baseline
ranges of 2.5--147 k$\lambda$ and 2.5--120 k$\lambda$ for \targeta\ and
\targetb, respectively, where $\lambda$ is the observed wavelength.

\begin{table*}[ht] 
 \caption{Summary of ALMA observations.} \label{tab-obs}
  \begin{center}
    \begin{tabular}{c|cc|cc}
     \hline
     Object & \multicolumn{2}{c|}{\targeta}  & \multicolumn{2}{c}{\targetb}  \\
     Array & 12-m & 7-m & 12-m & 7-m \\ \hline
     Project code &  \multicolumn{2}{c|}{2019.1.00673.S}  &
     \multicolumn{2}{c}{2018.1.00680.S}  \\
     Field center &
	 \multicolumn{2}{c|}{(\timeform{23h19m53.280s},~\timeform{0D38'13.400''})}  &
     \multicolumn{2}{c}{(\timeform{9h47m58.565s},~\timeform{-1D20'05.780''})}  \\
     Number of pointings & 7  & 7 & 7 & 7\\               
     Observation Start & 2019-11-15 & 2019-10-22 & 2019-01-14&
		     2018-11-28 \\
     Observation End  & 2019-11-22 & 2020-01-02 & 2019-01-20 &
     2019-05-13 \\
     Total on-source time [hr] & 12.1 &  78.6  & 8.0 & 64.5 \\
     Number of antennas & $43 - 47$ & $9 - 11$ & $46-51$ & $9-12$\\
     Central frequency [GHz] & 92 & 92 & 92 & 92\\
     Band widths [GHz] & 7.5 & 7.5  & 7.5 & 7.5\\  
     Baseline coverage [k$\lambda$]
     & $4.2 - 147$ & $2.5 - 16.3$ & $4.2-120$ & $2.5-16.3$\\
      \hline
    \end{tabular}
  \end{center}
\end{table*}
\begin{table*}[h] 
 \caption{Properties of the synthesized images from a range of baselines.
 }
 \label{tab-image}
  \begin{center}
    \begin{tabular}{c|cccc|cccc}
     \hline
     Object & \multicolumn{4}{c|}{\targeta}  & \multicolumn{4}{c}{\targetb}  \\
     Array or baseline range& $>15$k$\lambda$ &12-m & 
	     7-m & all$^*$ & $>15$k$\lambda$ & 12-m & 
	     7-m & all$^*$ \\ \hline
     Beam major axis FWHM [arcsec] &$3.14$ & $3.60$
	     & $18.7$ & $3.76$ & $3.16$ &$3.63$
	     & $17.8$ & $3.77$ \\ 
     Beam minor axis FWHM [arcsec] & $2.82$ & $3.25$ 
	     & $11.9 $ & $3.38$ & $2.71$ &$3.11$ 
	     & $11.2 $ & $3.22$ \\ 
     Beam position angle [deg] & $82.2$ & $82.6$ & $-86.1$  &$82.5$
     & $-90.0$ &$-89.1$ & $-84.4$  &$-89.1$  \\  
     Average 1$\sigma$ noise [$\mu$Jy/beam] 
     &$5.6$ & $5.0$ & $19.4$ & $4.8$
     & $5.8$ & $5.1$ & $21.0$ & $5.0$ \\ 
      \hline
    \end{tabular}
  \end{center}
   \begin{tabnote}
 $^*$ The 1$\sigma$ noise for all baselines after smoothing to $5''$
 FWHM is  5.8 $\mu$Jy/beam and 5.9 $\mu$Jy/beam for \targeta\ and \targetb, respectively. 
\end{tabnote}
\end{table*}

For both objects, we used the visibility data produced by the second
stage of ALMA's Quality Assurance process (QA2).  Imaging was done with
the Common Astronomy Software Applications package (CASA)
\citep{McMullin07,CASA2022} version 6.4.0. The procedure was similar to that
adopted by \citet{Kitayama16} and \citet{Kitayama20}. First, we
identified compact sources in the observing field using only the
baselines longer than 15k$\lambda$; the position and the flux density were determined in the $uv$ plane by the CASA task {\it uvmodelfit}.  To improve the signal-to-noise ratio, all
spectral channels were fitted together\footnote{ We checked that this simplification has no apparent effect on the results of the present paper. For the brightest source (C1 in table \ref{tab-source1}), adopting the flux density fitted separately at 85, 87, 97, and 99 GHz for subtraction would change the residual signal at the source position by less than 1/5 of the noise level.},  excluding the frequency ranges affected by line emissions; separation of line emissions is described in Appendix \ref{sec-lines}. 
All the compact sources detected at $>5\sigma$ (tables
\ref{tab-source1} and \ref{tab-source2}) were subtracted from the entire
visibility data.  Secondly, we performed image deconvolution with the
Multi-Scale CLEAN algorithm \citep{Cornwell08,Rich08,Steeb19} using the
CASA task {\it tclean}. Given similar sensitivity and source size
between \targeta\ and \targetb, we adopted a circular mask region with
a radius $50''$ and a flux threshold of 0.01 mJy (corresponding to $\sim
2\sigma$) for both objects. The choice of other parameters were
identical to \citet{Kitayama20}; we adopted [0, $4''$, $8''$, $16''$,
$32''$, $64''$] as the FWHMs of the Gaussian components, the
multi-frequency synthesis mode in joint mosaic imaging, and natural
weighting. All the ALMA images presented in this paper are corrected for primary beam attenuation (e.g., \cite{Mason20}) and have the pixel size of $0.5"$. 

Table \ref{tab-image} lists the parameters of the synthesized beams as
well as the $1\sigma$ noise levels of the synthesized image within
$45''$ from the field center. To eliminate
large-scale variation of the data caused by the SZE, the noise levels
were measured on difference maps created after subtracting the compact
sources  (tables \ref{tab-source1} and \ref{tab-source2});  the visibility data was divided into the first and the second half along the time
sequence of observations, the sign of the latter was flipped, concatenated with the former, and the resultant was inverse Fourier transformed into the image domain.

For display purposes, we also present the images smoothed by a Gaussian
filter to an effective beam size of $5''$ FWHM. The root-mean-square (RMS) noise level
measured on the above mentioned difference map smoothed to the $5''$
resolution is  5.8 $\mu$Jy/beam and 5.9 $\mu$Jy/beam for \targeta\ and \targetb, respectively.   These 
RMS values are only used for characterizing the significance levels of the
SZE signal on the smoothed images. Unless otherwise stated, quantitative
analysis in this paper is done on the unsmoothed images created
from all baselines whose characteristics are listed in table~\ref{tab-image}.

\subsection{Millimeter: ACT}

We extracted the ACT DR5 data of \targeta\ and
\targetb.  The ACT DR5 cluster search used the 98, 150 GHz maps made
from all ACT data obtained between 2008 and 2018, including both day and
night time observations (see \cite{Naess20} for details of the data
products used). The approximate beam FWHMs are $2.2'$ and $1.4'$ at 98
GHz and 150 GHz, respectively. The S/N of the ACT maps presented in this
paper is measured at the fixed filter scale of $2.4'$ as described in
\citet{Hilton21}.

\subsection{X-ray: \Chandra \ ACIS-S}

Of the two clusters studied in this paper, deep X-ray data are
 available only for \targeta. We extracted four datasets taken
 in 2005 by
 \Chandra\ ACIS-S for this cluster (ObsID: 5750, 7172, 7173, and 7174).
 After excluding the periods with high background rates, the total net
 exposure time is 69.7 ks. The data were processed with CIAO version
 4.13 \citep{Fruscione06} and the Calibration database (CALDB) version
 4.9.6.  The backgrounds were estimated from the off-center region at
 $\theta > 3.2'$ from the emission peak of this cluster, where the ICM
 emission is negligible. Exposure-corrected and background-subtracted
 data at observed energies $E_{\rm obs} = 0.4- 7.0$ keV were used
 throughout our analysis; for display purposes only,
 we applied adaptive smoothing to the brightness image including
 backgrounds using the task {\it fadapt} implemented in FTOOLS\footnote{https://heasarc.gsfc.nasa.gov/ftools/} 
 \citep{Ftools95,Ftools99,Ftools14}. Spectral fitting was done with XSPEC version 12.12.0
 \citep{Arnaud96}, assuming that the ICM is in collisional ionization
 equilibrium and the metal abundance $Z$ is 0.3 times the solar value
 given by \citet{Anders89}. The source redshift and the Galactic
 hydrogen column density were fixed at $z=0.90$ and $N_{\rm H}=4.2\times
 10^{20}$ cm$^{-2}$ \citep{HI4PI16}, respectively. We fixed the helium
 mass fraction at $Y=0.25$, which is nearly unchanged between the
 primordial gas and the solar photosphere (e.g.,
 \cite{Asplund09,Planck20}).

\subsection{Optical: Subaru Hyper Suprime-Cam}

\targeta\ and \targetb\ were both observed by five broad-band
filters $grizy$ \citep{Kawanomoto18} in the HSC-SSP. We used the CAMIRA
cluster catalogue \citep{Oguri18} updated for the third public data
release of HSC-SSP \citep{Aihara22} based on the associated bright star
masks \citep{Coupon18}.

   \begin{figure*}[ht]
    \begin{center}
       \includegraphics[width=16.0cm]{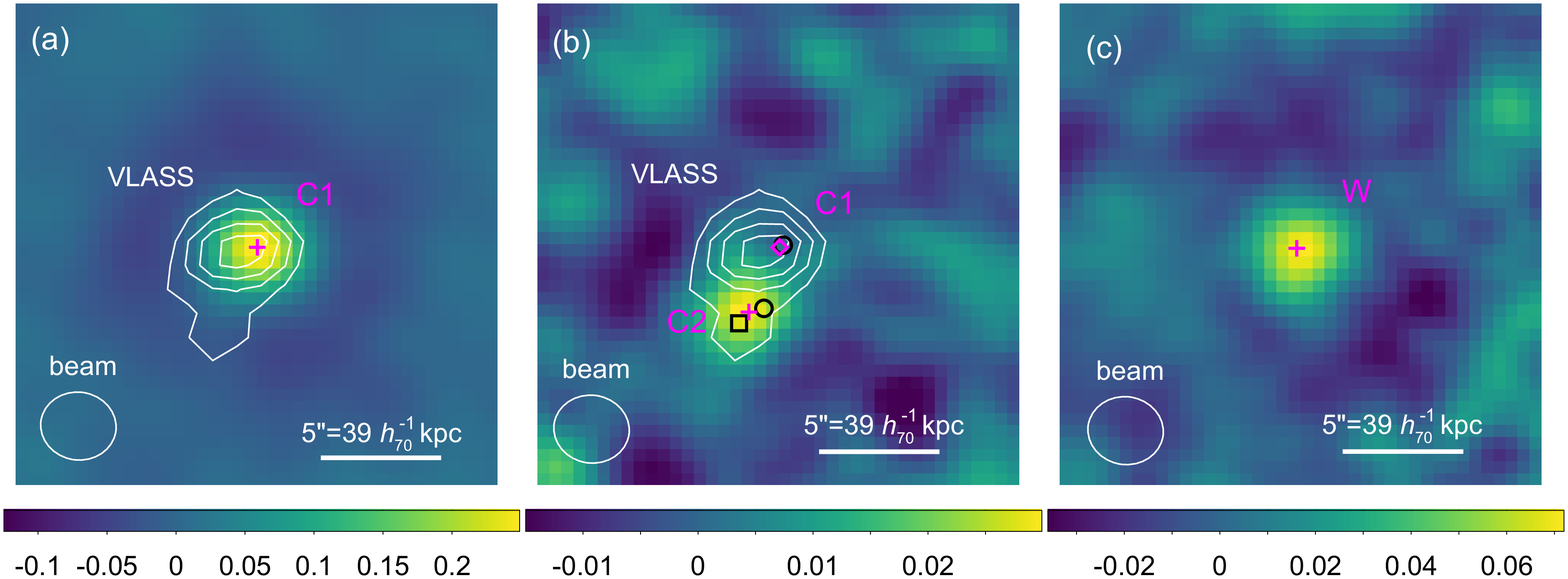}    
    \end{center}       
	 \begin{center}
	  \includegraphics[width=16.8cm]{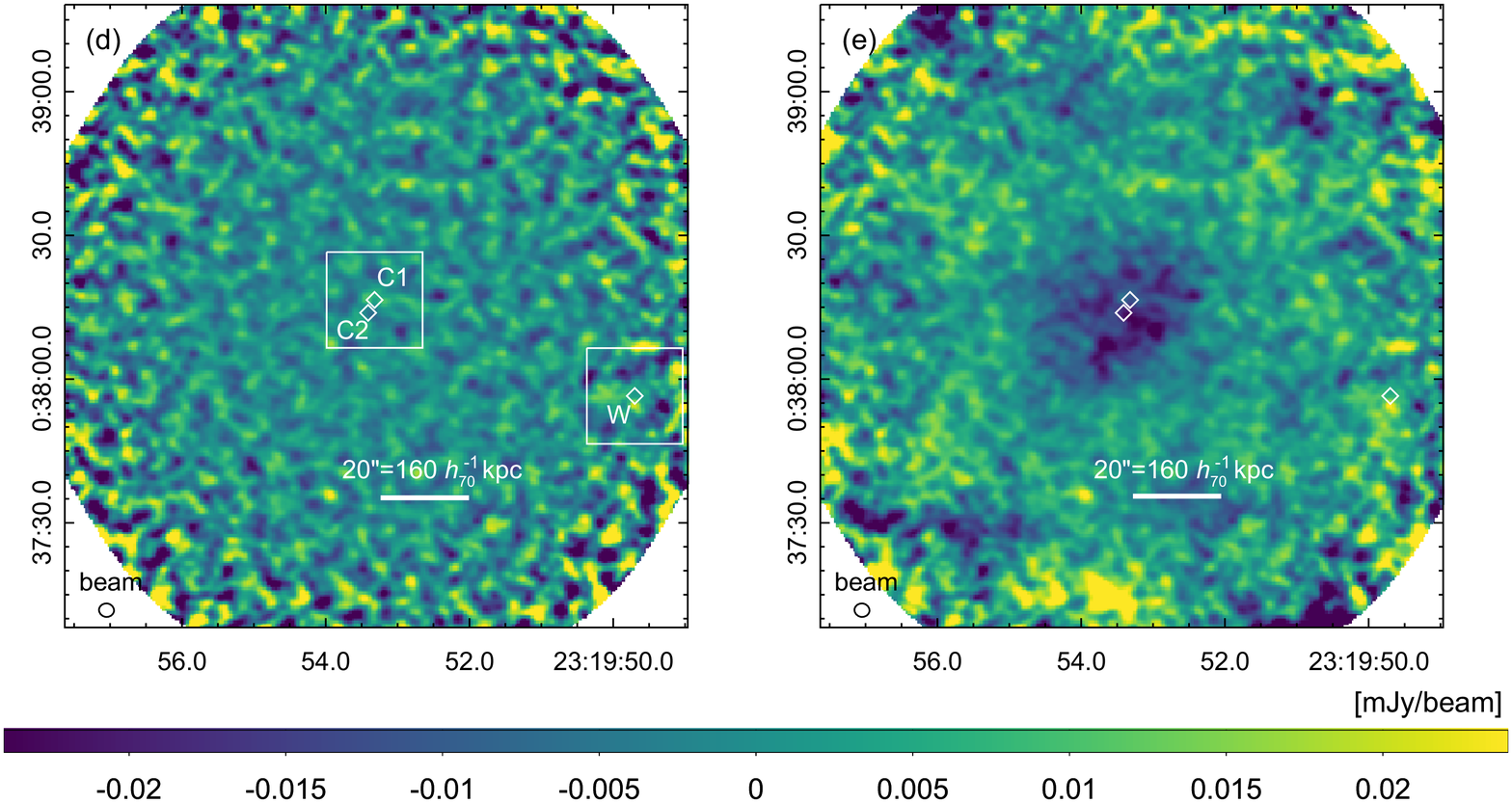}
	 \end{center}
   \vspace*{3mm}
    \caption{Dirty maps toward \targeta\ from only long-baselines at
    $>15$k$\lambda$ (panels a--d) and all baselines at $>2.5$k$\lambda$
    (panel e).  The regions shown in panels a--c are indicated by boxes
    in panel d.  The positions of $>5\sigma$ sources before and after
    subtraction are marked by crosses and diamonds, respectively.  The
    synthesized beam shape of ALMA is shown at the bottom-left in each
    panel.  (a) The central $20''\times 20''$ region before the sources
    are subtracted. Contours show the 3.0 GHz intensity (90, 70,
    50, 30 \% of the peak value) from the VLASS Epoch 1.1 image with the
    synthesized beam FWHMs of $2.8'' \times 2.2''$ and the position
    angle of $7.5^\circ$.  (b) Same as panel a, but after the brightest
    source, C1, is subtracted.  The position of an
    optical counterpart candidate (see text) is marked by a circle
    (or a box if the candidate is the BCG).  (c) The region
    around source W.  (d) Long-baseline image after sources C1, C2, and
    W are subtracted. (e) Similar to panel d, but produced from all
    baselines. } \label{fig-source}
   \end{figure*}

Weak lensing analysis was done on the HSC-SSP S19A
data\footnote{S19A is an internal data release between the second and
the third public data releases of HSC-SSP.}, following the method
described in \citet{Okabe19} and \citet{Okabe21}.  The galaxy
shapes were measured using the re-Gaussianization method
\citep{Hirata03} implemented in the HSC pipeline
\citep{Mandelbaum18,Li22}. The background galaxies behind each cluster
were selected using the colour--colour selection following
\citet{Medezinski18}. The S/N of the resulting surface mass density is
computed as in \citet{Okabe19}. We adopted the NFW density profile
\citep{Navarro96} for estimating the deprojected halo mass, from the
tangential shear profiles; given the low S/N ($\sim 2-3 $) of the
lensing signal, the halo concentration is linked to the mass by the
relation of \citet{Diemer15}.

  \begin{table*}[ht]
  
     \caption{Positions and continuum flux densities of point sources toward
   \targeta\ obtained with the CASA task {\it uvmodelfit}. For source W, the frequency ranges affected by line emissions (table \ref{tab-lines}) are excluded in the fit. The errors in the positions are less than $0.2''$.}
  \begin{center}
    \begin{tabular}{cccc}
     \hline
     Source ID & RA (J2000) & Dec (J2000) & 92 GHz flux density [$\mu$Jy]\\ \hline 
     C1 & \timeform{23h19m53.32s} &\timeform{0D38'16.52''}
	     & $262.9 \pm 3.9$ \\
     C2 & \timeform{23h19m53.41s} &\timeform{0D38'13.83''}
		 & $36.6 \pm 3.9$ \\
     W & \timeform{23h19m49.70s} &\timeform{0D37'56.48''}
	     & $76.8 \pm 6.1$ \\     
     \hline
    \end{tabular}
  \end{center}
 \label{tab-source1}
  \end{table*}
  \begin{table*}[h]
  
   \caption{Same as Table \ref{tab-source1}, but for \targetb. For source W1, the frequency range affected by a line emission (table \ref{tab-lines}) is excluded in the fit.} 
  \begin{center}
    \begin{tabular}{cccc}
     \hline
     Source ID & RA (J2000) & Dec (J2000) & 92 GHz flux density
     [$\mu$Jy]\\ \hline 
     W1 & \timeform{9h47m55.46s} &\timeform{-1D20'26.98''}
	     & $35.3 \pm 4.9$ \\
     W2 & \timeform{9h47m57.50s} &\timeform{-1D19'51.77''}
		 & $42.2 \pm 4.5$ \\
     \hline
    \end{tabular}
  \end{center}
 \label{tab-source2}
  \end{table*}

\section{Results}

\subsection{Compact millimeter sources}
\label{sec-source}

\subsubsection{\targeta}
\label{sec-source1}

There are three compact sources above the $5\sigma$ significance level
at 92 GHz in our target field toward \targeta\ as shown in figure~\ref{fig-source}(a)--(d). Only compact sources, not the SZE, are visible on the images
from long-baselines. All the detected sources are consistent with being
point-like and their properties are summarized in table~\ref{tab-source1}.

   \begin{figure*}[h]
    \begin{center}
 \includegraphics[width=11.6cm]{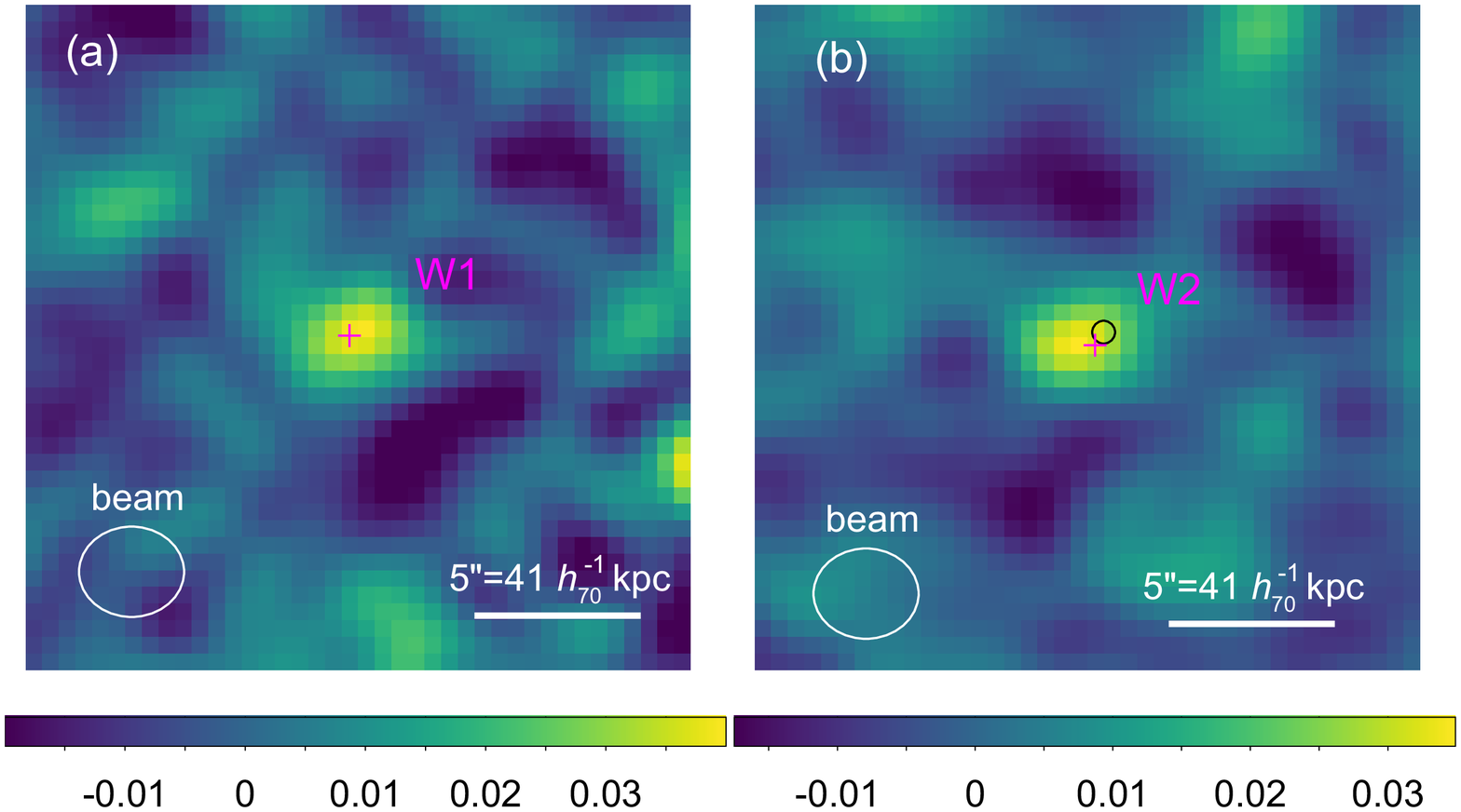}
    \end{center}       
	 \begin{center}
	  \includegraphics[width=16.8cm]{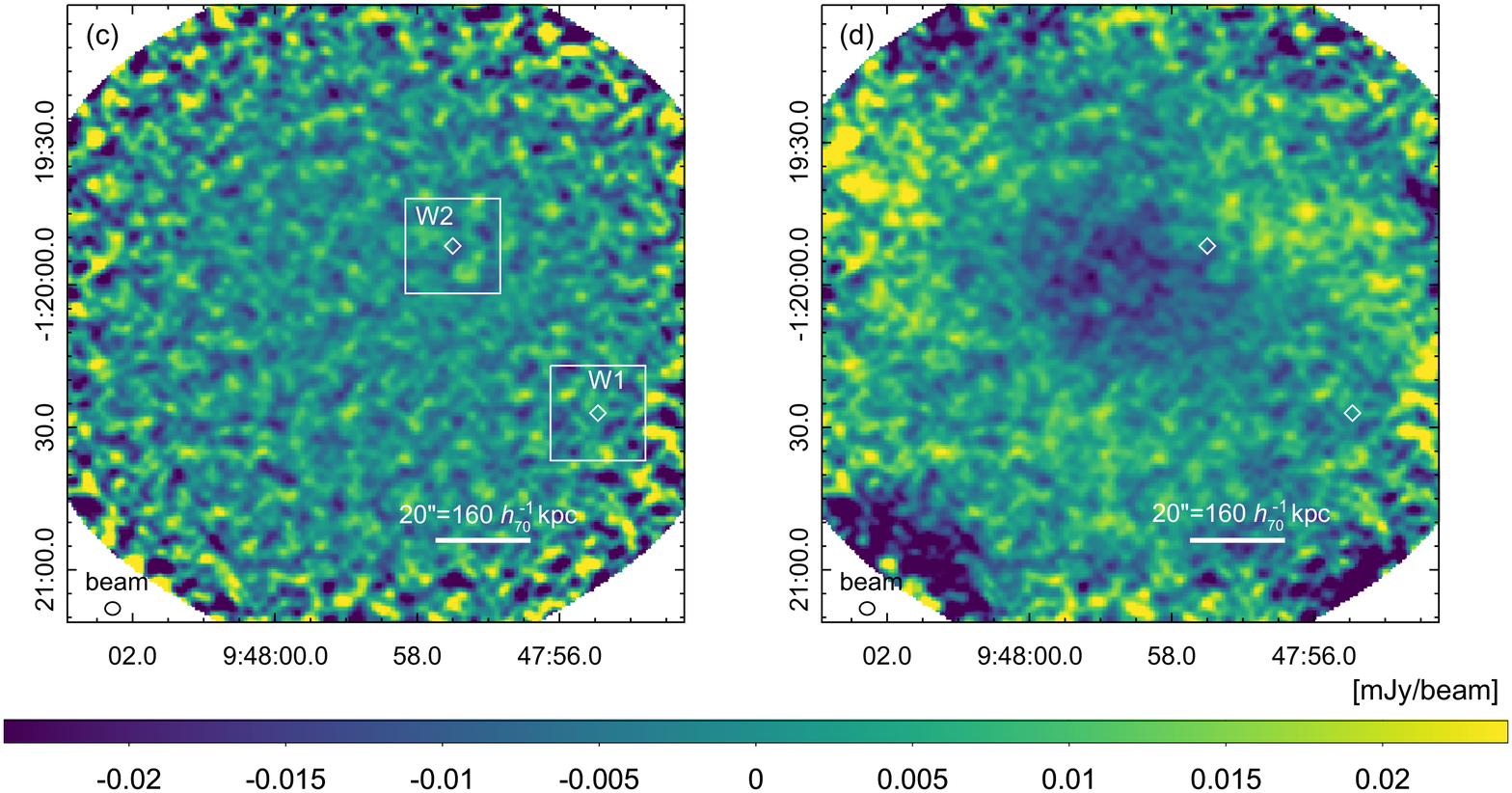}
	 \end{center}
   \vspace*{3mm}
    \caption{Similar to figure \ref{fig-source} but toward \targetb. The
    regions shown in panels a and b are indicated by boxes in panel c.
    (a) Long-baseline ($>15$k$\lambda$) image of the region around
    source W1. (b) Similar to panel a, but for the region around source
    W2.  The position of an optical counterpart
    candidate (see text) is marked by a circle.  (c) Long-baseline
    image after sources W1 and W2 are subtracted. (d) Similar to panel
    c, but produced from all baselines. } \label{fig-source2}
   \end{figure*}

Two sources (C1 and C2) are located within $5''$ from the X-ray center
of \targeta. There is a galaxy detected by HSC at the photometric
redshift of $z_{\rm phot}= 0.92 \pm 0.04$ within $0.2''$ from C1.  Two
galaxies at $z_{\rm phot}= 0.91 \pm 0.03$ and $0.92 \pm 0.02$ lie at the
projected distance of  $0.6''$ from C2; 
the latter is the Brightest Cluster Galaxy (BCG) of
\targeta. Both C1 and C2 are bright at lower frequencies and
detected in the Faint Images of the Radio Sky at Twenty-cm (FIRST;
\cite{Becker95}) as well as the Very Large Array Sky Survey (VLASS;
\cite{Gordon21}).  Their combined flux listed in the FIRST Catalog
Database\footnote{http://sundog.stsci.edu/} is $4.59
\pm 0.22$ mJy at 1.4 GHz, where we have estimated the error from the
image of this region available at the FIRST Cutout
Server\footnote{https://third.ucllnl.org/cgi-bin/firstcutout}.  The
flux measured on the VLASS Epoch 1.1
image\footnote{http://cutouts.cirada.ca/} within a diameter of $15''$
around the position of C1 is $4.9 \pm 0.5$ mJy at 3.0 GHz. Figure
\ref{fig-source}(a)(b) indicates that the peak intensity ratio between
C1 and C2 is about $2.3 : 1$ and $ 7.4 : 1$ at 3.0 GHz and 92 GHz,
respectively, implying that C2 has a much steeper spectrum than C1.
Note that the dust emission can dominate the source flux at 92 GHz,
corresponding to the rest-frame frequency of $175$ GHz at $z=0.90$.  The
luminosity of C1 and C2 is $\nu L_\nu = (9.73 \pm 0.14) \times 10^{41}
h_{70}^{-2}$ erg/s and $(1.32 \pm 0.14) \times 10^{41}h_{70}^{-2}$
erg/s, respectively, at $\nu=175$ GHz in the rest-frame of the cluster.

Another source (W) is located at $\sim 1'$ from the X-ray center of
\targeta.  This source hosts bright line emissions as described in Appendix \ref{sec-lines}.  It has no obvious optical
counterpart within $1''$ and is undetected at 1.4 GHz in the FIRST
and VLASS images.

The above sources are removed from the visibility data in our subsequent
analysis. Figure \ref{fig-source}(d)(e) shows that the residuals at long
baselines are consistent with noise, whereas the extended signal from
the ICM becomes apparent once shorter baselines are included.

\subsubsection{\targetb}
\label{sec-source2}

There are two compact sources above the $5\sigma$ significance level at
92 GHz in our target field toward \targetb\ (figure \ref{fig-source2}
and table \ref{tab-source2}).
All the detected sources are consistent with being point-like and
removed from the visibility data in our subsequent analysis.

One source (W1) lies at $\sim 1'$ south-west of the ALMA SZE center. This source appears to host a bright line emission as described in Appendix \ref{sec-lines}. 
There is a compact object classified as a star in
the SDSS Data Release 17
\citep{SDSS17}\footnote{http://skyserver.sdss.org/dr17} at $0.6''$ from
this source. There is no other object detected by HSC, VLASS, or FIRST
within a projected distance of $1''$ from W1.

Another source (W2) is located at $\sim 20''$ north-west of the ALMA SZE
center. There is a galaxy detected by HSC at the
photometric redshift of $z_{\rm phot} = 0.9 \pm 0.5$ lying at $0.5''$
from this source. It is undetected by VLASS or FIRST.
The large uncertainty in $z_{\rm phot}$ of this
object is due to degeneracy of the spectral energy distribution (SED)
modeling of low-$z$ and high-$z$ galaxies in the five-band photometric
space \citep{Tanaka15}.

Figure \ref{fig-source2}(c)(d) further demonstrates that the above sources
are subtracted successfully and do not affect the extended signal from
the ICM.

\begin{figure*}[tp]
 \begin{center}
\includegraphics[height=9.3cm]{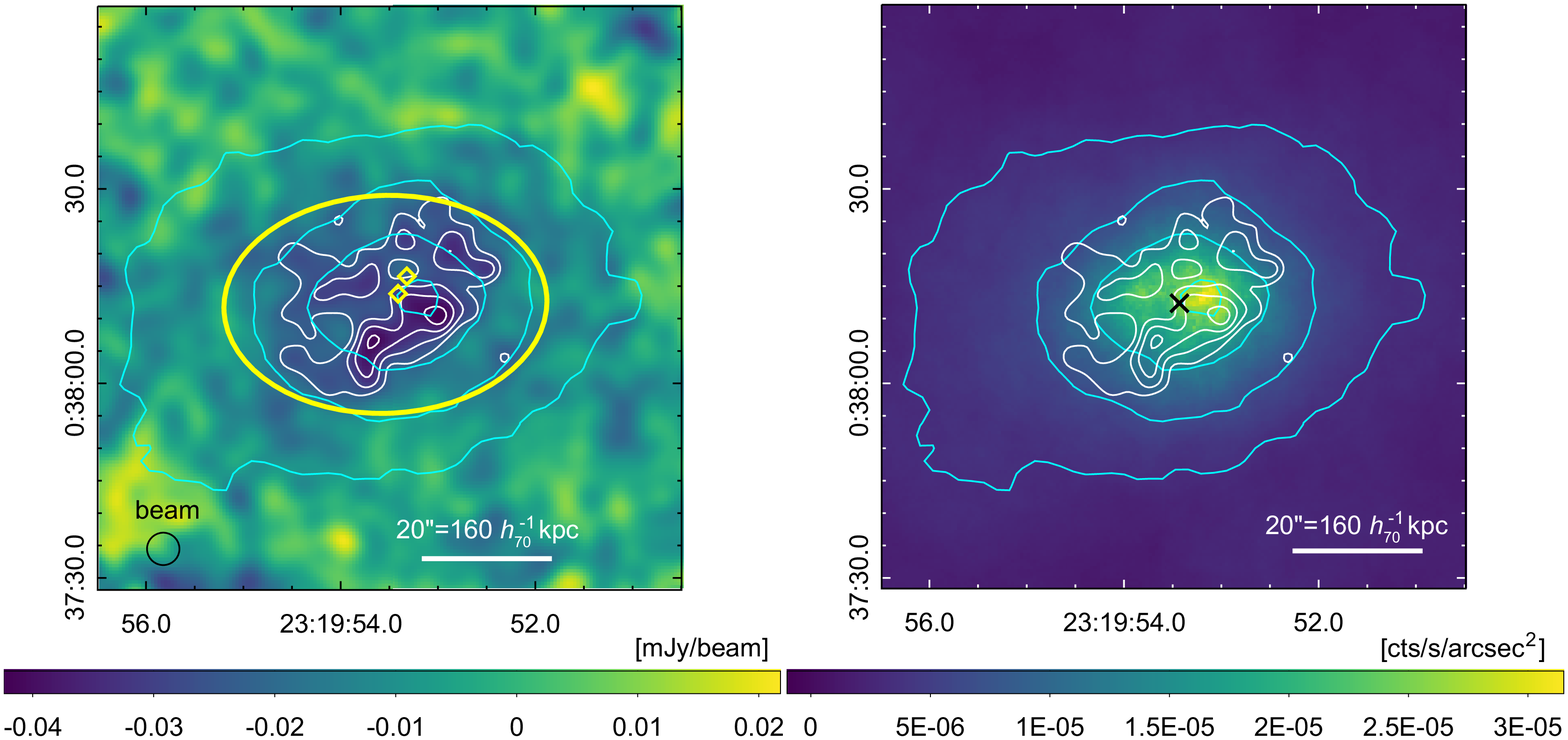}  
  \includegraphics[width=8.4cm]{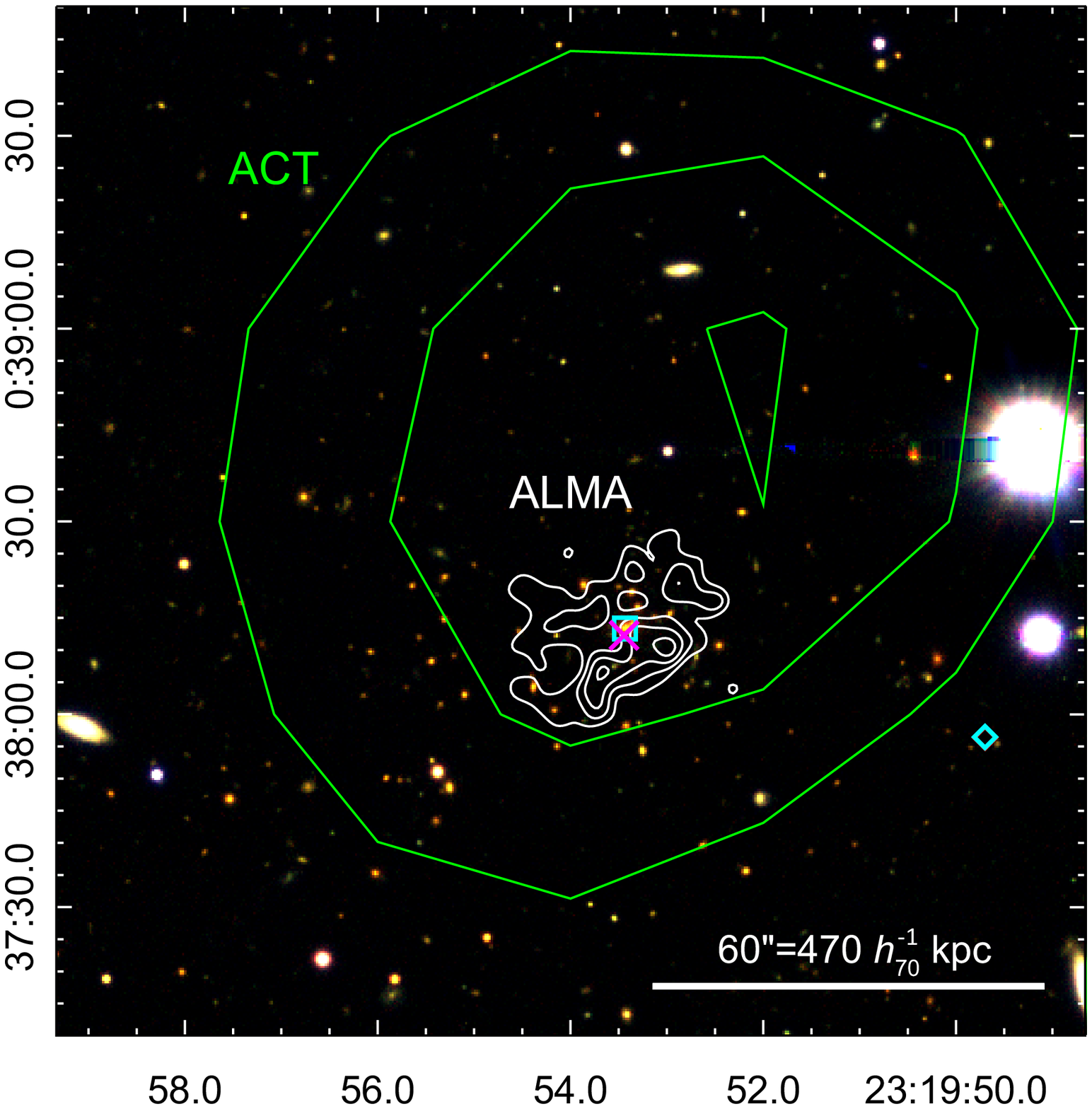}    
  \includegraphics[width=8.4cm]{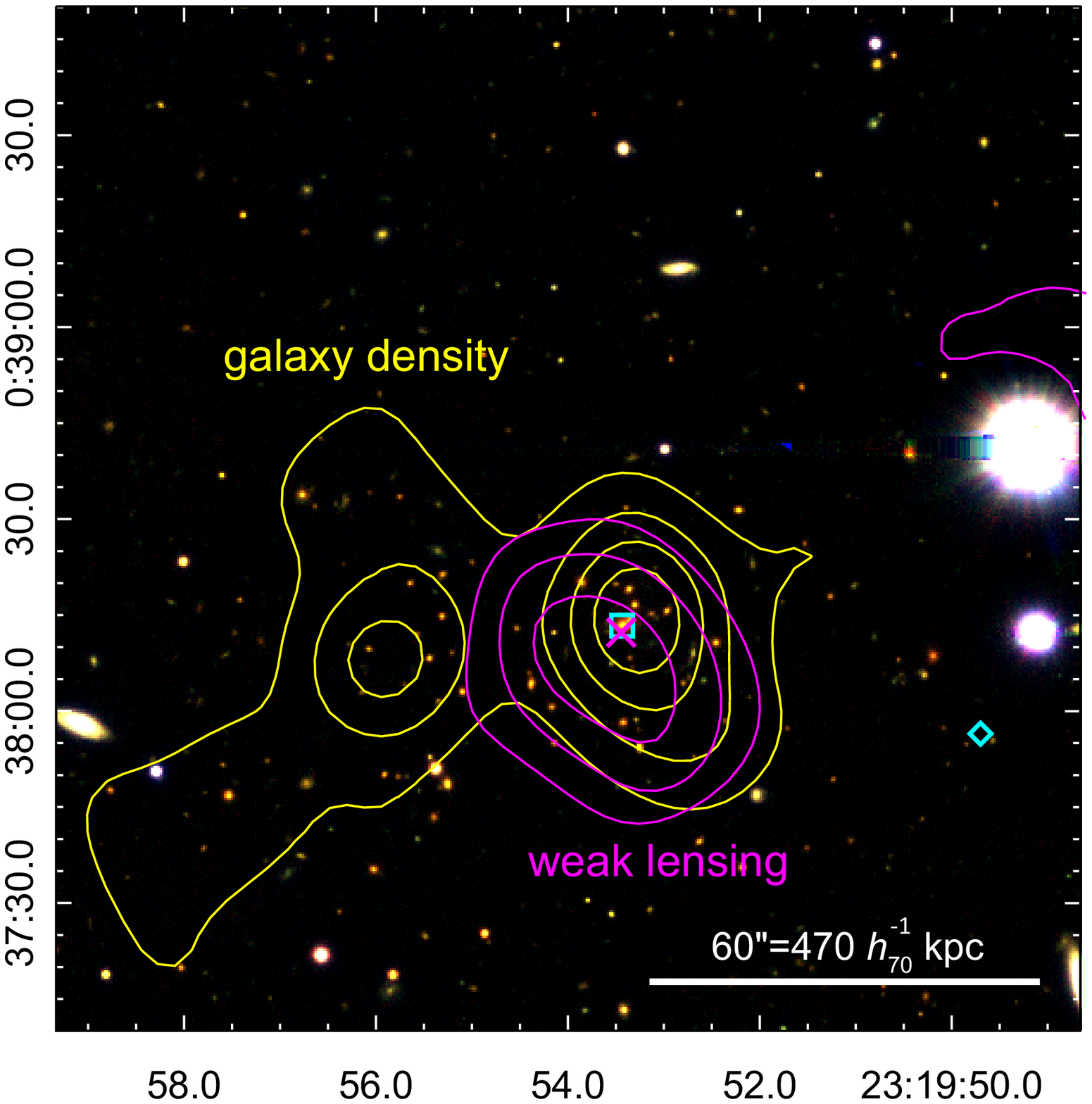}
 \end{center}
 \caption{Multi-wavelength view of \targeta\ at $z=0.90$.  Top left:
 Deconvolved ALMA SZE image at the central frequency of 92 GHz smoothed
 to have a beam size of $5''$ FWHM.  The ellipse shows the FWHM
 location of the best-fit elliptical Gaussian profile (table
 \ref{tab-gaussfit}), and the diamonds indicate the positions of
 subtracted sources.  Top right: \Chandra\ 0.4--7.0 keV X-ray brightness
 image adaptively smoothed with a circular top-hat filter that contains
 at least 100 photons. The cross marks the X-ray center defined in the
 text. Bottom left and bottom right: A wider-field HSC-SSP optical
 $riz-$color images with the X-ray center marked by a cross.
 The positions of the BCG and the off-center source W
 are indicated by a box and a diamond, respectively. Wherever
 plotted, white contours show the significance levels of the ALMA SZE
 image ($4-7\sigma$ in increments of $1\sigma = 5.8~\mu$Jy/beam), cyan 
 contours the brightness of the \Chandra\ X-ray image ($80\%,~40\%, ~
 20\%, ~10\%$ of the peak value), green contours the significance
 levels of the SZE map from the ACT DR5 cluster
 catalogue (S/N = 3, 4, and 5), yellow contours the surface density of
 probable member galaxies averaged by a Gaussian with $20''$ FWHM
 ($80\%,~60\%, ~ 40\%, ~20\%$ of the peak value), and magenta contours the
 projected weak lensing mass smoothed by a Gaussian with $30''$ FWHM
 (S/N = 1.5, 2, 2.5, and 3).}  \label{fig-xszJ2319}
\end{figure*}

  \begin{table*}[ht]
   \caption{The results of an elliptical Gaussian fit to the ALMA image
   within a radius of $50''$ from the field center by the CASA task {\it imfit}. The errors in the
   central positions are less than $2.1''$.}

  \begin{center}
    \begin{tabular}{ccccccc}
     \hline
      &  \hsp RA \hsp & \hsp \hsp Dec \hsp \hsp &  \hsp \hsp  major axis FWHM \hsp &
     \hsp minor axis FWHM  \hsp
     & \hsp position angle \hsp & \hsp Flux density \hsp \\
     & (J2000)& (J2000)& [arcsec] & [arcsec] & [deg] & [mJy]\\ \hline 
     \hsp \targeta \hsp & \timeform{23h19m53.54s} &\timeform{0D38'12.19''}
	     & $49.8 \pm 4.8$ &  $33.6 \pm 3.2$ & $92.0 \pm 9.7$ & $-2.04 \pm 0.20$\\
     \targetb & \timeform{9h47m58.74s} &\timeform{-1D19'58.37''}
		 & $58.6 \pm 3.5$ & $38.7 \pm 2.3$ & $72.3 \pm 5.5$ & $-4.12 \pm 0.24$\\
     \hline
    \end{tabular}
  \end{center}
 \label{tab-gaussfit}
  \end{table*}
  \begin{table*}[tp]
   \caption{The results of an elliptical $\beta$ model fit using equations (\ref{eq-ellipbeta}) and (\ref{eq-ellipbeta2}) to the
   \Chandra\ brightness data of \targeta\ with the Sherpa package.  The errors in the central
   positions ($x_0$ and $y_0$) are less than $0.6''$.}
  \begin{center}
    \begin{tabular}{ccccccc}
     \hline
     $x_0$ & $y_0$ & $S_{\rm 0}$ & $\beta$ 
     & $\theta_{\rm c}$ 
     & $q$  & $\psi$
     \\
     (J2000)& (J2000) 
	     &[counts/s/arcsec$^2$] & 
		 & [arcsec] & & [deg]  \\ \hline 
      \timeform{23h19m53.45s} &\timeform{0D38'12.29''}
		 & $(2.64^{+0.21}_{-0.20})\times 10^{-5}$ 
	     & $0.673^{+0.034}_{-0.032}$
	 & $13.2^{+1.4}_{-1.2}$   		 
			 & $0.694^{+0.036}_{-0.034}$ 
			     & $96.7 \pm 3.9$ \\
     \hline
    \end{tabular}
  \end{center}
 \label{tab-betafit}
  \end{table*}

\subsection{The intracluster medium}
\label{sec-icm}

\subsubsection{\targeta}

Figure \ref{fig-xszJ2319} shows the deconvolved ALMA image of \targeta\
after compact sources listed in table~\ref{tab-source1} are removed.
The image has been smoothed to an effective beam size of $5''$ FWHM for
display purposes. To obtain the overall morphology of the signal, we
fit the unsmoothed ALMA image with an elliptical Gaussian using the CASA task {\it imfit}, varying its
center, major and minor axis FWHMs, and the position angle\footnote{In
this paper, the position angle is measured for the major axis of an
ellipse from north ($0^\circ$) through east ($90^\circ$).} (table
\ref{tab-gaussfit}); errors of fitted parameters are estimated based on \citet{Condon97}, assuming a constant noise level within a radius of $50"$ from the field center. The extended SZE signal is
detected at high significance with ALMA with a mean size of $\sim
40''$ FWHM and an axis ratio of $\sim 0.7$. The integrated flux density
within a radius of $45''$ from the position shown in table
\ref{tab-gaussfit} is $-1.98 \pm 0.10$ mJy;
for reference, the flux inferred from the elliptical Gaussian fit is
$-2.04 \pm 0.20$ mJy.

The figure also shows the \Chandra\ X-ray brightness image of \targeta\
adaptively smoothed with a circular top-hat filter that contains at
least 100 photons. We fit  
the unsmoothed X-ray brightness with an elliptical $\beta$ model of the form
\begin{eqnarray}
 S(\vec{\theta})
  = S_0 \left[ 1 + \left(\frac{\bar{\theta}}{\theta_{\rm
		    c}}\right)^2 
		       \right]^{-3\beta + \frac{1}{2}},  
  \label{eq-ellipbeta}
\end{eqnarray}
with the position vector on the sky $\vec{\theta} \equiv (x,y)$ related
to the ``circular mean distance'' $\bar{\theta} \equiv \sqrt{\bar{x}^2 +
\bar{y}^2}$ from the emission center $(x_0, y_0)$ by
\begin{eqnarray}
 \pmatrix{\bar{x} \cr \bar{y}} &\equiv &\pmatrix{\frac{1}{\sqrt{q}} & 0 \cr 0 & \sqrt{q}}
\pmatrix{\cos \psi &  \sin \psi \cr 
 - \sin \psi & \cos \psi} \pmatrix{x-x_0 \cr y-y_0},  
  \label{eq-ellipbeta2}
\end{eqnarray}
where $S_0$ is the brightness at $(x_0, y_0)$, $\theta_{\rm c}$ is the
angular core radius, $q$ is the minor-to-major axis ratio ($q \le
1$) on the sky, $\psi$ is the position angle, and $\bar{x}$
is intended to align with the minor axis of the ellipse. We used the Sherpa modeling package ({\it beta2d}) in CIAO \citep{Sherpa01,Sherpa07,Sherpa09} for this purpose. 
The best-fit values of $x_0$, $y_0$, $S_0$, $\beta$, $\theta_{\rm c}$,
$q$, and $\psi$ are listed in table~\ref{tab-betafit}.

The overall morphology of the SZE on the sky is in agreement with
that of X-rays; the emission center also matches within $1.5''$ between
two images. Given this agreement and the higher angular resolution
($\sim 0.5''$) of \Chandra, we refer to the best-fit position of the
X-ray emission center in table~\ref{tab-betafit} as the center of
\targeta\ in the rest of this paper.  The intrinsic
X-ray luminosity within $15''$ and $150''$ from the center is $L_{\rm
X}=(2.17 \pm 0.09)\times 10^{44} h_{70}^{-2}$ erg/s and $(4.42 \pm
0.30)\times 10^{44} h_{70}^{-2}$ erg/s, respectively, at $E_{\rm obs} =
0.4-7.0$ keV.  Further comparison by means of three dimensional gas
model will be discussed in sections \ref{sec-sim} and \ref{sec-yprof}.

Figure \ref{fig-xszJ2319} also shows that the centroid of the ACT
SZE map of \targeta\ is offset from the centers of the ALMA and
\Chandra\ images by $\sim 30''$ (230 $h_{70}^{-1}$kpc).  We will discuss the
possible origin of this offset in section \ref{sec-act}.

We plot in figure~\ref{fig-szprof} azimuthally averaged intensity
profiles around the cluster center in four quadrants with position
angles of $315^{\circ} \sim 45^{\circ}$ (north), $45^{\circ} \sim 135^{\circ}$
(east), $135^{\circ} \sim  225^{\circ}$ (south), and $225^{\circ} \sim 
315^{\circ}$ (west). The statistical error in each bin is computed using
equation (1) of \citet{Kitayama16}.

Figures \ref{fig-xszJ2319} and \ref{fig-szprof} suggest that the ICM is
disturbed near the center of \targeta. The SZE signal tends to be
stronger in the south and in the west of the center; the significance of
departures from the azimuthal average is at $\sim 2 \sigma$ (see section
\ref{sec-diff}). At $\theta \simgt 20''$,
elongation in the east-west direction becomes more obvious. The X-ray
emission peak is also offset from the cluster center,
defined by fitting the global emission profile, by
$\sim 4''$ or $\sim 32 h_{70}^{-1}$ kpc.

\begin{figure*}[tp]
 \begin{center}
 \includegraphics[height=9.0cm]{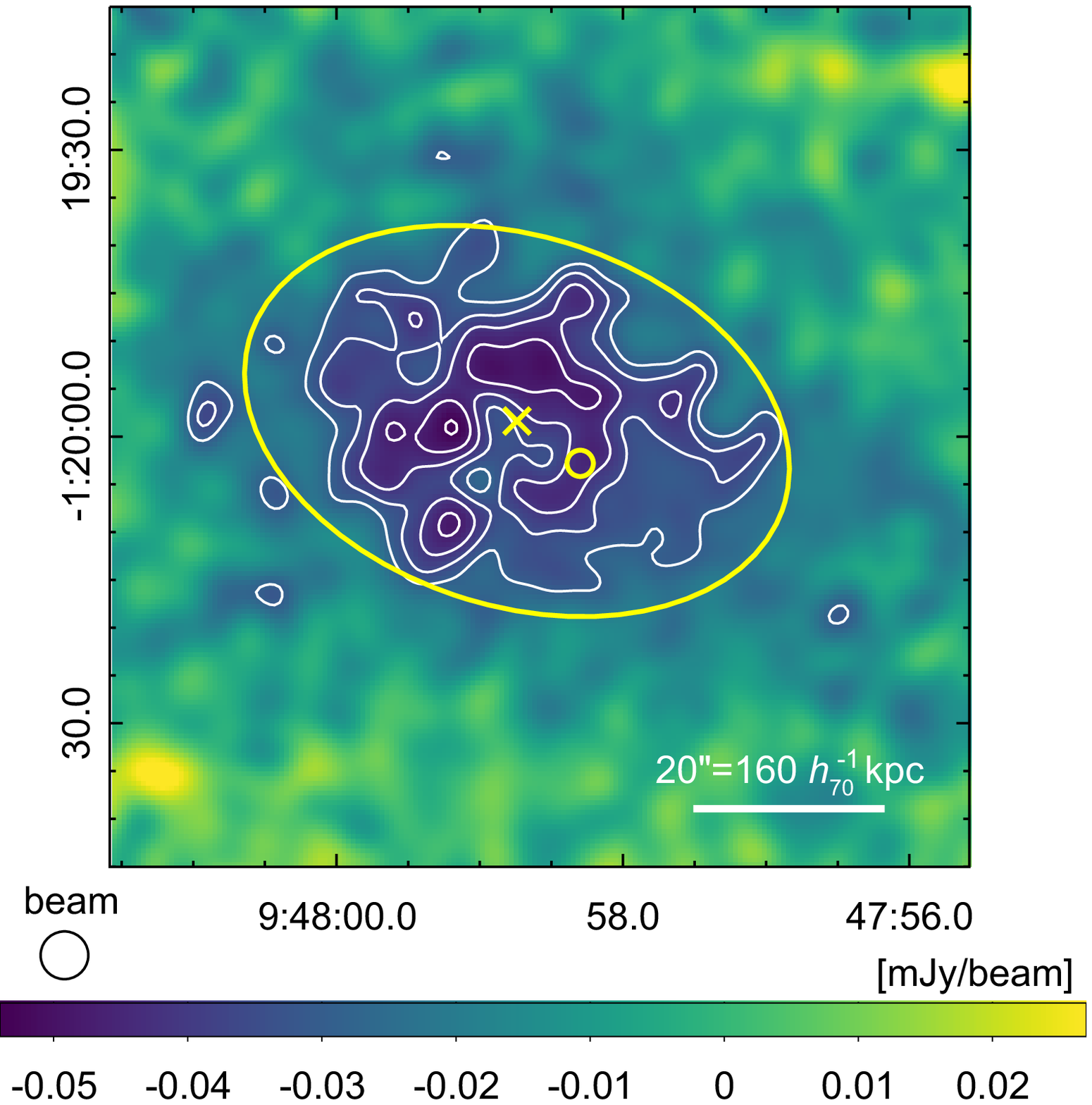} \\
  \includegraphics[width=8.4cm]{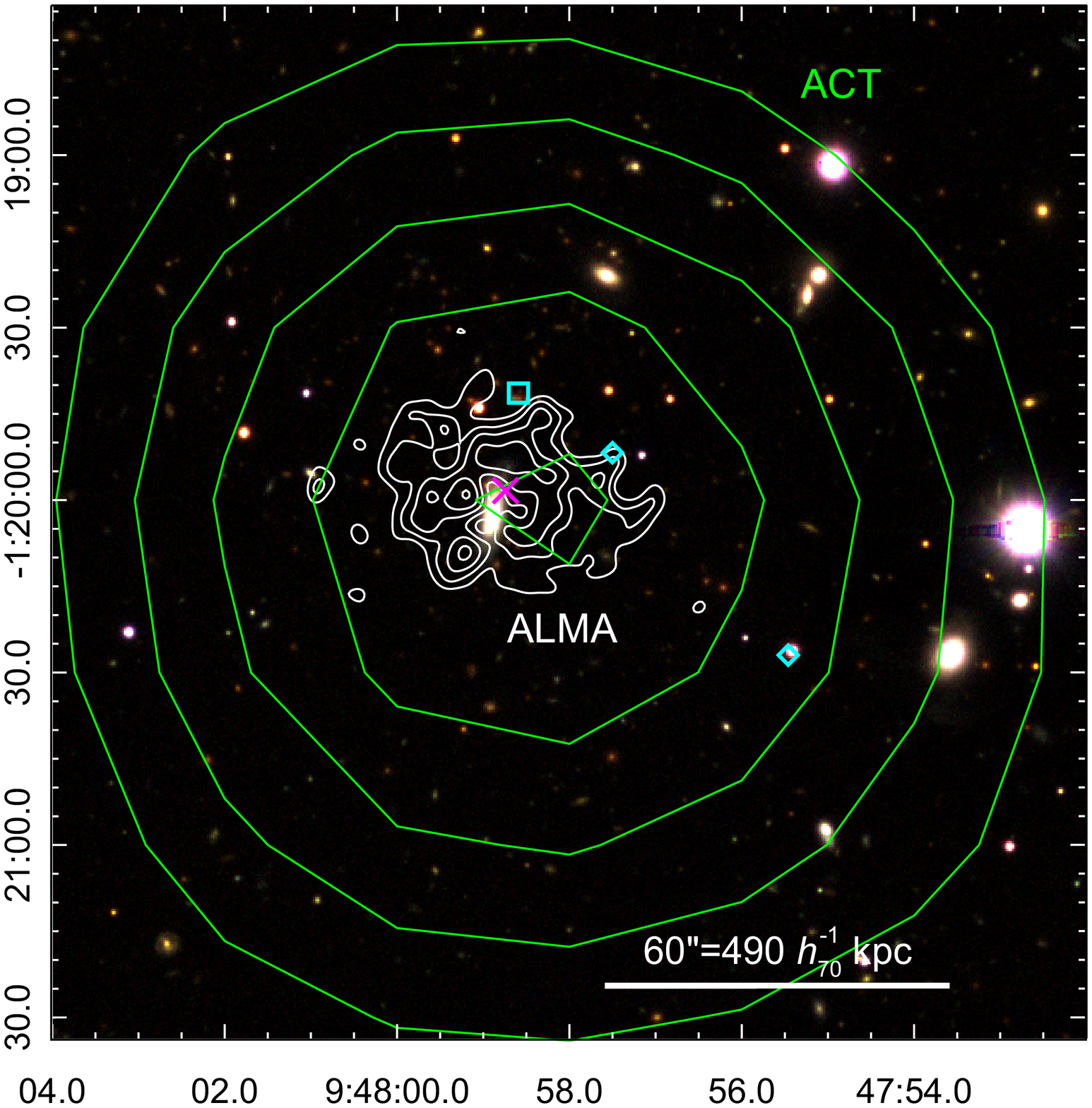}  
  \includegraphics[width=8.4cm]{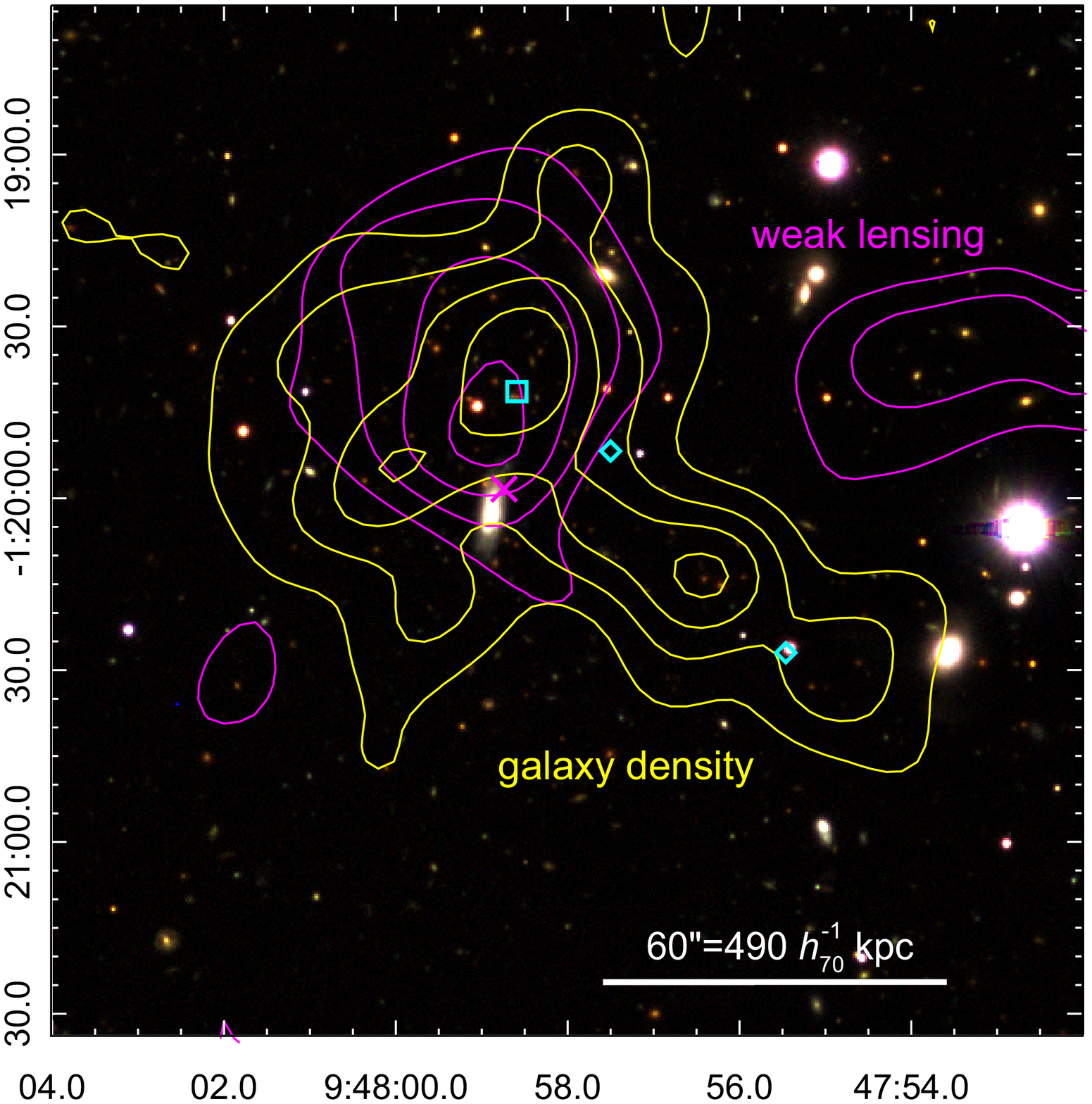}
 \end{center}
 \caption{Multi-wavelength view of \targetb\ at $z=1.11$.  Top:
 Deconvolved ALMA image at 92 GHz smoothed to have a beam size of $5''$
 FWHM.  Symbols indicate the centers of the SZE signal measured by
 ALMA (cross) and ACT (circle). The ellipse shows the FWHM location of
 the best-fit elliptical Gaussian profile (table \ref{tab-gaussfit}).
 Bottom left and bottom right: A wider-field HSC-SSP optical $riz-$color
 image with the ALMA SZE center marked by a cross.  The
 positions of the BCG and the off-center sources (W1 and W2) are
 indicated by a box and diamonds, respectively.  Wherever
 plotted, white contours show the significance levels of the ALMA SZE
 image ($5-9\sigma$ in increments of $1\sigma = 5.9~\mu$Jy/beam),
 green contours the significance levels of the SZE map from
 the ACT DR5 cluster catalogue (S/N = 3, 4, and 5),
 yellow contours the surface density of probable member galaxies
 averaged by a Gaussian with $20''$ FWHM ($80\%,~60\%, ~ 40\%, ~20\%$ of
 the peak value), and magenta contours the projected weak lensing mass
 smoothed by a Gaussian with $30''$ FWHM (S/N = 1.5, 2, 2.5, and 3).}
 \label{fig-szJ0947}
\end{figure*}
\begin{figure*}[tp]
 \begin{center}
  \includegraphics[width=8.4cm]{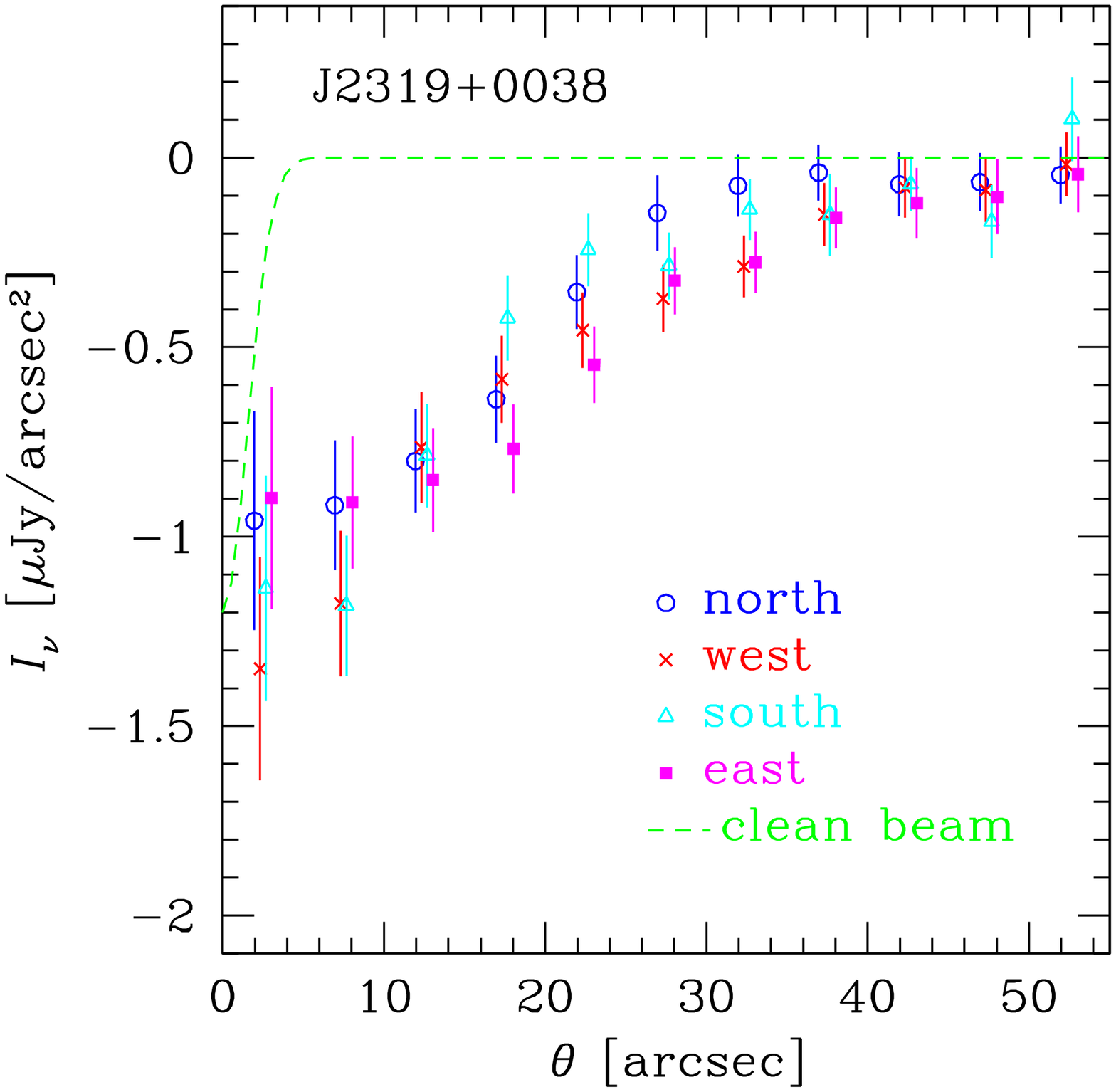}
  \includegraphics[width=8.4cm]{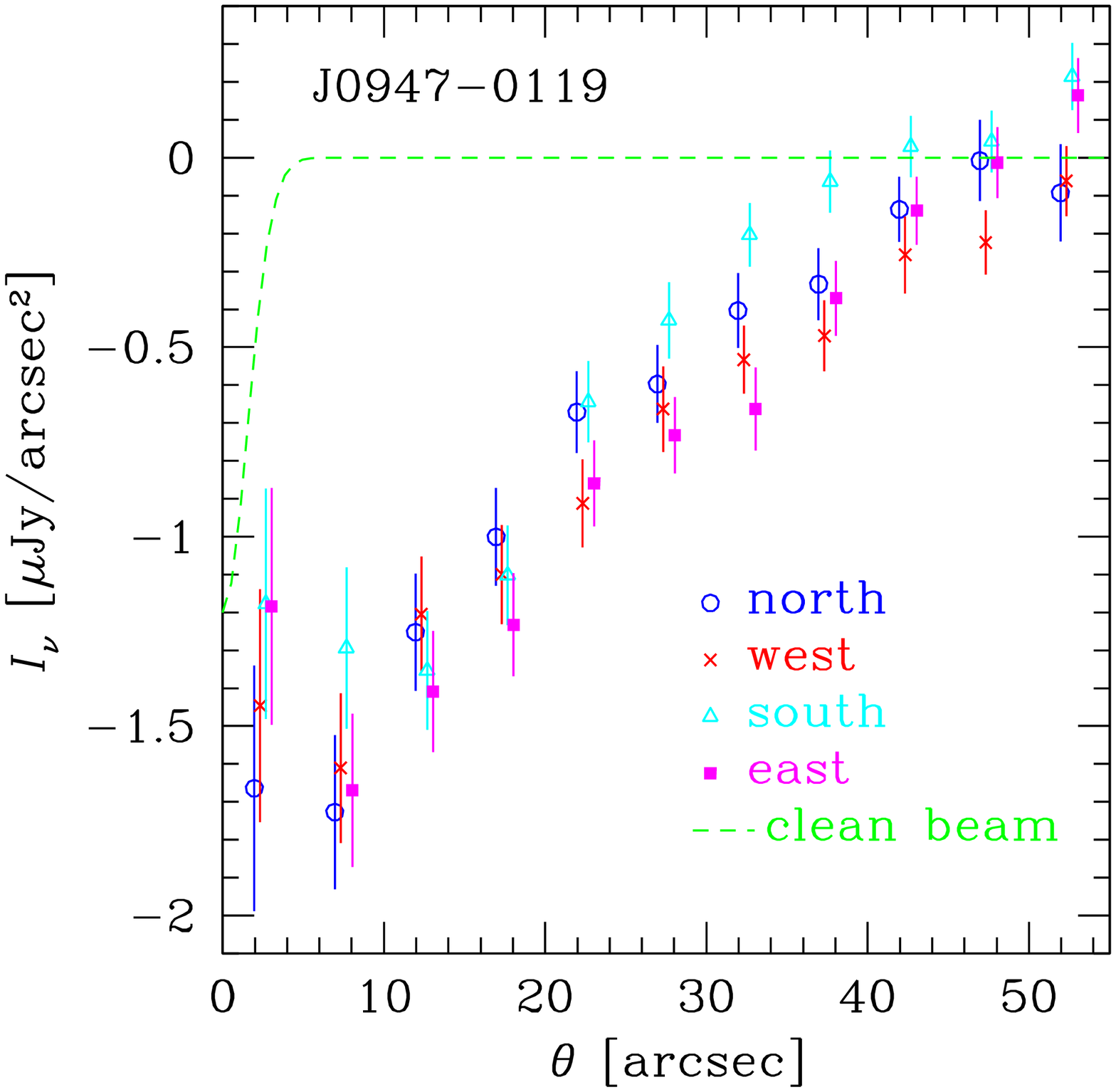} 
 \end{center}
 \caption{Azimuthally averaged SZE intensity profiles \targeta\ (left
 panel) and \targetb\ (right panel) as a function of the projected
 distance from the cluster center in the four quadrants; north
 (circles), west (crosses), south (triangles), and east (squares). For
 clarity, symbols are slightly shifted horizontally. The dashed lines
 show azimuthally averaged shapes of the synthesized beam.}
 \label{fig-szprof}
\end{figure*}

\subsubsection{\targetb}

Figure \ref{fig-szJ0947} shows the deconvolved ALMA image of \targetb\
after compact sources listed in table~\ref{tab-source2} are removed.  As
with the case of \targeta, the results of an elliptical Gaussian fit are
listed in table~\ref{tab-gaussfit} and azimuthally averaged intensity
profiles in four quadrants are plotted in figure~\ref{fig-szprof}.

The SZE signal of \targetb\ tends to be stronger and more extended than
that of \targeta, with a mean size of $\sim 45''$ FWHM and an
axis ratio of $0.66 \pm 0.06$ (table \ref{tab-gaussfit}). The integrated
flux density within $45''$ from the position shown in table~\ref{tab-gaussfit}
is $-3.66\pm 0.11$ mJy. For
reference, the flux inferred from the elliptical Gaussian fit is $-4.12
\pm 0.24$ mJy.

The best-fit center of the ALMA image (\timeform{9h47m58.74s},
\timeform{-1D19'58.37''}) is  $8''$ north-east  of the centroid position
of the ACT image (\timeform{9h47m58.30s},~\timeform{-1D20'02.80''}) and represents reasonable agreement considering the angular resolution (FWHM
$\sim 1.4'$) and the significance (S/N$=13.2$) of
the ACT data (see also section \ref{sec-act}).  Given the lack of
\Chandra\ data for this cluster, we refer to the former position as the
center of \targetb\ in the rest of this paper.

Figures \ref{fig-szJ0947} and \ref{fig-szprof} suggest that the
ICM in \targetb\ is disturbed. The SZE signal tends to be
stronger along  the north-east and the south-west  direction across the
emission center. As will be discussed in section \ref{sec-diff}, the
statistical significance of disturbance is at $\sim 2 \sigma$, and  tends to be larger in \targeta.

\subsection{Galaxy density and weak lensing mass}
\label{sec-hsc}

Figures \ref{fig-xszJ2319} and \ref{fig-szJ0947} also show the
HSC-SSP optical images of \targeta\ and \targetb, respectively.
The richness of these clusters is $N_{\rm mem}=31.6$ and
$67.5$, respectively, where $N_{\rm mem}$ is the number of galaxies with
stellar mass larger than $10^{10.2}M_\odot$ taking into account
membership probability (see \citet{Oguri14} for definition). The galaxies in both clusters show bimodal density peaks indicative of subcluster mergers, if they are not subject 
to a chance projection effect. 
The BCG lies in the main peak of
galaxy density in each cluster.

The weak lensing signal is only marginally detected with the peak
S/N of 2.8 and 3.2 for \targeta\ and \targetb, respectively, after being
smoothed by a Gaussian with $30''$ FWHM.  While the significance is low,
the peak of the surface mass density is in agreement with the main
peak of galaxy concentration. Assuming the NFW density profile
\citep{Navarro96} and the mass--concentration relation of
\citet{Diemer15}, the values of the characteristic mass,
estimated from the weak lensing measurements, are $M_{500}=3.6
^{+4.6}_{-2.2} \times 10^{14} h_{70}^{-1}M_\odot$ and $2.6 ^{+3.2}_{-1.4} \times
10^{14} h_{70}^{-1}M_\odot$ for \targeta\ and \targetb,
respectively\footnote{$M_{500}$ is defined as the total mass enclosed in
the radius $R_{500}$, within which the average matter density is 500
times the critical density of the Universe.}. Note that the errors on $M_{500}$ from weak lensing are very large for these distant clusters,  
owing to the limited number of background galaxies.

Figure \ref{fig-xszJ2319} further indicates that the ALMA SZE and
\Chandra\ X-ray centers of \targeta\ coincide with the highest peak of
galaxy density and the weak lensing mass. The BCG also
lies within $1''$ from the X-ray center. On the other hand, no
concentration of the ICM is detected around the second peak of galaxy
density, which lies at $\sim 40''$ east of the main
peak. The ICM morphology is still elongated toward the second peak of
galaxy density.

On the other hand, the SZE center of \targetb\ (figure
\ref{fig-szJ0947}) is persistently offset from the BCG
and the main peak of galaxy density by about $17''$ (140 $h_{70}^{-1}$kpc),
suggesting the presence of a strong disturbance in this cluster. The ICM
morphology is also elongated toward the second peak of galaxy density.

The above results imply that highly asymmetric
galaxy distributions seen in optical images (figures \ref{fig-xszJ2319}
and \ref{fig-szJ0947})  could be associated with relatively small values
($\simlt 0.7$) of the axis ratio found for the ICM in both clusters
(tables \ref{tab-gaussfit} and \ref{tab-betafit}); average values of the
axis ratio reported for massive clusters at low $z$ are $0.8\sim 0.9$
(e.g., \cite{Kawahara10,Donahue16}).

\section{Interpretation and implications}

\subsection{Imaging simulations}
\label{sec-sim}

To quantify the missing flux of the
ALMA data and to test the fidelity of the image reconstruction algorithm, we performed imaging simulations as follows.

\subsubsection{Method}

\label{sec-simmethod}

We first
specified the Compton $y$-parameter (i.e., projected electron pressure)
\begin{equation}
  y(\vec{\theta}) =  \int \sigma_{\rm T}
   n_{\rm e} \frac{kT_{\rm e}}{m_{\rm e}c^2}  dl,  
\end{equation}
 for each cluster in consideration, where $\sigma_{\rm T}$ is the
 Thomson cross section, $k$ is the Boltzmann constant, $m_{\rm e}$ is
 the electron mass, $c$ is the speed of light,
 and $l$ denotes the physical length along the line-of-sight toward the
  sky position $\vec{\theta}$.  Details of the models on $n_{\rm e}$ and
  $T_{\rm e}$ will be described  in section \ref{sec-simmodel}.

  Input model
  images of the SZE were created from the above Compton $y$-parameter map separately at
  four spectral windows centered at 85, 87, 97, and 99 GHz with an
  effective bandwidth of 1.875 GHz each. A relativistic correction to
  the SZE intensity by \citet{Itoh04} was taken into account. 
  
  Visibility data were then produced using the CASA task {\it simobserve}
  including both instrumental and atmospheric thermal noise in each
  spectral window; this procedure was repeated 10 times adopting
  different noise seeds. The pointing directions, the array
  configuration, the hour angle, the total effective integration time,
  and the average precipitable water vapor were set to match those of
  each executing block of real observations for each cluster.  
  
  Finally,
  the mock visibility was deconvolved in the same way as the real data
  as described in section \ref{sec-almadata}. For
  reference, we also performed ``noise-only'' runs in which the input
  SZE signal is set to zero.  We have checked that the RMS levels of dirty
  maps created from such noise-only runs are consistent with the
  observed values in table~\ref{tab-image}.  To take into account any
  bias in producing an image at a single frequency from the data taken
  over finite bandwidths, the simulation outputs are compared with an
  input model evaluated at the central frequency 92 GHz.

\subsubsection{Input models}

\label{sec-simmodel}

The model Compton $y$-parameter map of \targeta\ was constructed from
the available X-ray data as follows. Fitting the 0.4--7.0 keV spectrum
within $45''$ from the cluster center yielded the average electron
temperature of $kT_{\rm e}=5.90^{+0.79}_{-0.62} $ keV.
Assuming that the gas is isothermal\footnote{Our subsequent analysis is insensitive to this simplification. We checked that adopting a mean temperature profile for high redshift clusters \citep{McDonald14b} in the input model would change the value of $c_1$ in table \ref{tab-conv} only within its error range. We
 will also show in section \ref{sec-deproj} that a weak temperature gradient
 is inferred from the joint analysis of the SZE and the
 X-ray data of \targeta, whereas the gas density profile still  follows equation (\ref{eq-nbeta}).}, a triaxial electron density profile
 consistent with an elliptical $\beta$ model specified by equations
 (\ref{eq-ellipbeta}) and (\ref{eq-ellipbeta2}) is
 \begin{equation}
  n_{\rm e}(\vec{r}) =
   n_{\rm e0}
   \left[1+\frac{\bar{\theta}^2 + \left(\phi/\eta\right)^2}{\theta_{\rm
			      c}^2}\right]^{-\frac{3}{2}\beta},
 \label{eq-nbeta}
 \end{equation}
 where $n_{\rm e0}$ is the central electron density, $\phi$ denotes the
 line-of-sight angular displacement between the three dimensional
 position $\vec{r}$ and the cluster center (i.e., physical distance
 divided by the angular diameter distance to the cluster $d_{\rm A}$),
 $\bar{\theta}$ is the mean angular radius on the
 sky as in equations (\ref{eq-ellipbeta}) and  (\ref{eq-ellipbeta2}), and
 $\eta$ is the elongation factor along the line-of-sight.  In other
 words, we assumed that the electron density profile has the axis ratio
 of
 \begin{equation}
   \sqrt{q} : \frac{1}{\sqrt{q}} :\eta
\label{eq-3daxis}
 \end{equation}
  where the first axis corresponds to $\bar{x}$ in equation
  (\ref{eq-ellipbeta2}) and the third axis is effectively along the
  line-of-sight; we neglected the inclination of the
  third axis from the line-of-sight, given the lack of data to constrain
  it.  Adopting the fitted values of $S_0$, $\beta$, $\theta_{\rm c}$,
  $q$, and $\psi$ from table \ref{tab-betafit} as well as $kT_{\rm e}=5.90$ keV, we obtained $n_{\rm e0}= 1.43 \times
 10^{-2} h_{70}^{1/2} \eta^{-1/2}$ cm$^{-3}$.  Integrating the product
 of $n_{\rm e}$ and $kT_{\rm e}$ over the line-of-sight gives a Compton
 $y$-parameter map with the peak value of $y_{\rm peak}=1.1\times
 10^{-4} h_{70}^{-1/2} \eta^{1/2}$.  To specify the absolute value of
 the model intensity in the simulations, we assumed fiducially
 $h_{70}/\eta=1$ to obtain $y_{\rm peak}=1.1\times 10^{-4}$; we will
 discuss the impact of varying $h_{70}/\eta$ in section \ref{sec-yprof}.
 To take into account uncertainties of this model parameterization, we will also
 examine the cases in which the overall normalization of the
 $y$-parameter is doubled or halved, i.e., $y_{\rm peak}=2.2 \times
 10^{-4}$ and $0.55 \times 10^{-4}$.  We will show in sections
 \ref{sec-diff} and \ref{sec-yprof} that the case with $y_{\rm
 peak}=1.1\times 10^{-4}$ indeed agrees well with the observed ALMA
 data.

  Given the lack of X-ray data for \targetb, we used the average pressure
 profile measured from a statistical sample of X-ray clusters at $0.6 <
 z < 1.2$ by \citet{McDonald14b}. We also assumed a triaxial gas profile
 consistent with the axis ratio and the position angle observed on the
 sky for this cluster (table \ref{tab-gaussfit}). As with \targeta, the axis ratio was assumed to be given by equation (\ref{eq-3daxis}).
  As discussed in sections \ref{sec-diff} and \ref{sec-yprof}, the
  observed ALMA data are well reproduced by this profile with 
  $M_{500}=5.7 \times 10^{14}$ M$_\odot$ assuming $h_{70}=1$ and
  $\eta=1$.  We hence adopted this profile for the present simulations
  and estimated the mean temperature to be $kT_{\rm e}=9.4$ keV  using
  the scaling relation of \citet{Reichert11}.  The inferred peak value
  of the $y$-parameter is $y_{\rm peak}=1.5 \times 10^{-4}$. The sensitivity of the result to
  $y$-parameter map is examined by considering limiting cases where the overall map normalization is doubled or halved (i.e., $y_{\rm peak}=3.0 \times
  10^{-4}$ and $0.75 \times 10^{-4}$).

\begin{figure*}[tp]
 \begin{center}
    \includegraphics[height=6.8cm]{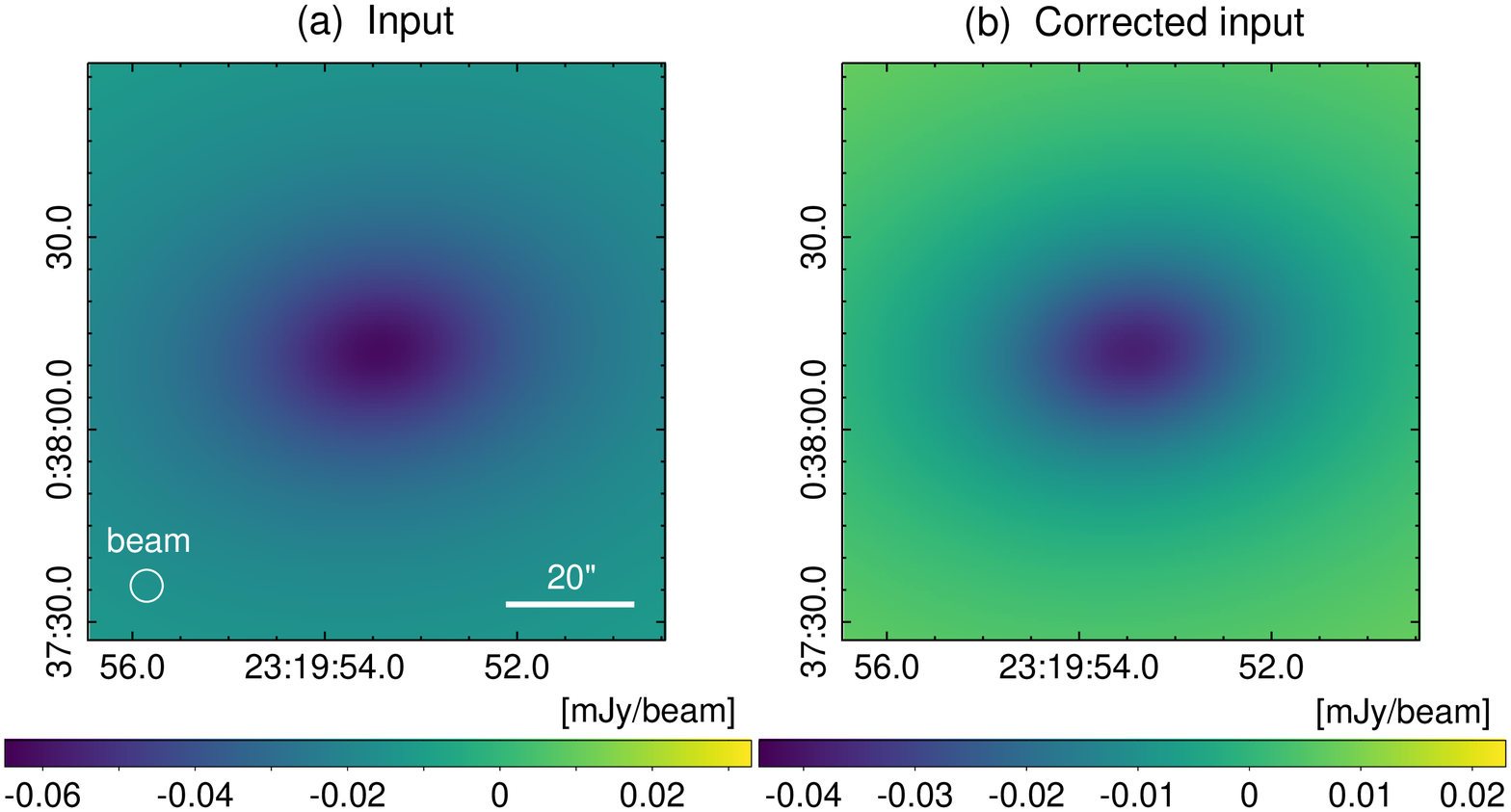}
  \includegraphics[height=7.0cm]{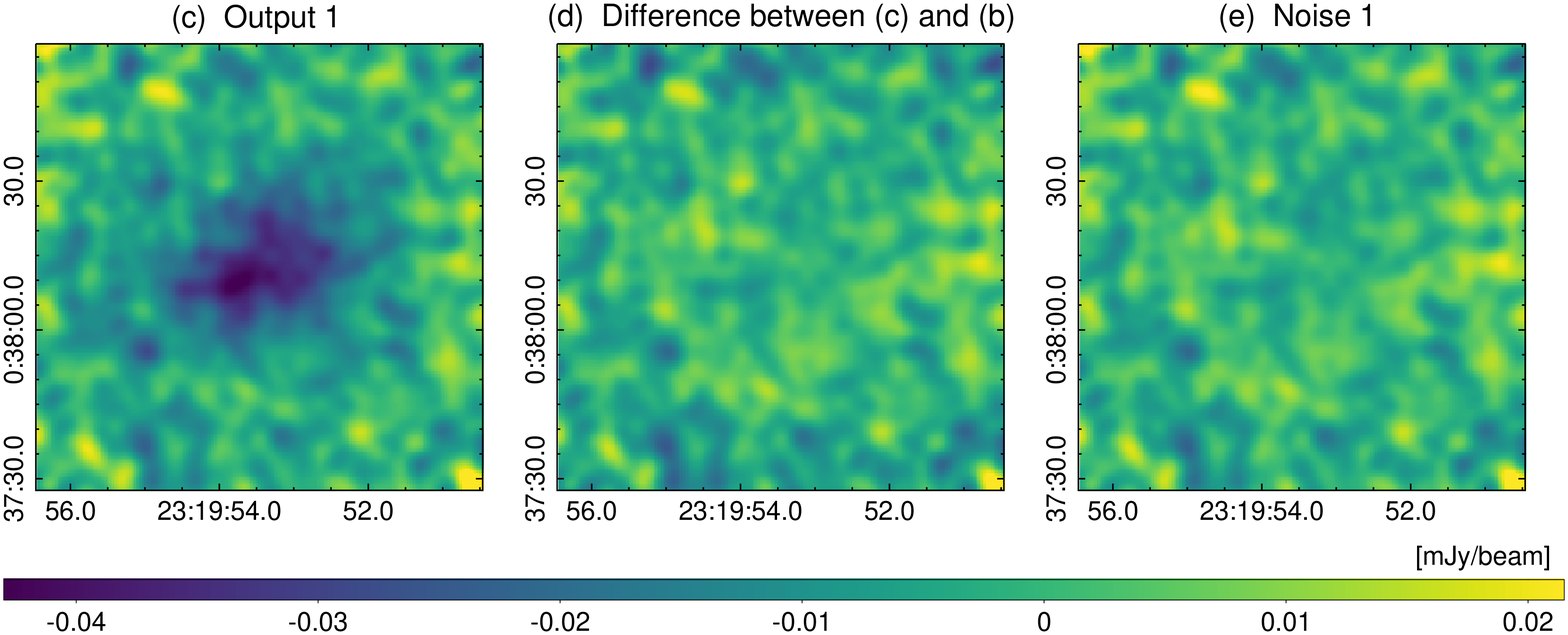}
  \includegraphics[height=7.0cm]{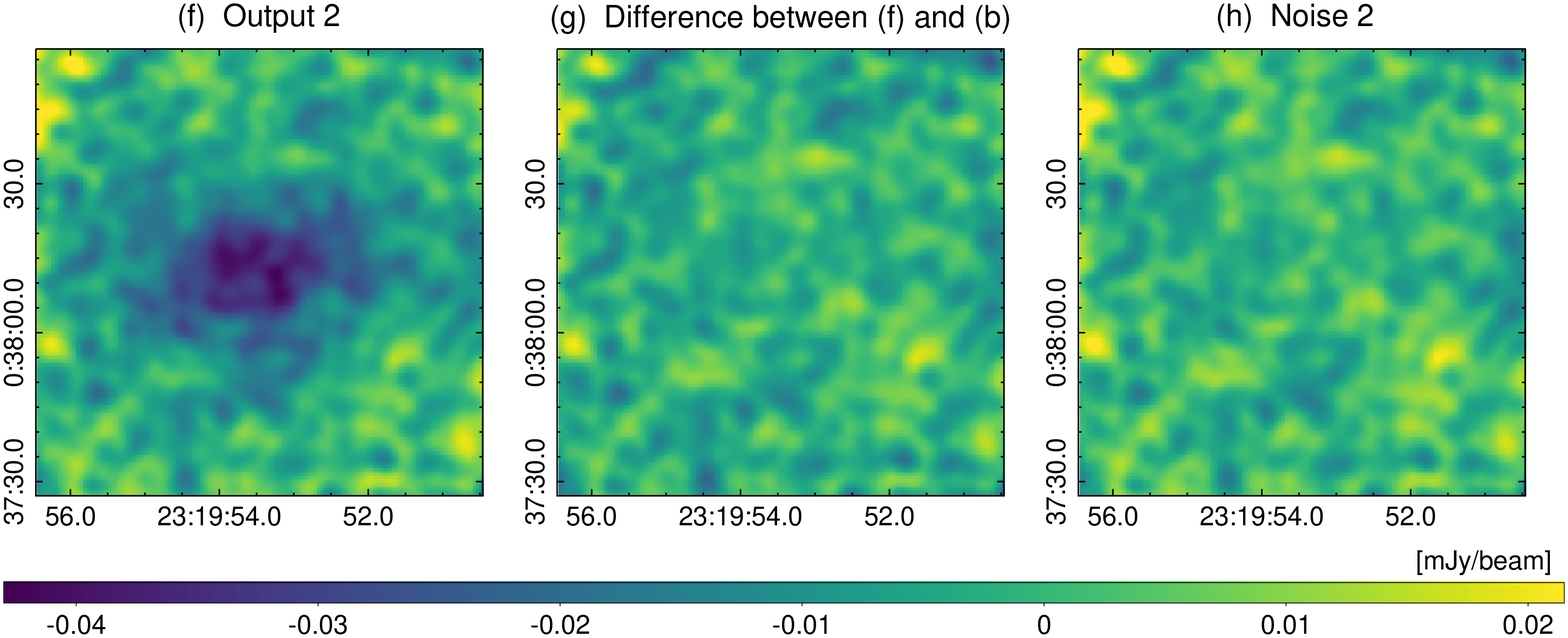}
 \end{center}
 \caption{Mock images of \targeta\ at 92 GHz with $y_{\rm peak}=1.1
 \times 10^{-4}$. All the images have been smoothed to 5$''$ FWHM. (a)
 Input model. (b) Input model to which the correction  encoded by equation
 (\ref{eq-conv}) has been applied. (c) Simulation output including noise
 shown in panel e. (d) Difference between panels c and b. (e) Noise-only
 output for the run shown in panel c.  (f) Same as panel c but with a
 different noise realization shown in panel h. (g) Difference between
 panels f and b. (h) Noise-only output for the run shown in panel f.
 }  \label{fig-simJ2319}
\end{figure*}

\begin{figure*}[tp]
  \begin{center}
    \includegraphics[height=6.8cm]{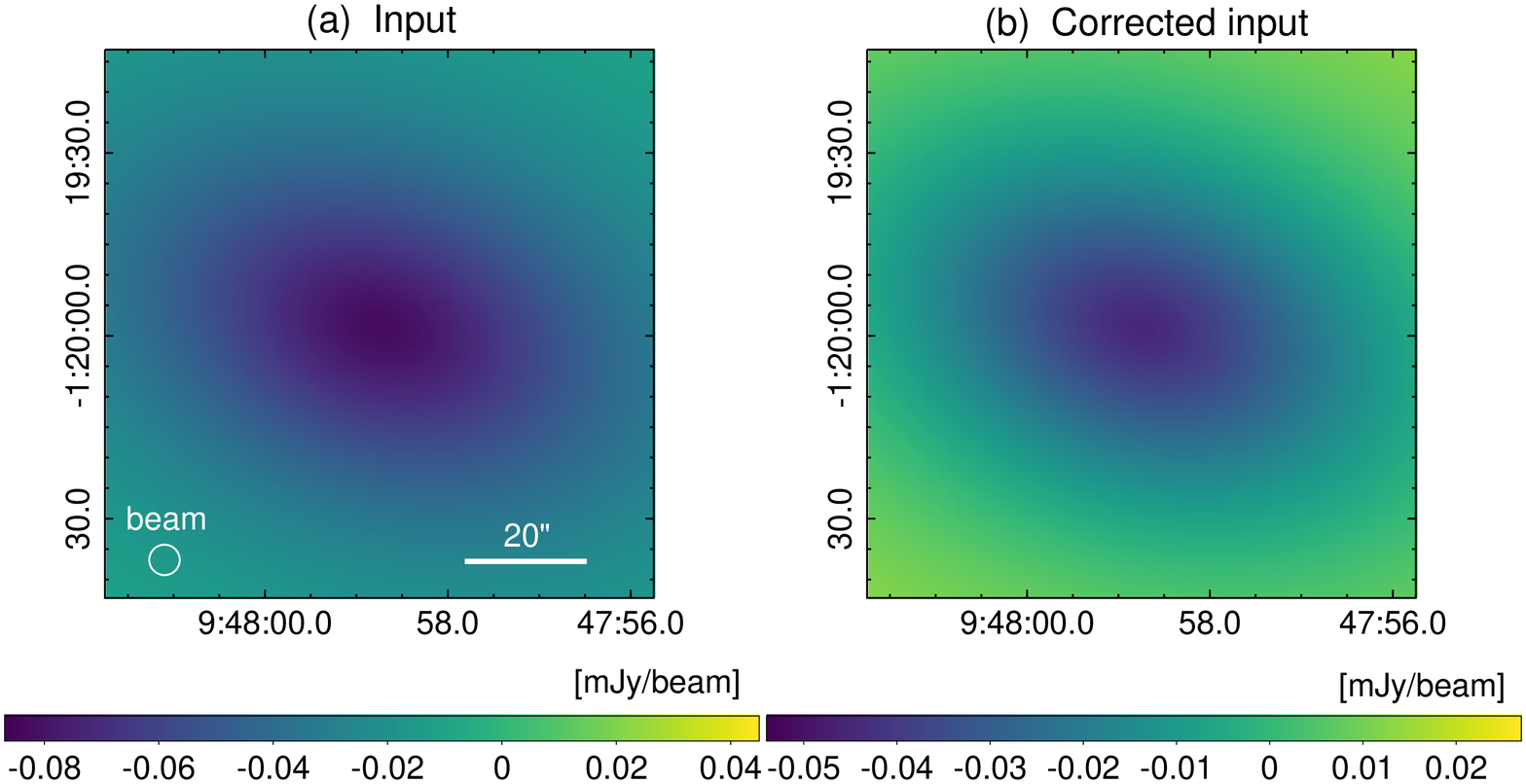}
  \includegraphics[height=6.8cm]{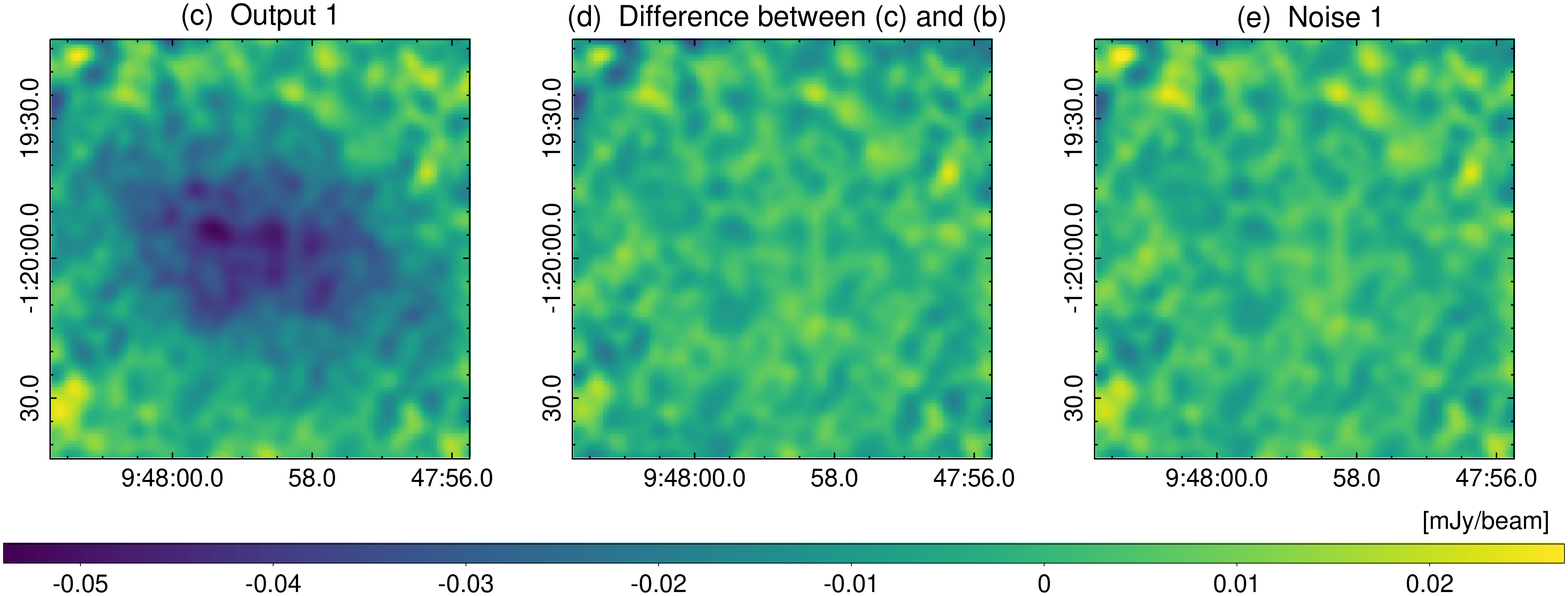}
   \includegraphics[height=6.8cm]{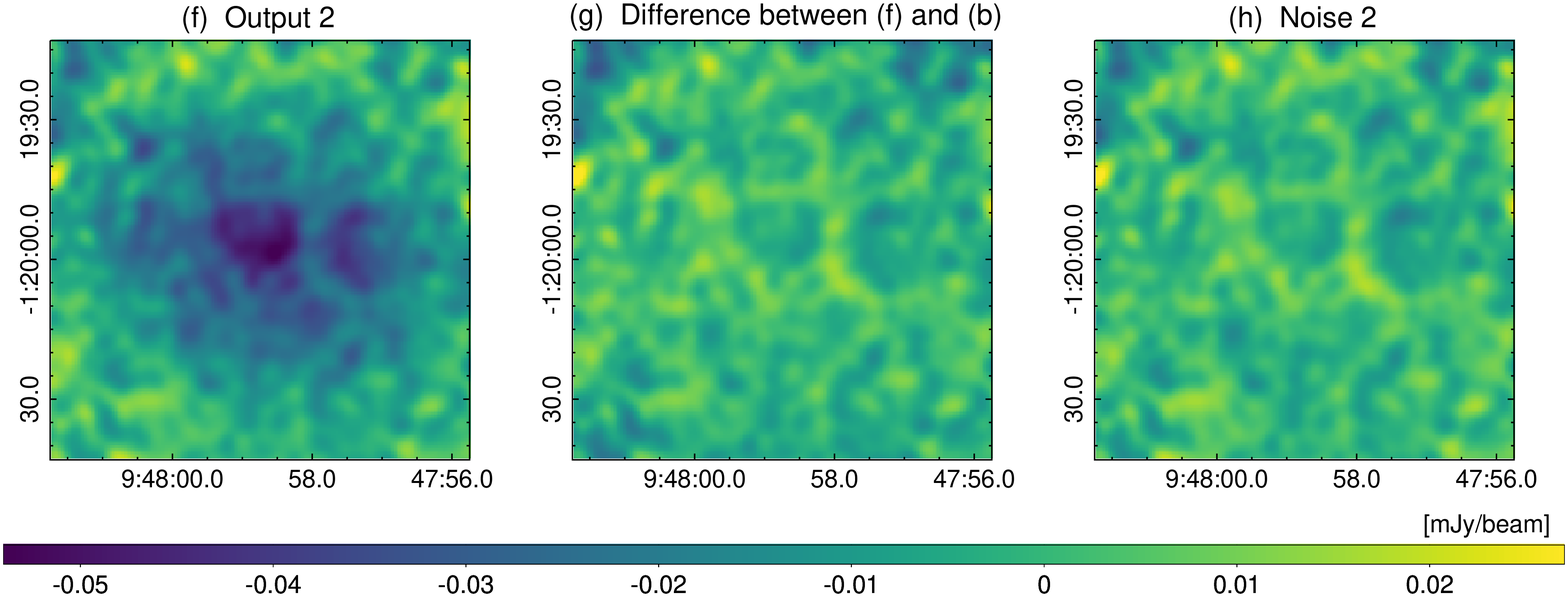}
  \end{center}
 \caption{Same as \ref{fig-simJ2319}, except for showing mock images of
 \targetb\ with $y_{\rm peak}=1.5 
 \times 10^{-4}$. }  \label{fig-simJ0947}
\end{figure*}

\begin{figure*}[tp]
 \begin{center}
  \includegraphics[width=8.4cm]{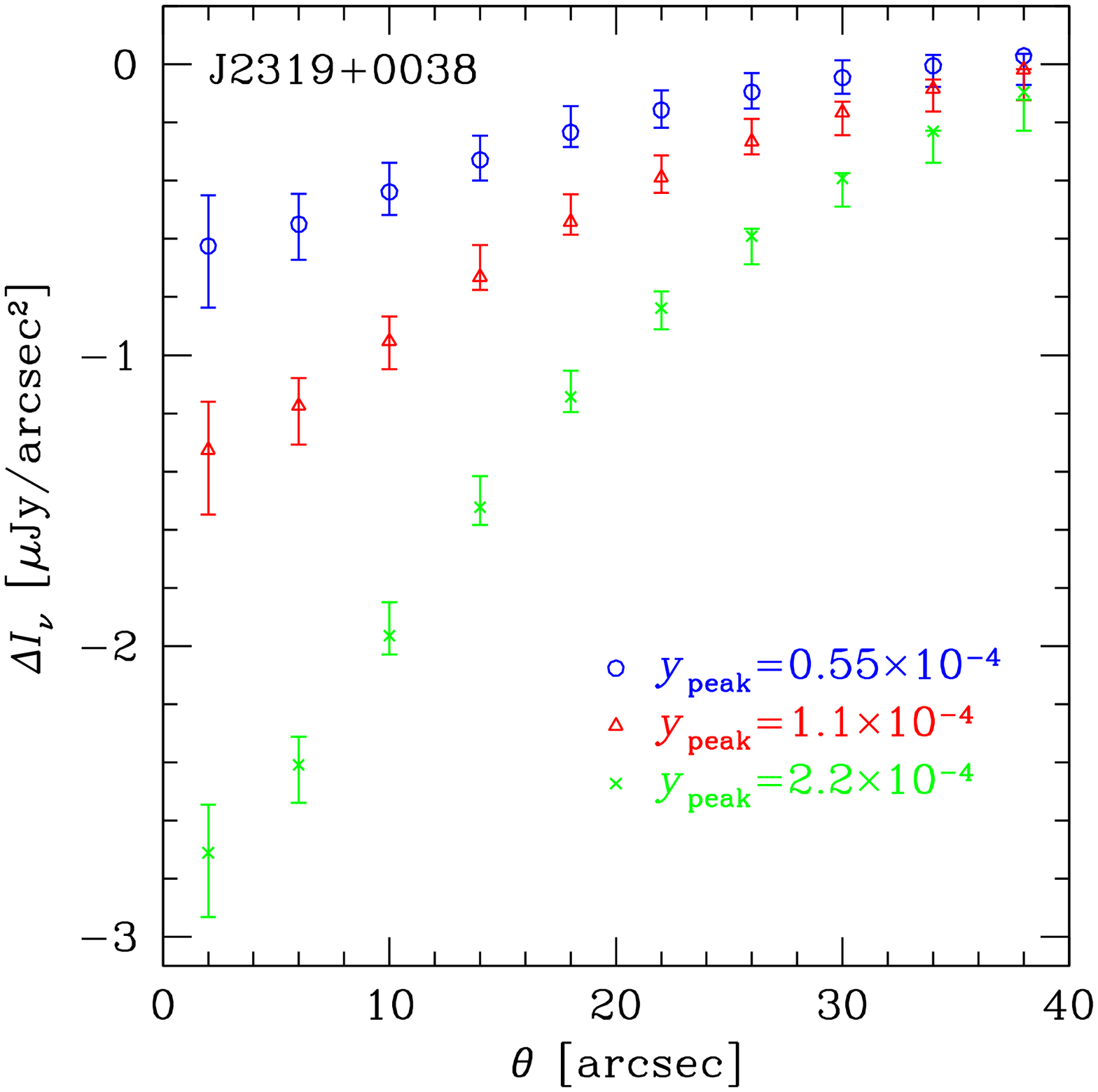}
  \includegraphics[width=8.4cm]{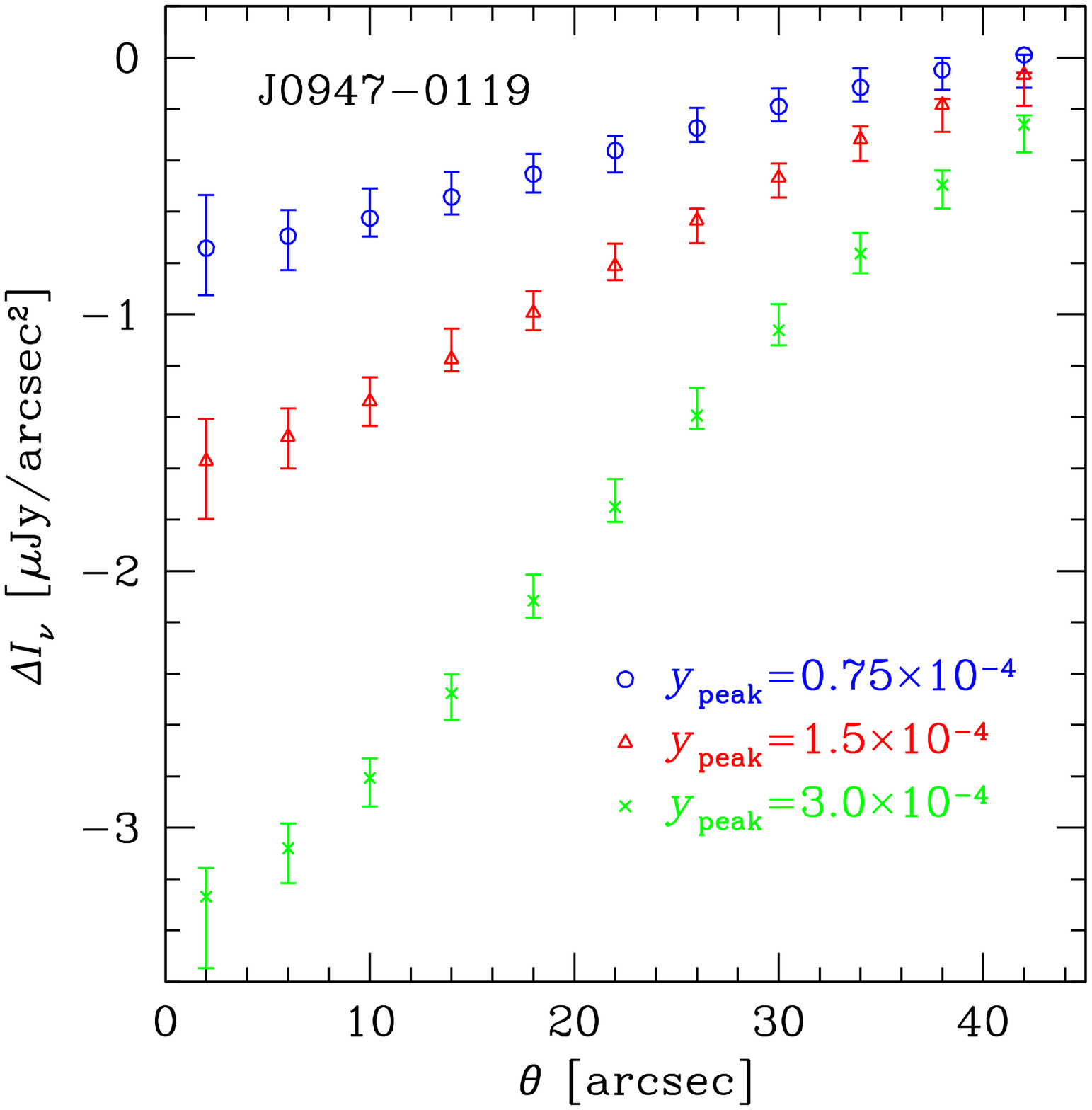}
 \end{center}
 \caption{Results of imaging simulations for \targeta\ (left panel) and
 \targetb\ (right panel). Error bars show azimuthally averaged intensity
 profiles from the simulations. Symbols show the same quantity from the
 input model to which the correction of by equation (\ref{eq-conv}) has
 been applied for each value of $y_{\rm peak}$ as indicated in the figure.
 }  \label{fig-szprofsim}
\end{figure*}
  \begin{table*}[h]
   \caption{The coefficients of the linear relation (equation
   \ref{eq-conv}) from imaging simulations. The coefficient $c_1$ is
   assumed to be common for each cluster. }
  \begin{center}
    \begin{tabular}{cccc}
     \hline
     cluster & $y_{\rm peak}$ [$10^{-4}$]& $c_1$ & $c_0$ [$\mu$Jy/arcsec$^2$]\\ \hline
     & $0.55$ &  $0.89 \pm 0.03$ & $0.35\pm 0.04$ \\
     \targeta\ & $1.1$&  $0.89\pm 0.03$ & $0.62 \pm 0.06$\\
     & $2.2$ & $0.89\pm 0.03$ & $1.19 \pm 0.11$ \\ \hline 
      & $0.75$ &  $0.79 \pm 0.03$&  $0.42 \pm 0.05$\\
     \targetb\ & $1.5$& $0.79 \pm 0.03$ & $0.75 \pm 0.07$\\
     & $3.0$ & $0.79 \pm 0.03$& $1.38 \pm 0.12$\\    
     \hline
    \end{tabular}
  \end{center}
 \label{tab-conv}
  \end{table*}
\begin{figure*}[ht]
 \begin{center}
  \includegraphics[width=8.4cm]{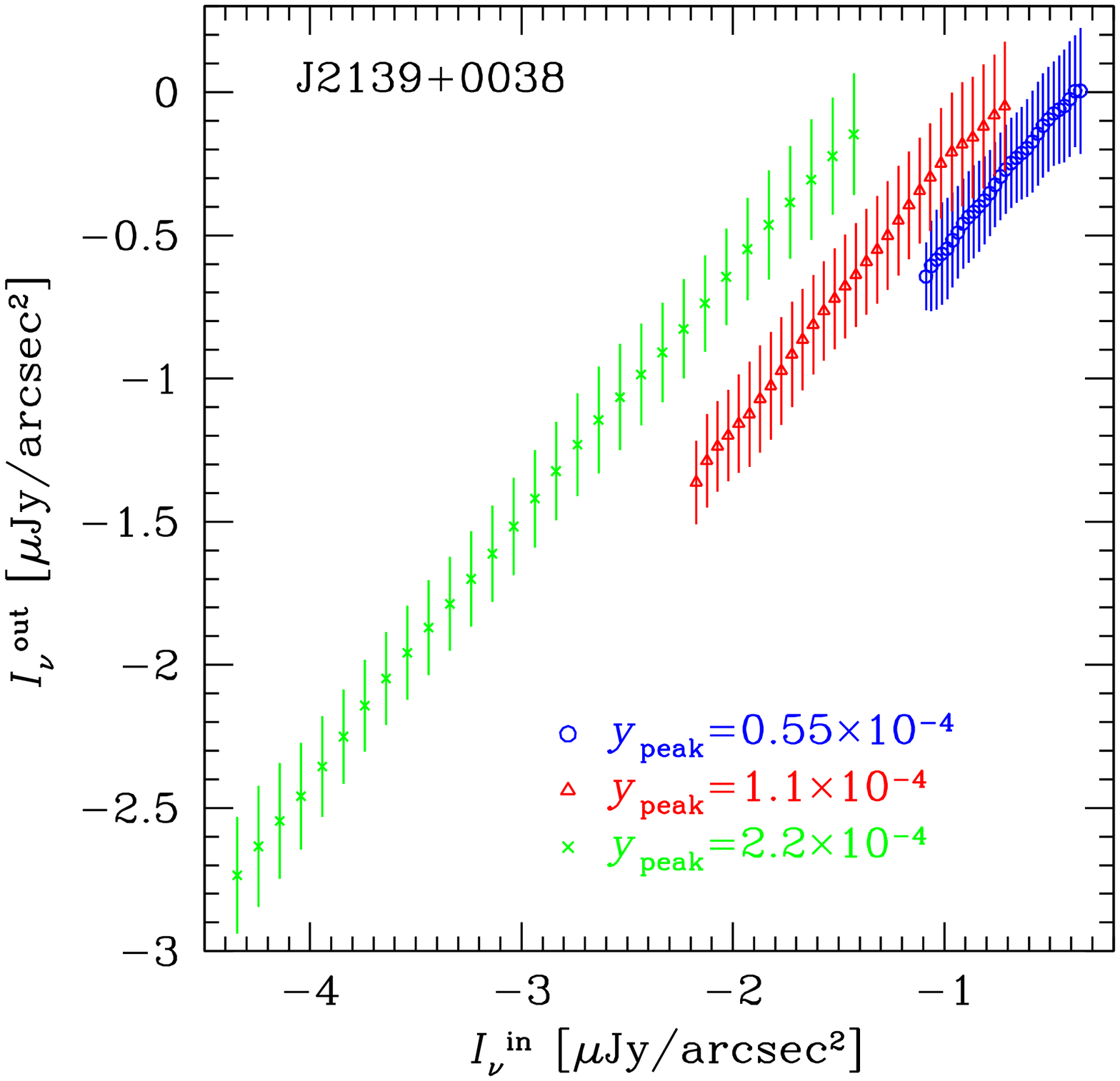}
  \includegraphics[width=8.4cm]{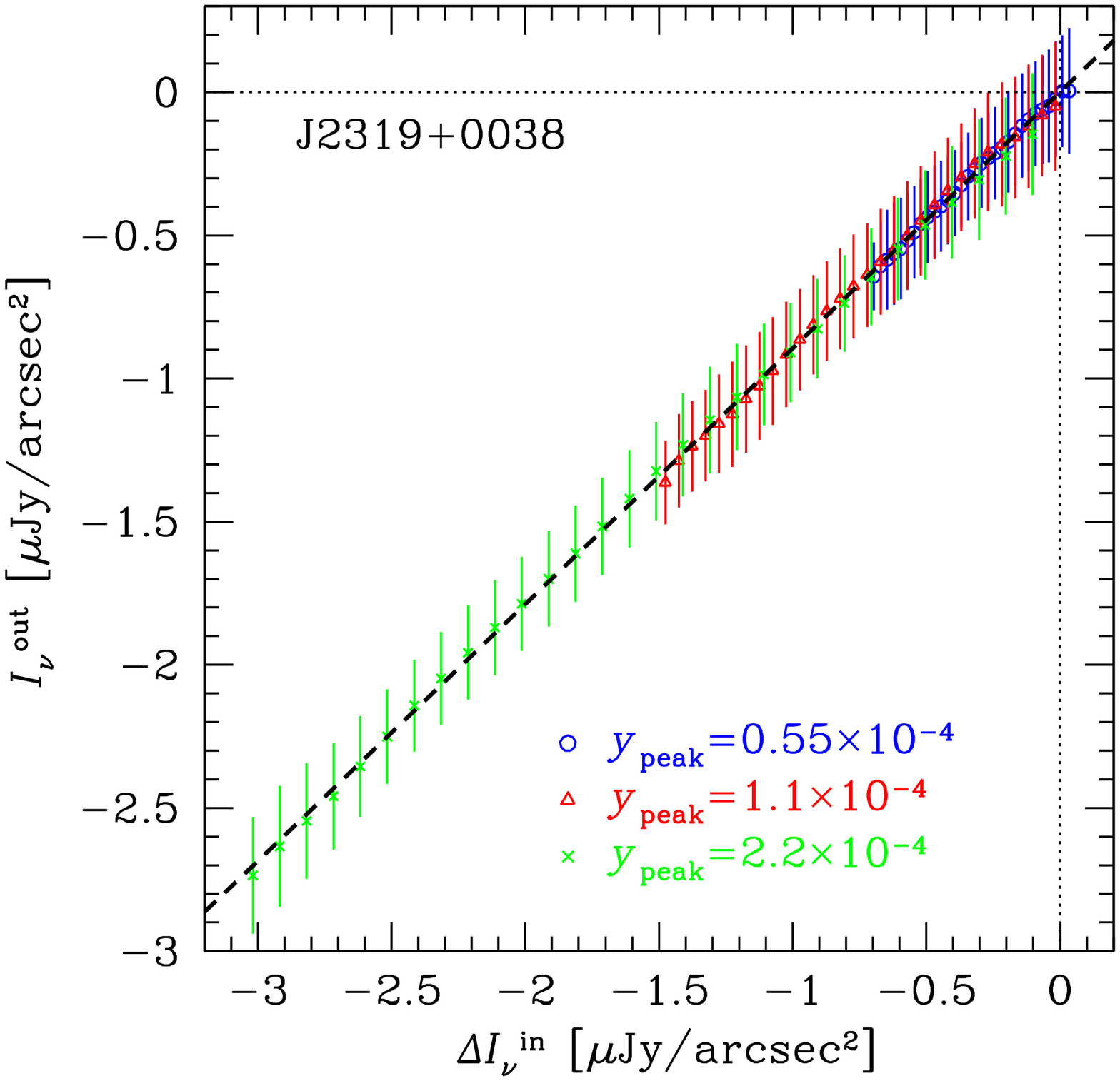}
 \end{center}
 \caption{Left: Relation between the output intensity and the input
 intensity at 92~GHz from the imaging simulations for \targeta.  Error
 bars and symbols denote the standard deviation and the mean, respectively, in each bin for $y_{\rm peak}= 0.55
 \times 10^{-4}$ (circles), $1.1\times 10^{-4}$ (triangles), and $2.2\times
 10^{-4}$ (crosses). Right: Same as the left panel, except that the fitted
 value of $c_0/c_1$ in equation (\ref{eq-conv}) has been added to
 $I_{\nu}^{\rm in}$ to give zero intercept in each case. The thick
 dashed line shows the best fitting relation with $c_1=0.89$ for this cluster (table
 \ref{tab-conv}).}  \label{fig-conv}
\end{figure*}
\begin{figure*}[ht]
 \begin{center}
    \includegraphics[width=8.4cm]{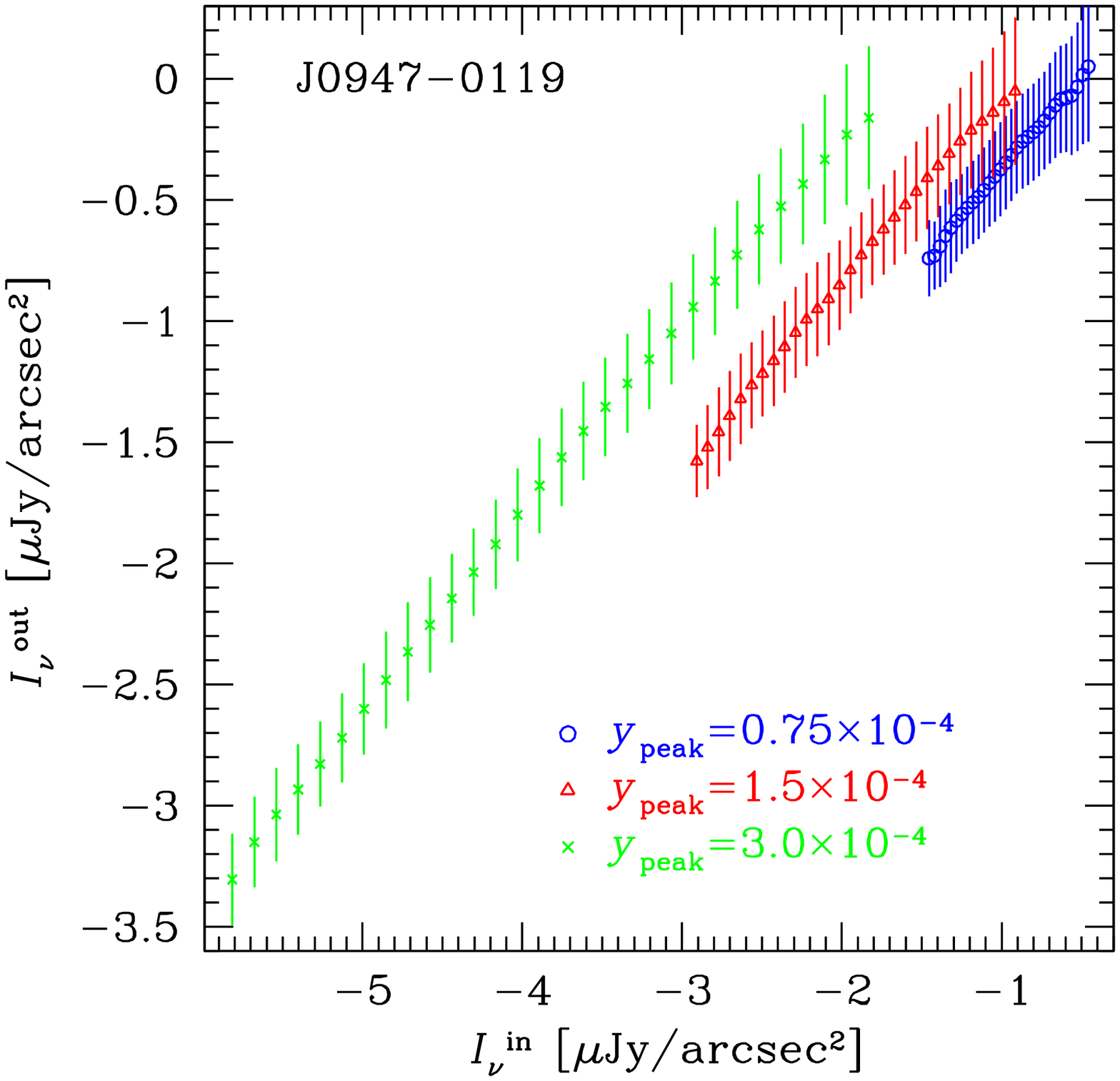}
  \includegraphics[width=8.4cm]{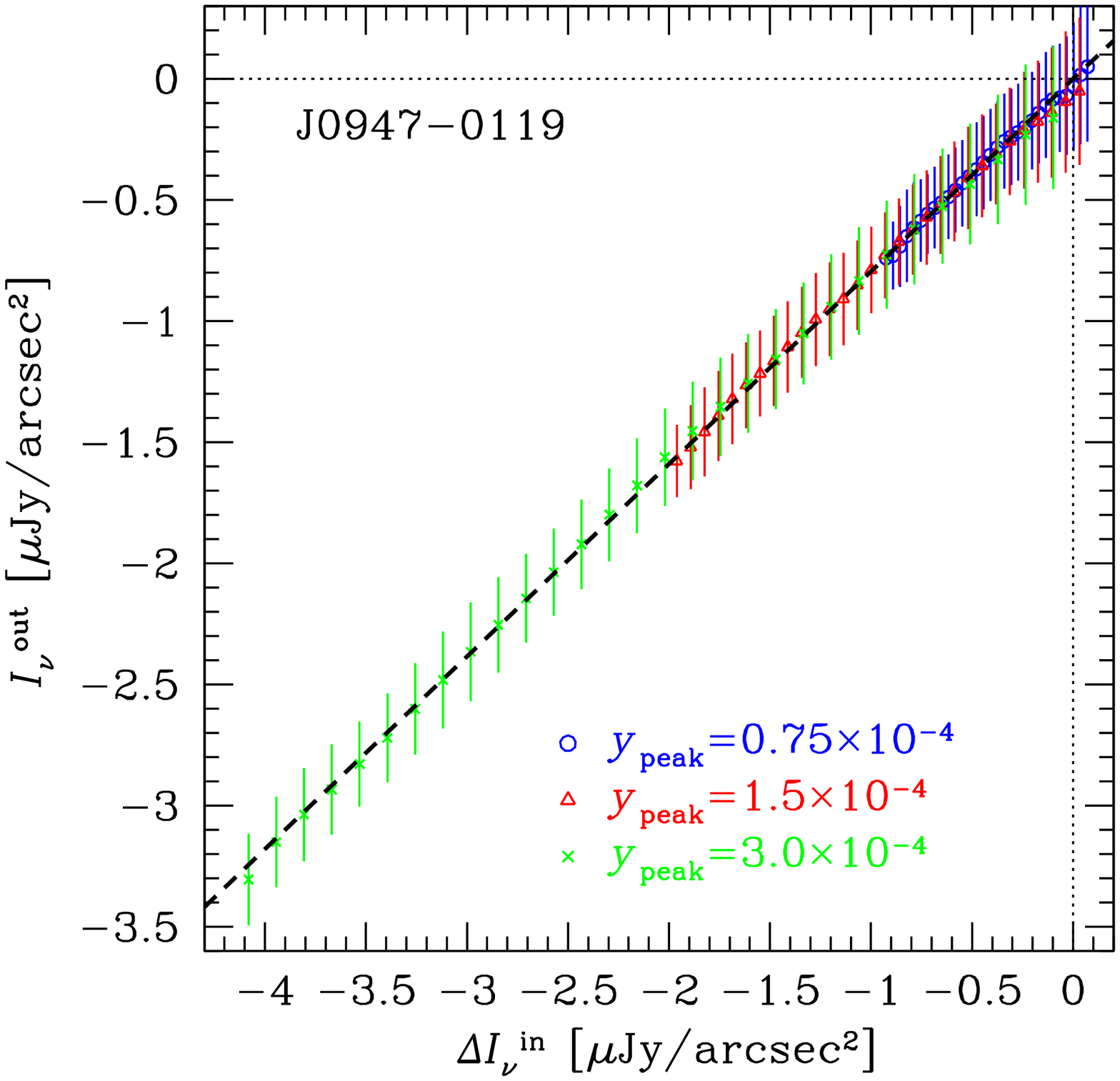}     
 \end{center}
 \caption{Similar to figure \ref{fig-conv} but showing the results
 for \targetb. The thick dashed line shows the best fitting relation
 with $c_1=0.79$ for this cluster (table \ref{tab-conv}).
 } \label{fig-conv2}
\end{figure*}
\begin{figure*}[tp]
 \begin{center}
  \includegraphics[width=16.5cm]{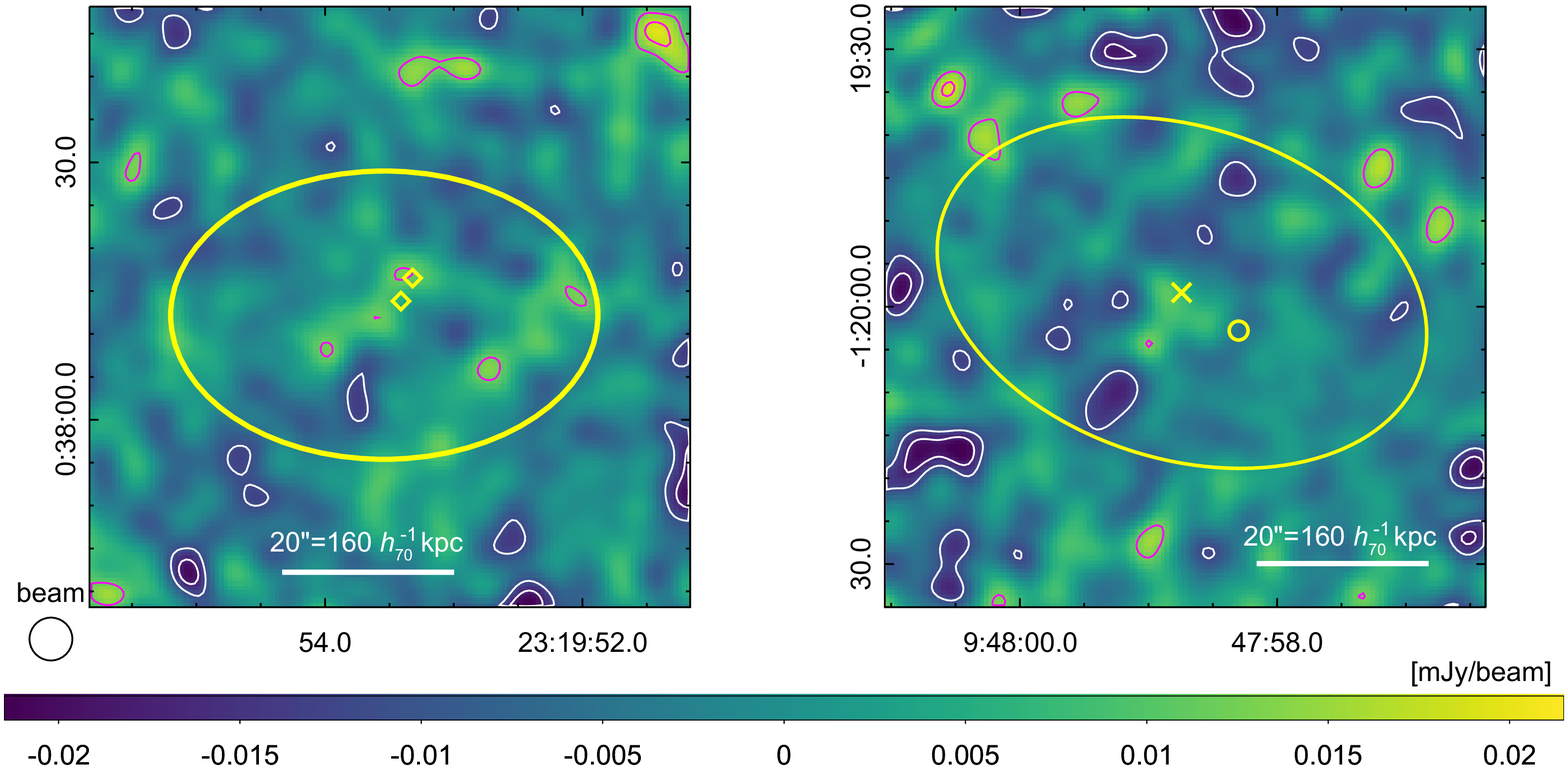}
 \end{center}
 \caption{Residual SZE images of \targeta\ (left) and \targetb\ (right)
 after the model used in the simulation described in section
 \ref{sec-sim} is subtracted. Contours show 2, 3, and 4 $\sigma$
 significance levels,  for  positive (magenta) or negative (white) values of residuals. 
  Meanings of ellipses and symbols are the
 same as in figures~\ref{fig-xszJ2319} and \ref{fig-szJ0947}.}
 \label{fig-diff}
\end{figure*}

\subsubsection{Missing flux correction using simulation results}

\label{sec-simimage}

Figures \ref{fig-simJ2319} and \ref{fig-simJ0947} compare arbitrarily
 chosen realizations of the simulated images and the input models for
 \targeta\ and \targetb, respectively. The azimuthal average of all
 realizations are shown in figure~\ref{fig-szprofsim}.  The simulated
 images show similar amplitude and spatial extension to the real data
 plotted in figures~\ref{fig-xszJ2319} and \ref{fig-szJ0947} for $y_{\rm
 peak}=1.1 \times 10^{-4}$ (\targeta) and $y_{\rm peak}=1.5 \times
 10^{-4}$ (\targetb), respectively.
 
As noted in \citet{Kitayama16} and \citet{Kitayama20}, the following
linear relation holds on average for the simulated ALMA images 
\begin{eqnarray}
I_{\nu}^{\rm out}(\vec{\theta}) = c_1 I_{\nu}^{\rm in}(\vec{\theta}) + c_0, 
\label{eq-conv}
\end{eqnarray}
where $I_{\nu}^{\rm out}$ and $I_{\nu}^{\rm in}$ are respectively the
intensities of output and input images at the same sky position $\vec{\theta}$. This equation relates the object's observed to intrinsic intensity and is used in sections \ref{sec-yprof} and \ref{sec-deproj} to correct for the missing flux.
To obtain $c_1$ and $c_0$ in equation (\ref{eq-conv}), a set of data ($I_{\nu}^{\rm in}$, $I_{\nu}^{\rm out}$)
was created for each sky position in each simulated image and then binned in descending order of $I_{\nu}^{\rm in}$. Each bin contains at least 1000 pixel data after accumulating 10 realizations for each adopted value of $y_{\rm peak}$.   
The mean and the standard deviation of $I_{\nu}^{\rm out}$ in each bin are plotted against $I_{\nu}^{\rm in}$ in figures~\ref{fig-conv} and \ref{fig-conv2}. These figures indicate that 
equation (\ref{eq-conv}) holds well and $c_1$ is insensitive to 
the adopted value of $y_{\rm peak}$ (i.e., the overall normalization of the emission); $c_1$ depends primarily on the shape and the gradient of the emission as well as the observing conditions (e.g., the $uv$ coverage). 

Table
\ref{tab-conv} lists the fitted values of $c_0$ and $c_1$, assuming that $c_1$ is common among different values of 
$y_{\rm peak}$ for each cluster. The obtained value of $c_1$ 
is close to unity but takes a smaller value for \targetb, whose $y$-parameter profile is shallower than \targeta\ (figure \ref{fig-yprof}). The specific value of 
$c_0$ will not affect our real data analysis as it will be eliminated in equation (\ref{eq-conv2}) described below.

Figures
\ref{fig-simJ2319}--\ref{fig-szprofsim} further illustrate that the
simulated images are in agreement with the input models once corrected by equation (\ref{eq-conv}), as described above. The
RMS values at $\theta < 45''$ in figures~\ref{fig-simJ2319}(d)(g) and
\ref{fig-simJ0947}(d)(g) are 6.2, 5.9, 5.7, and 6.3 $\mu$Jy/beam,
respectively, and fully consistent with noise.  

The above results imply
that the relative intensity with respect to some reference point
$\vec{\theta}_{\rm ref}$ follows
\begin{eqnarray}
 I_{\nu}^{\rm out}(\vec{\theta})
  - I_{\nu}^{\rm out}(\vec{\theta}_{\rm ref})
  = c_1 \left[
	 I_{\nu}^{\rm in}(\vec{\theta})
	 - I_{\nu}^{\rm in}(\vec{\theta}_{\rm ref})\right]
  \label{eq-conv2}
\end{eqnarray}
nearly independently of the underlying value of $y_{\rm peak}$.  The
coefficient $c_1$ denotes how much of the intrinsic intensity is
retained on average in the deconvolved image for each cluster. In the
rest of this paper, we refer to the conversion 
from $I_{\nu}^{\rm out}(\vec{\theta})
  - I_{\nu}^{\rm out}(\vec{\theta}_{\rm ref})$ to 
$I_{\nu}^{\rm in}(\vec{\theta})
	 - I_{\nu}^{\rm in}(\vec{\theta}_{\rm ref})$ using 
	 the best-fit values of $c_1$ from table \ref{tab-conv} 
as the ``missing flux correction". 
We will apply this correction to the real data taking  
off-center positions as $\vec{\theta}_{\rm ref}$; specifically, 
the reference points are taken at the mean radius
$\bar{\theta}=35''$ from the cluster center, which lie at the envelope
of the emission profiles from real data (figure \ref{fig-szprof}) and
simulation outputs (figure \ref{fig-szprofsim}).

\subsubsection{Systematic errors}

\label{sec-syst}

We estimated systematic errors associated with the above 
missing flux correction as follows. The errors of $I_{\nu}^{\rm out}$ plotted in
figures \ref{fig-conv} and \ref{fig-conv2} are dominated by statistical
ones; each bin contains more than 1000 pixel data (i.e., 100 per
realization).  The systematic deviation from equation
(\ref{eq-conv}) apart from such statistical errors is represented by the
RMS deviation of the mean values of $(I_{\nu}^{\rm in},I_{\nu}^{\rm
out})$ from the best-fitting result; $\Delta I_\nu^{\rm in}=0.021
~\mu$Jy/arcsec$^2$ and  $0.029$ $\mu$Jy/arcsec$^2$  for \targeta\ and
\targetb, respectively. We regard $\sqrt{2}$ times this value (i.e.,
$\Delta I_\nu^{\rm in}=0.030 ~\mu$Jy/arcsec$^2$ and  $0.042$
$\mu$Jy/arcsec$^2$  for \targeta\ and \targetb, respectively) as the
1$\sigma$ systematic uncertainty in the difference $I_{\nu}^{\rm in}(\vec{\theta})
	 - I_{\nu}^{\rm in}(\vec{\theta}_{\rm ref})$ recovered using 
	 equation (\ref{eq-conv2}).

We also examined if the subtraction of compact sources as described in section \ref{sec-source} affects the measurements of the SZE by adding sources C1 and C2 
(table \ref{tab-source1})  to the simulations with $y_{\rm peak}=1.1\times 10^{-4}$ 
for \targeta; the other sources are irrelevant to the measured SZE profiles. 
 Repeating the simulations 10 times varying noise seeds, we found that the errors associated with the subtraction of C1 and C2 are fully statistical with no seizable systematic effects on $I_\nu^{\rm out}$. At arbitrary positions in the cluster, we take the difference of $I_\nu^{\rm out}$ with respect to the point-source-free run with the same noise seed, regard its RMS value over the 10 realizations as an additional error in $I_\nu^{\rm out}$ due to the source subtraction, and add it to the errors of the measured SZE profiles in sections \ref{sec-yprof} and \ref{sec-deproj}.

Table \ref{tab-syst} summarizes the systematic errors mentioned above, together with those included in the subsequent analysis.

  \begin{table*}[h]
  
   \caption{List of systematic errors included in the analysis of sections \ref{sec-yprof} and \ref{sec-deproj}.} 
  \begin{center}
    \begin{tabular}{lll}
     \hline 
      Error source & \targeta\ & \targetb\ 
      \\ \hline
     Missing flux correction & $0.030~\mu$Jy/arcsec$^2$ & $0.042~\mu$Jy/arcsec$^2$\\
     Point source subtraction & $0.117~\mu$Jy/arcsec$^2$ ~($0<\theta<4"$) &  $-$\\
     & $0.023~\mu$Jy/arcsec$^2$ ~($4"<\theta<8.4"$)& $-$\\
     & $0.021~\mu$Jy/arcsec$^2$ ~($0<\theta<10"$)& $-$\\
     Flux calibration of ALMA at 92GHz & $6\%$ of the SZE intensity & $6\%$ of the SZE intensity\\
     Effective area of \Chandra\ ACIS-S & $4\%$ of the X-ray intensity& $-$\\
     \hline
    \end{tabular}
  \end{center}
 \label{tab-syst}
  \end{table*}
\subsection{Residual SZE images}
\label{sec-diff}

Figure \ref{fig-diff} shows residual images after the elliptical models
described in section \ref{sec-sim} are subtracted from the real ALMA
images in figures~\ref{fig-xszJ2319} and \ref{fig-szJ0947}.  The
subtracted models are essentially the same as those plotted in panel (b)
of figures \ref{fig-simJ2319} and \ref{fig-simJ0947}; the intrinsic
intensity has been corrected by equation \ref{eq-conv} and smoothed to
$5''$ FWHM.

The residual images exhibit moderate ($2 \sim 3\sigma$) levels of
deviations, which should comprise real departures from symmetry and
noise. The RMS value of the residuals for \targetb, 
 6.8 $\mu$Jy/beam, is
slightly larger than that for \targeta, 6.5 $\mu$Jy/beam.  Apart from
these deviations, the overall morphology of the real ALMA image is
well reproduced by the elliptical model for each cluster.

\subsection{Inner pressure profiles}
\label{sec-yprof}

The SZE images provide a direct probe of integrated electron pressure,
i.e., the Compton $y$-parameter. We used the results of section
\ref{sec-sim} to reconstruct the Compton $y$-parameter profile as
follows.

\subsubsection{Method}

The observed intensity was averaged over elliptical bins
to take into account elongated morphology of the clusters; the axis
ratio and the position angle were fixed at the best-fit values listed in
tables \ref{tab-betafit} and \ref{tab-gaussfit} for \targeta\ and
\targetb, respectively. The bins are geometrically spaced with the
inner-most bin at $ 0 < \bar{\theta} < 4''$ and the bin width increasing by a factor of 1.1, so that the statistical error in the final $y$-parameter profile is less than about 20\% in each bin
(an increasing bin width alleviates the decreasing S/N with radius).

The intrinsic SZE
intensity at 92 GHz was then computed for each bin using equation
(\ref{eq-conv2}) and taking the 7th bin containing $\bar{\theta}=35''$
(centered at $\bar{\theta}=34.4''$) as the reference points
($\vec{\theta}_{\rm ref}$).  
The systematic errors associated with
the missing flux correction ($\Delta I_\nu^{\rm in}=0.030 ~\mu$Jy/arcsec$^2$ and
$0.042$ $\mu$Jy/arcsec$^2$ for \targeta\ and \targetb, respectively, estimated in section \ref{sec-syst}) and the absolute calibration uncertainty of ALMA (6\%;
\cite{Kitayama16}) were added in quadrature to the statistical error in
each bin.  For \targeta, the error from the source subtraction (section \ref{sec-syst}, table \ref{tab-syst}), estimated as $0.117$ and 
$0.023~\mu$Jy/arcsec$^2$ for the inner-most and the second inner-most bins, respectively,  was also added in quadrature.  

Finally, the intrinsic SZE intensity is proportional to the
Compton $y$-parameter times the temperature-dependent relativistic
correction factor $c_{\rm rel}$, defined to be the ratio between the
true SZE intensity and the SZE intensity in the non-relativistic
limit. For a range of temperatures considered in this paper, $3< kT
< 10$ keV, $c_{\rm rel}$ is in the range $0.94 < c_{\rm rel} < 0.98$ at
92 GHz (e.g., \cite{Itoh04}).  To eliminate the dependence
of the data points on the temperature, $c_{\rm rel}$ is used for
correcting the model predictions and not the data  in section \ref{sec-yprofres}. 

\begin{figure*}[tp]
 \begin{center}
    \includegraphics[width=8.4cm]{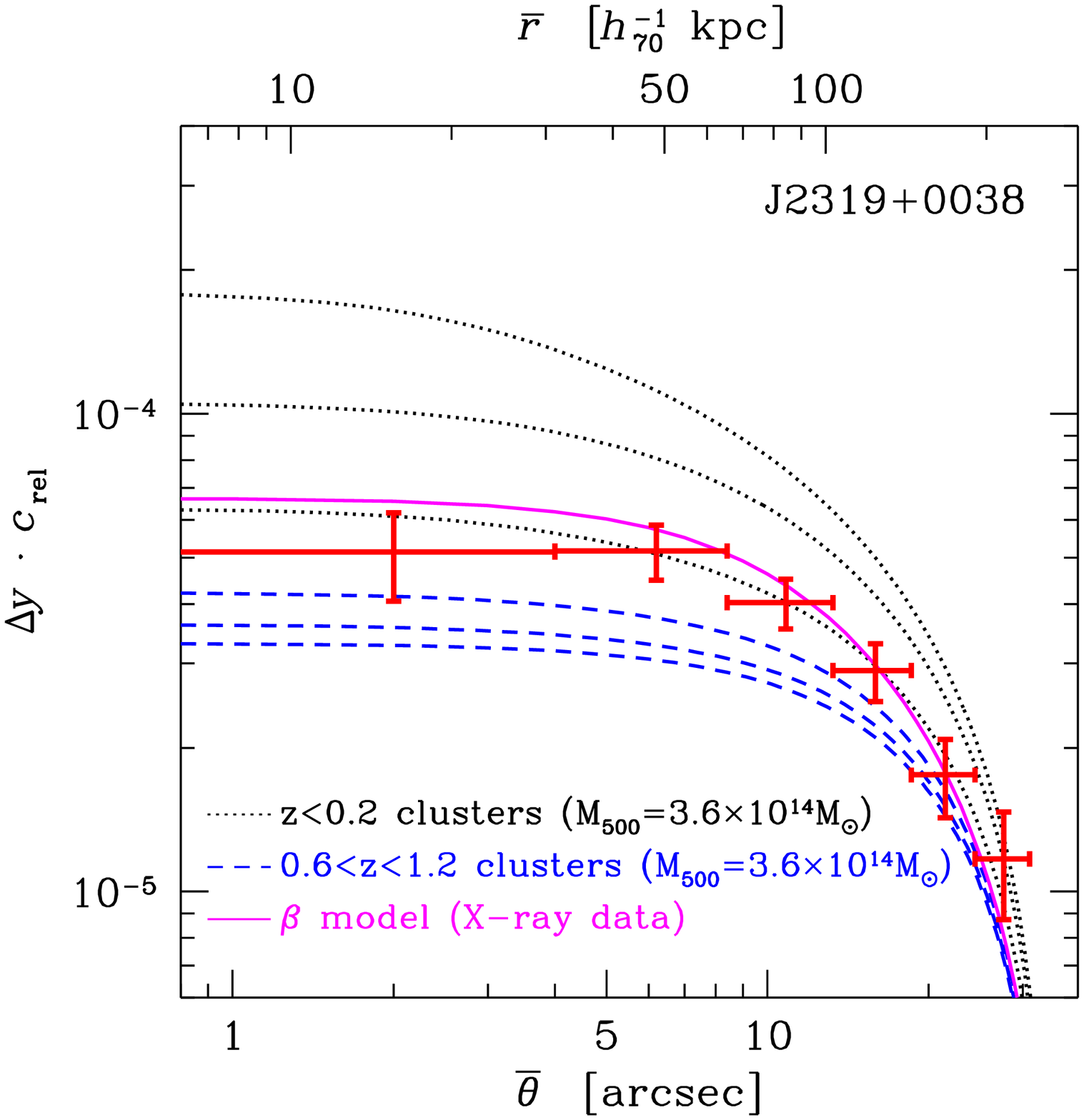}
  \includegraphics[width=8.4cm]{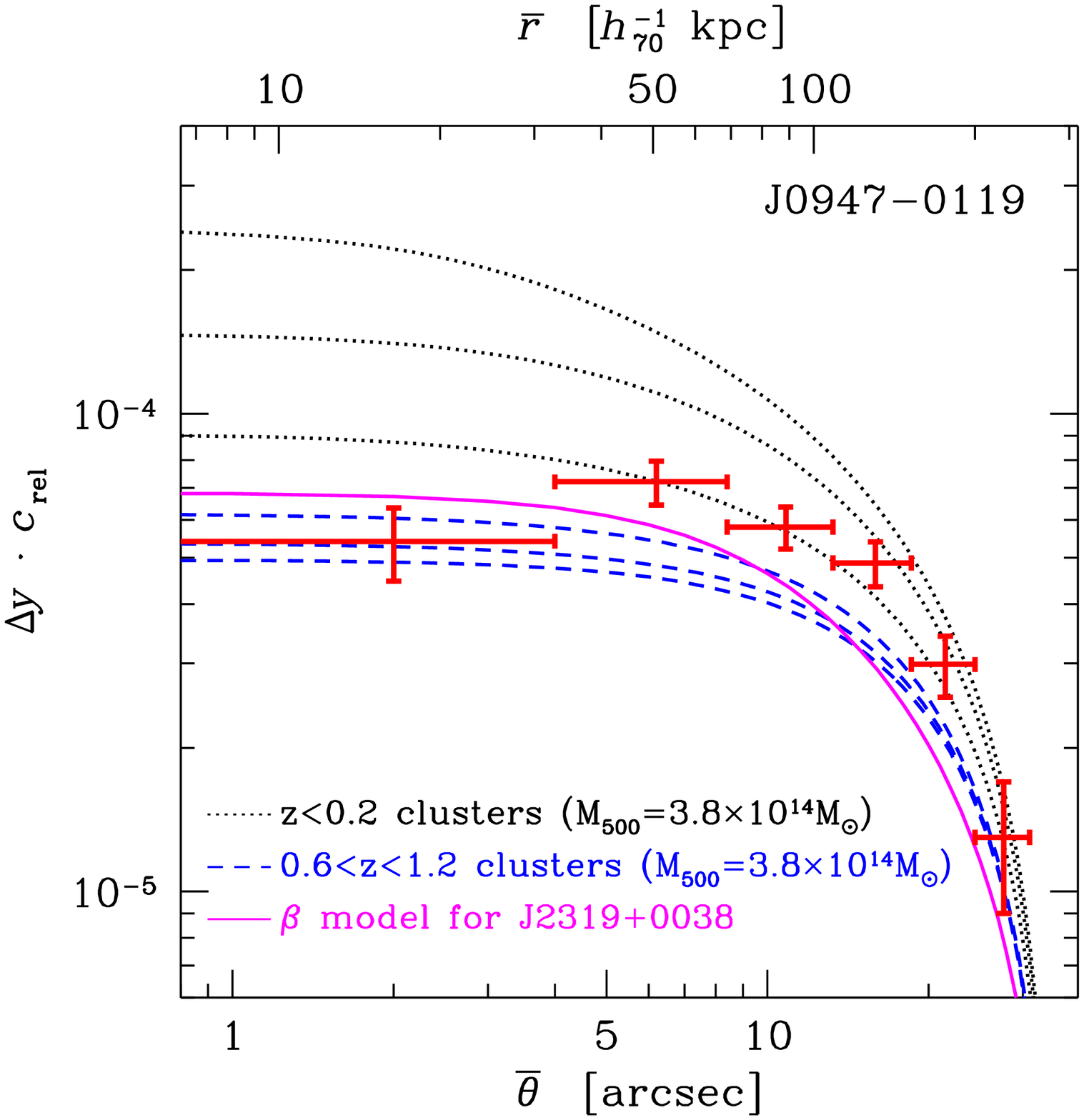}   
  \includegraphics[width=8.4cm]{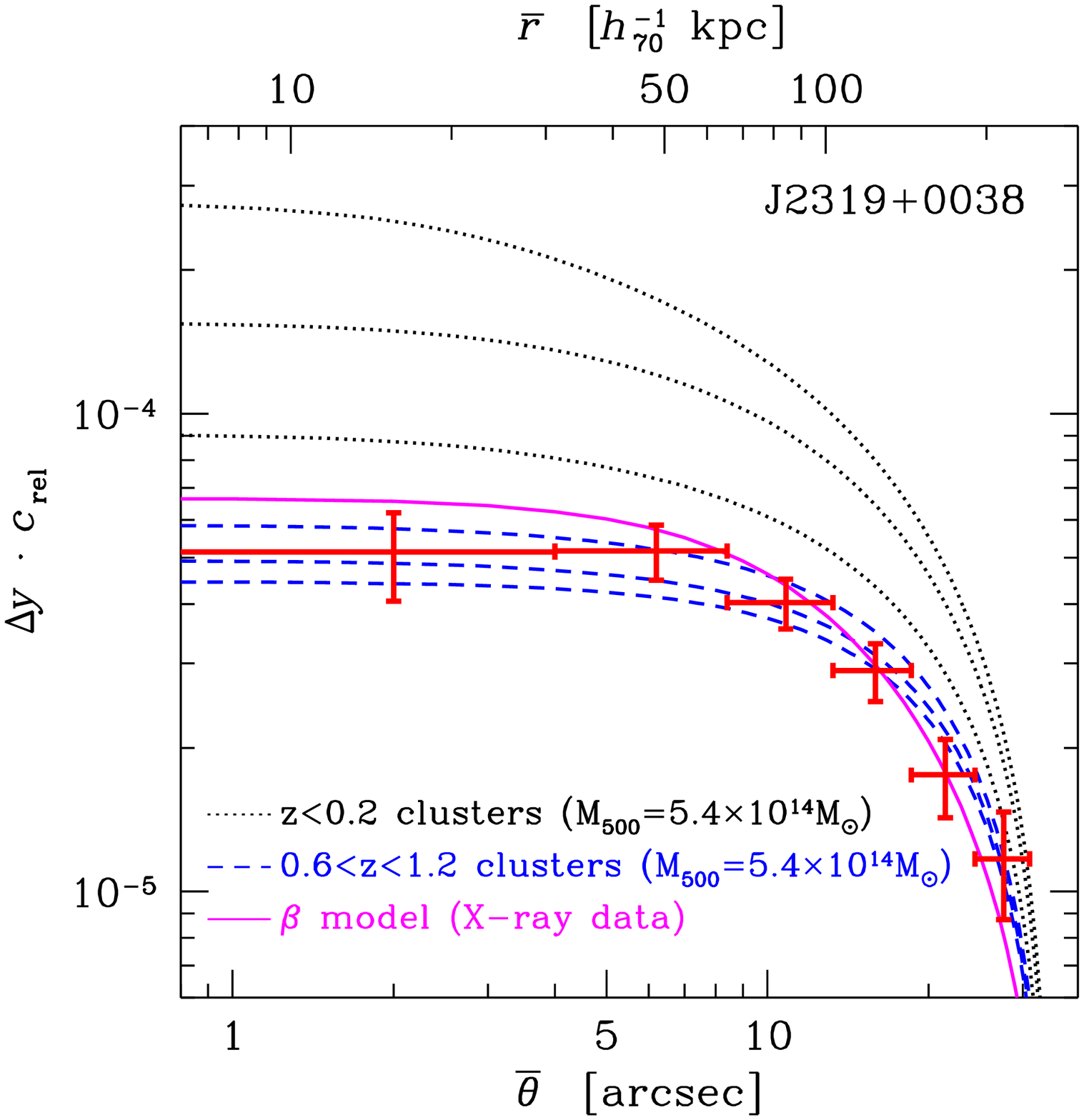}
  \includegraphics[width=8.4cm]{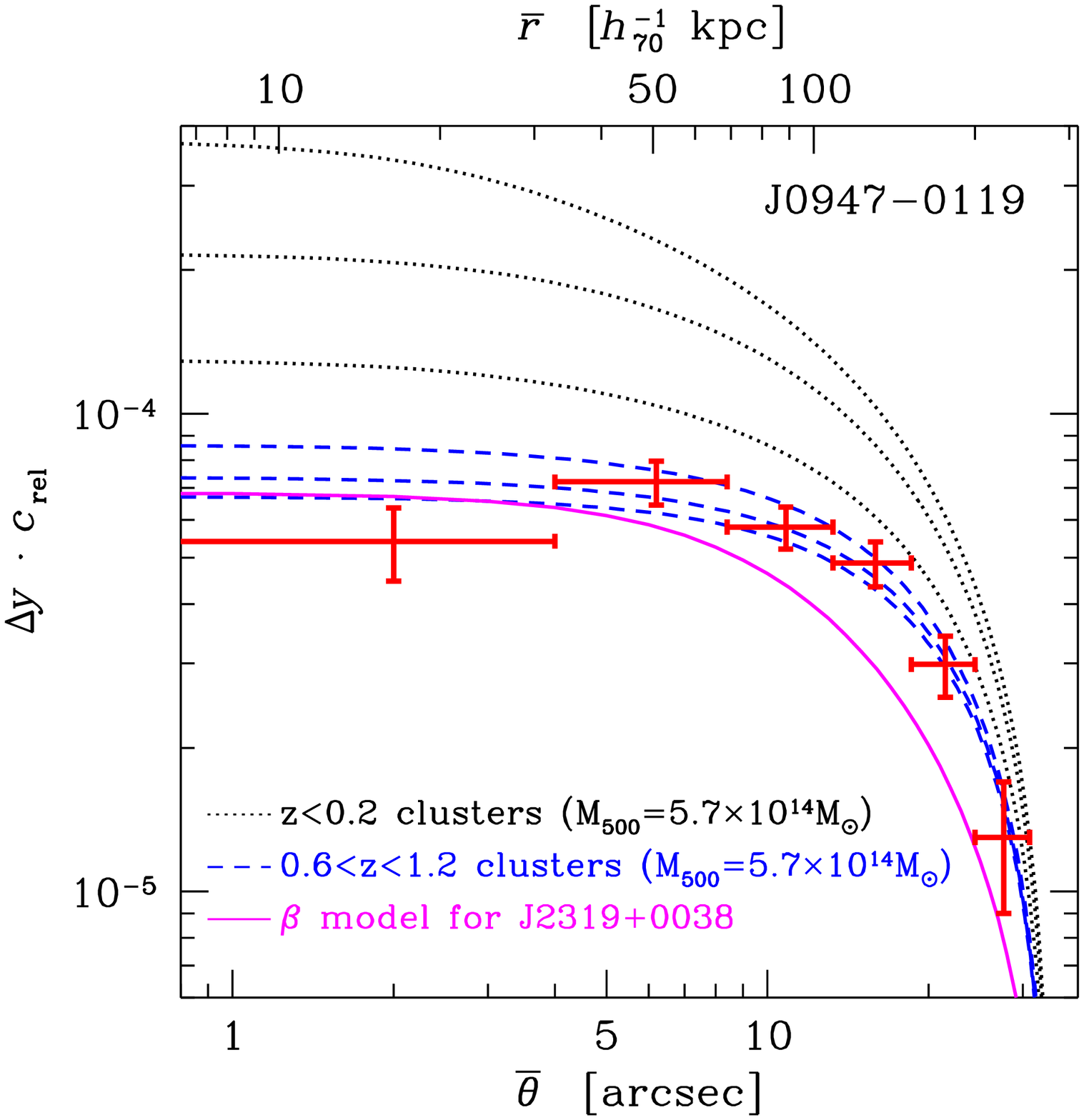}   
 \end{center}
 \caption{Azimuthally averaged Compton $y$-parameter of \targeta\ (left
 panels) and \targetb\ (right panels) as a function of the mean angular radius $\bar{\theta}$, or the corresponding physical size $\bar{r} = \bar{\theta} d_{\rm A}$.  To eliminate the dependence of the
 data points (error bars) on the temperature, the $y$-parameter
 times the relativistic correction factor $c_{\rm rel}$ is plotted (see
 text for details).  Overlaid are the expectations from the generalized
 NFW pressure profile for fixed $M_{500}$ of X-ray selected clusters at $z<0.2$ (dotted lines) by \citet{Arnaud10} and SZE selected clusters at
 $0.6<z<1.2$ (dashed lines) by \citet{McDonald14b}; three lines
 correspond to ``cool-core'', ``all'', and
 ``non-cool-core'' clusters in each sample from top to bottom. 
 The value of $M_{500}$ is fixed 
 at $3.6 \times 10^{14}M_\odot$ (top-left),  $3.8 \times
 10^{14}M_\odot$ (top-right), $5.4 \times 10^{14}M_\odot$ (bottom-left), and $5.7 \times
 10^{14}M_\odot$ (bottom-right) }as marked by an asterisk in 
 table \ref{tab-yfit}. The solid line shows the $\beta$ model profile
 inferred from the X-ray data of \targeta. All the lines adopt
 $h_{70}=1$ and $\eta=1$.  Both the data points and the
 expectations are relative to the positions at $\theta=34.4''$.
 \label{fig-yprof}
\end{figure*}

\subsubsection{Reconstructed $y$-parameter profile}
\label{sec-yprofres}

Figure \ref{fig-yprof} compares the Compton $y$-parameter profile so
obtained with various model predictions. By construction the model predictions, not the data points presented, are influenced by the relativistic correction via the assumed temperature and the values of $h_{70}$ and $\eta$ (i.e., introduced in equations \ref{eq-nbeta} and \ref{eq-3daxis}, respectively). 
All the model predictions are
smoothed to the same beam size and averaged over the same radial bins as
the data in the analysis; unbinned predictions are shown for display purposes only.

We first examine the isothermal $\beta$ model inferred from the X-ray
data of \targeta; $kT=5.90$ keV, $n_{\rm e0}=1.43 \times 10^{-2} h_{70}^{1/2}
\eta^{-1/2}$ cm$^{-3}$, $\beta=0.673$ and $r_{\rm c}= 103 \
h_{70}^{-1}$kpc. The observed $y$-parameter profile of \targeta\ is
consistent with this model and provides a useful limit on $h_{70}/\eta$.  If
$h_{70}/\eta$ is varied as a free parameter, the best-fit value is
\begin{equation}
 \frac{h_{70}}{\eta} = 1.07^{+0.15}_{-0.12} 
\end{equation}
with the minimum $\chi^2$ of 3.1 for 5 degrees of freedom (dof).
Taking $h_{70}/\eta=1$ as in the left panel of figure
\ref{fig-yprof} gives $\chi^2=3.4$ and is in reasonable agreement with
the data.

We also consider the generalized NFW pressure profile \citep{Nagai07}
with model parameters obtained from a sample of X-ray selected clusters
at $z<0.2$ by \citet{Arnaud10} or SZE selected clusters at $0.6<z<1.2$
by \citet{McDonald14b}.  Specifically, we used equation (13) of
\citet{Arnaud10} with the parameter values listed in table~5 of
\citet{McDonald14b} for either ``cool-core'', ``non-cool-core'', or
``all'' clusters in each sample and allowed the characteristic mass 
$M_{500}$ to vary. 
The relativistic correction was computed for the
temperature specified by $M_{500}$ and $z$ from the scaling relation of
\citet{Reichert11}. 
The SZE intensity predicted from the generalized NFW
pressure profile varies as $I_{\rm SZ} \propto c_{\rm
rel} (M_{500}h_{70})^{2/3 + 0.12} h_{70}^{1/2} \eta R_{500} \simeq
M_{500}^{1.10} h_{70}^{0.60} \eta$, where $R_{500} \propto M_{500}^{1/3}
h_{70}^{-2/3}$, $c_{\rm rel}\propto T_{\rm e}^{-0.03}$ for $3 < kT_{\rm
e}<10$ keV at 92 GHz, and $M_{500} h_{70} \propto T_{\rm e}^{1.62}$ from
equation (23) of \citet{Reichert11}.  Fitting the observed SZE intensity
hence gives approximately
\begin{equation}
 M_{500} \propto h_{70}^{-0.55}\eta^{-0.91}.
\end{equation}
With this variation in mind, we present the results with $h_{70}=1$ and
$\eta=1$ in what follows. 

  \begin{table*}[ht]
   \caption{The results of fitting the $y$-parameter profile by
   generalized NFW models with $h_{70}=1$ and $\eta=1$;
   the fitted mass varies approximately as $M_{500}
   \propto h_{70}^{-0.55} \eta^{-0.91}$ (see text). The model profiles
   are sorted so that, for given $M_{500}$ and $z$, the inner pressure
   gradient gets shallower in descending order. The values marked by an asterisk are used in figure~\ref{fig-yprof}.}
   
  \begin{center}
    \begin{tabular}{c|cc|cc}
     \hline
    Assumed pressure profile &  \multicolumn{2}{c|}{\targeta\ at
     $z=0.90$}  &   \multicolumn{2}{c}{\targetb\ at $z=1.11$}  \\
     & $M_{500}$ [$10^{14} M_\odot$]& $\chi^2/$dof & $M_{500}$ [$10^{14}M_\odot$]&
     $\chi^2/$dof  \\ \hline  
     $z<0.2$, cool-core & $2.1\pm 0.1$ & $11/5$ & $2.1\pm 0.1$ & $44/5$\\
     $z<0.2$, all & $2.5\pm 0.1$ & $3.8/5$ & $2.7 \pm 0.1$ & $24/5$\\
     $z<0.2$, non-cool-core & $^{*}3.6\pm 0.2$ & $1.3/5$ & $^{*}3.8 \pm 0.2$ & $11/5$\\
     $0.6 < z < 1.2$, cool-core & $4.8\pm 0.3$ & $2.3/5$ & $5.0 \pm 0.3$ &
     $5.5/5$ \\
     $0.6 < z < 1.2$, all& $^{*}5.4\pm 0.4$ & $3.3/5$  & $^{*}5.7 \pm 0.4$ &
     $4.7/5$ \\
     $0.6 < z < 1.2$, non-cool-core &$6.0\pm 0.5$ & $3.8/5$  & $6.1 \pm 0.4$ &
     $4.5/5$ \\               
     \hline
    \end{tabular}
  \end{center}
 \label{tab-yfit}
  \end{table*}

Table \ref{tab-yfit} lists the values of $M_{500}$ fitted 
to the observed $y$-parameter profile of each cluster. 
As representative cases, we plot 
in figure~\ref{fig-yprof} the model predictions for 
$M_{500}$ {\it fixed} at the best-fit value for ``non-cool-core clusters at $z<0.2$" (top panels) or ``all clusters at $0.6<z<1.2$" (bottom panels) as marked by 
an asterisk in table~\ref{tab-yfit}. Notice that the predictions for the other models, shown for reference, do {\it not} correspond to the overall best-fit. The agreement between the model and data improves upon allowing $M_{500}$ the freedom to vary.

The $y$-parameter profile of \targeta\ is consistent with the average
pressure profiles at $0.6<z<1.2$ or a slightly steeper profile of non-cool-core
clusters at $z<0.2$.  It is still much shallower
than the average of cool-core clusters at $z<0.2$.  The fitted
value of $M_{500}$ for non-cool-core clusters at $z<0.2$ or for
cool-core clusters at $0.6<z<1.2$ is in reasonable agreement with
$M_{500}=4.01^{+0.38}_{-0.39} \times 10^{14} M_\odot$ inferred from
the \Chandra\ X-ray data of \targeta\ assuming hydrostatic equilibrium
\citep{Hicks08}, or with $M_{500} = 3.6 ^{+4.6}_{-2.2} \times 10^{14}
M_\odot$ from our weak lensing analysis (section \ref{sec-hsc}). For
reference, \citet{Hilton21} reports the value\footnote{ The mass
quoted here is $M_{500c}^{\rm UPP}$ in \citet{Hilton21}.}  $M_{500} =
2.08^{+ 0.43}_{- 0.36} \times 10^{14} M_\odot$ from the ACT SZE data
assuming the average pressure profile at $z<0.2$ by \citet{Arnaud10},
whereas the value inferred from the mass-richness relation of
\citet{Okabe19} is $M_{500} \simeq 1.8 \times 10^{14} M_\odot$.

The $y$-parameter profile of \targetb\ appears to be even shallower than that of
\targeta\ and reproduced well by the average pressure profile of
non-cool-core clusters at $0.6<z<1.2$. It is inconsistent with either
average profiles of local clusters or the same $\beta$ model as
\targeta\ ($\chi^2/\mbox{dof}=39/6$).   For reference, $M_{500} =
2.6 ^{+3.2}_{-1.4} \times 10^{14} M_\odot$ is inferred from our weak
lensing analysis (section \ref{sec-hsc}), $M_{500} = 4.05^{+ 0.64}
_{- 0.56} \times 10^{14} M_\odot$ from the ACT SZE data of \targetb\
assuming the average pressure profile at $z<0.2$ \citep{Hilton21}, and
$M_{500} \simeq 5.7 \times 10^{14} M_\odot$ from the mass-richness
relation of \citet{Okabe19}.

Note that the above $y$-parameter profiles
are derived around the X-ray center of  \targeta\ and 
the SZE center of
and \targetb, given the lack of high-resolution X-ray data for the latter (section \ref{sec-icm}).  The shallow pressure profile of \targetb\ could hence be due partly to a mismatch, if any,  between the SZE center and the gas density peak, seen frequently in merging clusters. Future X-ray observations of \targetb\ will be useful for investigating this point further.

\subsection{Deprojected electron temperature and density of \targeta}
\label{sec-deproj}

Resolved SZE and X-ray brightness images of \targeta\ further allow us
to constrain temperature and density profiles, 
thereby relaxing the isothermal approximation. The procedure is an
extension of that described in \citet{Kitayama20} to an ellipsoidal gas
profile.

\subsubsection{Method}

The observed SZE or X-ray 
brightness was averaged over an elliptical annulus with the
axis ratio and the position angle fixed at the best-fit values in
table \ref{tab-betafit}. The bin size is $\Delta \bar{\theta} = 10''$
so that the statistical error in the brightness is less than 10\% in
each bin at $\bar{\theta} < 30''$.  The intrinsic SZE intensity was
computed for each bin using equation (\ref{eq-conv2}) and taking the 4th
bin centered at $\bar{\theta}=35''$ as the reference points
($\vec{\theta}_{\rm ref}$). 

Systematic errors from the flux calibration
of ALMA (6\%, \cite{Kitayama16}), the missing flux correction of the SZE
(0.030 $\mu$Jy/arcsec$^2$, section \ref{sec-syst}), and the effective area
of \Chandra\ ACIS-S (4\%) were added in quadrature to the
statistical error. The error from the source subtraction , estimated as $0.021~\mu$Jy/arcsec$^2$ for the 
inner-most bin (section \ref{sec-syst}, table \ref{tab-syst}), was also added in quadrature.   

We then fit the volume averaged brightness in
each elliptical annulus (in 2D) of the SZE at $\bar{\theta} < 30''$ and
of the 0.4--7.0 keV X-rays at $\bar{\theta} < 60''$ together varying the
temperature and the density in each ellipsoidal shell (in 3D); the
orientation of the axis was assumed to be the same as the triaxial
$\beta$ model described in section \ref{sec-simmodel}. As the temperature at
$\bar{\theta} > 30''$ ($\bar{r} > 230$ kpc) cannot be constrained by the
ALMA data, it was fixed at the projected mean value of 5.90 keV from the
X-ray spectral analysis described in section \ref{sec-sim}\footnote{ We checked
that the deprojected quantities at $\bar{\theta} < 20''$ (160 kpc) are insensitive to this assumption; they change within $\pm 3\%$ if the temperature at $\bar{\theta} > 30''$ is varied within the $1\sigma$ error range of $5.90^{+0.79}_{-0.62}$ keV. }. For the SZE, we modeled incremental brightness relative to the bin centered at
$\bar{\theta}=35''$ taking account of the temperature-dependent
relativistic correction in each ellipsoidal shell. The X-ray emissivity
was computed by SPEX version 3.0.6.01 \citep{Kaastra96,Kaastra20}.

From the reconstructed electron temperature and
density, we evaluated entropy $K$ and the radiative cooling time $t_{\rm
cool}$ in each shell by
\begin{eqnarray}
 K &=&\frac{kT_{\rm e}}{n_{e}^{2/3}}, \\
 t_{\rm cool} &=& \frac{3}{2}\frac{(n_{\rm e} + n_{\rm H} + n_{\rm
  He})kT_{\rm e}}{n_{\rm e} n_{\rm H}\Lambda_{\rm bol}},
\end{eqnarray}
where $n_{\rm H}$ and $n_{\rm He}$ are the number
densities of hydrogen and helium atoms, respectively, and $n_{\rm e}
n_{\rm H}\Lambda_{\rm bol}$ denotes the bolometric luminosity per unit
volume. We obtained $\Lambda_{\rm bol}$ by integrating the rest-frame
X-ray emissivity from $0.1$ eV to 1 MeV (e.g., \cite{Schure09}).

\begin{figure*}[tp]
 \begin{center}
  \includegraphics[width=8.4cm]{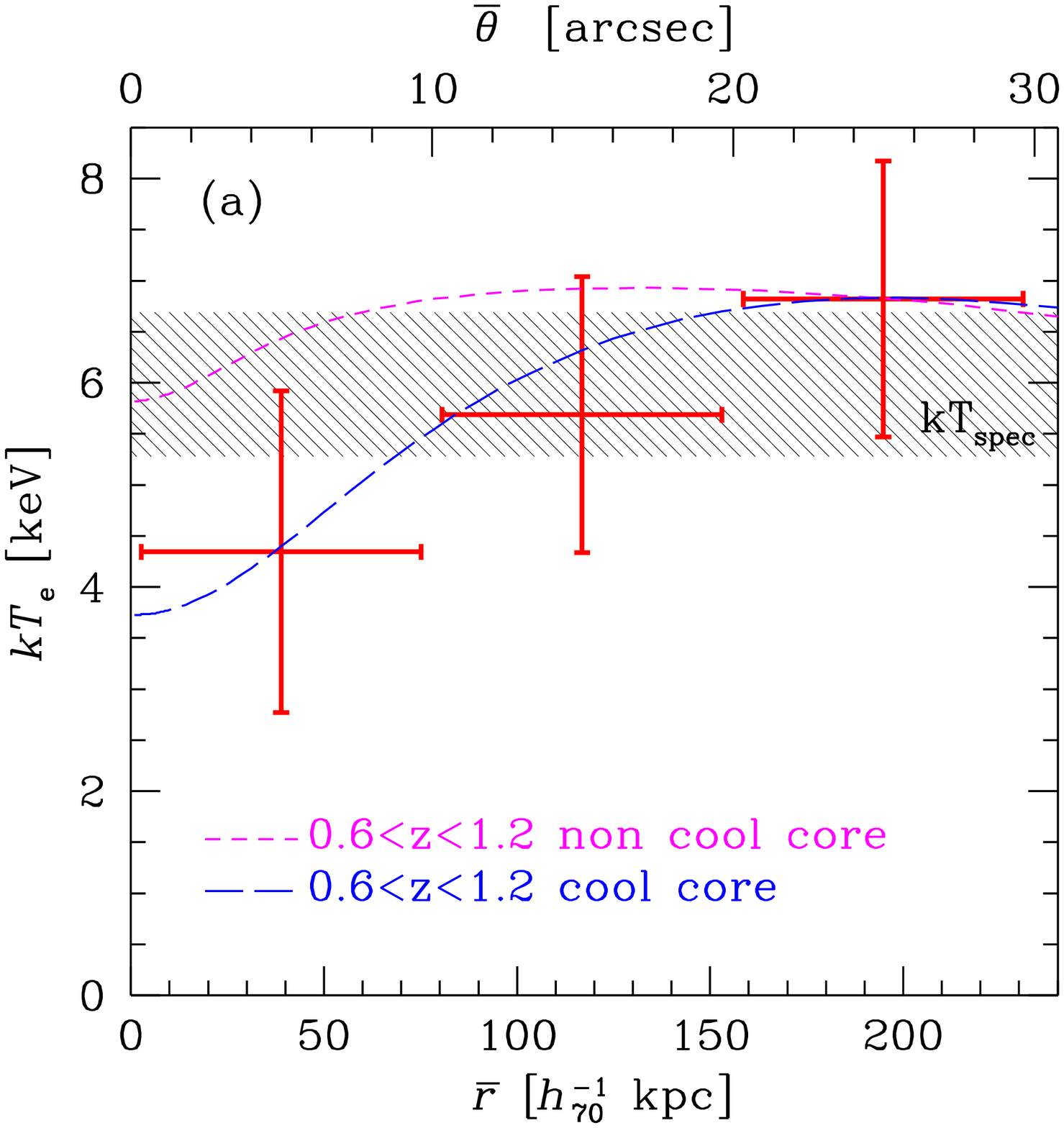}
  \includegraphics[width=8.4cm]{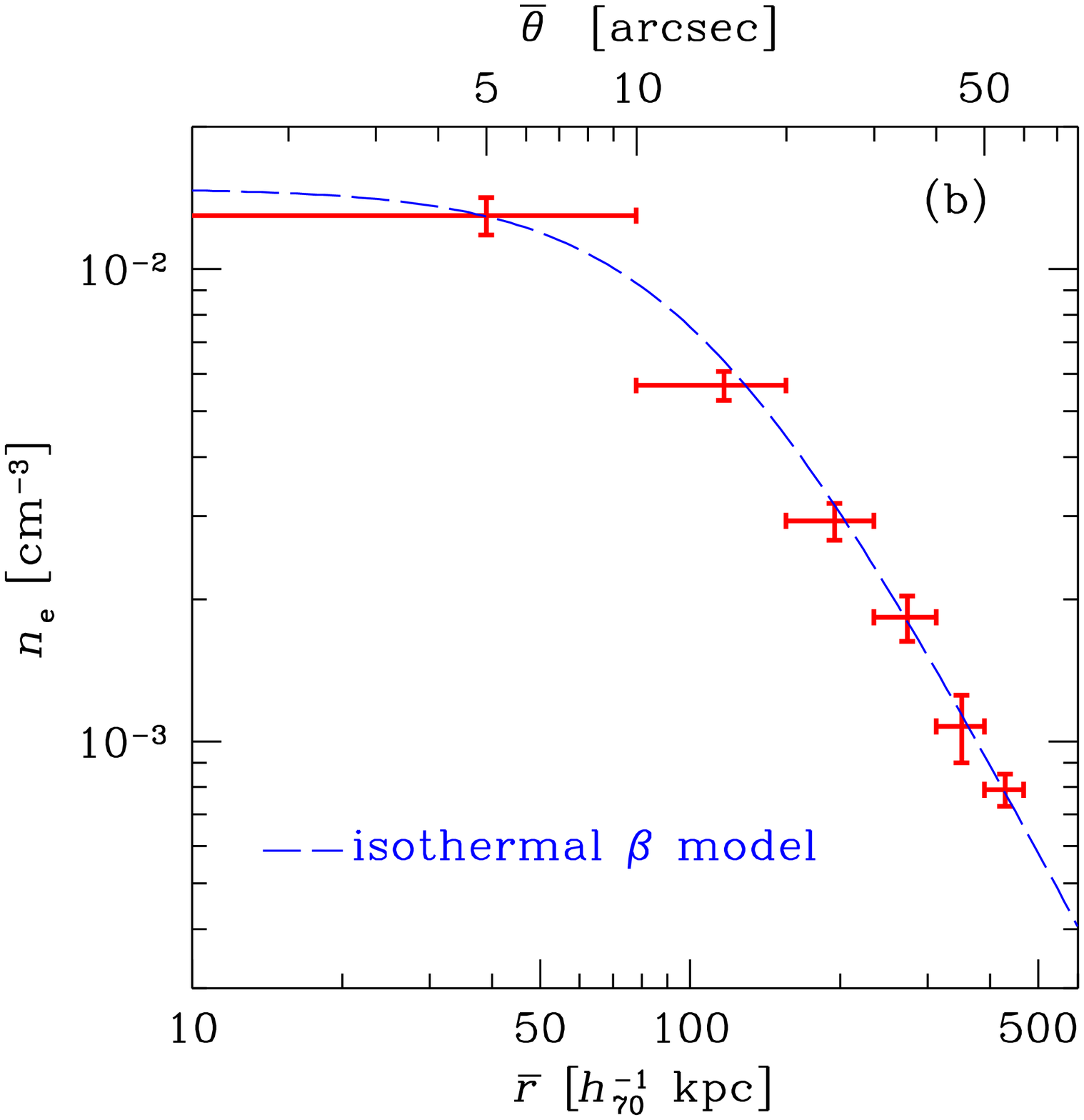}
  \includegraphics[width=8.4cm]{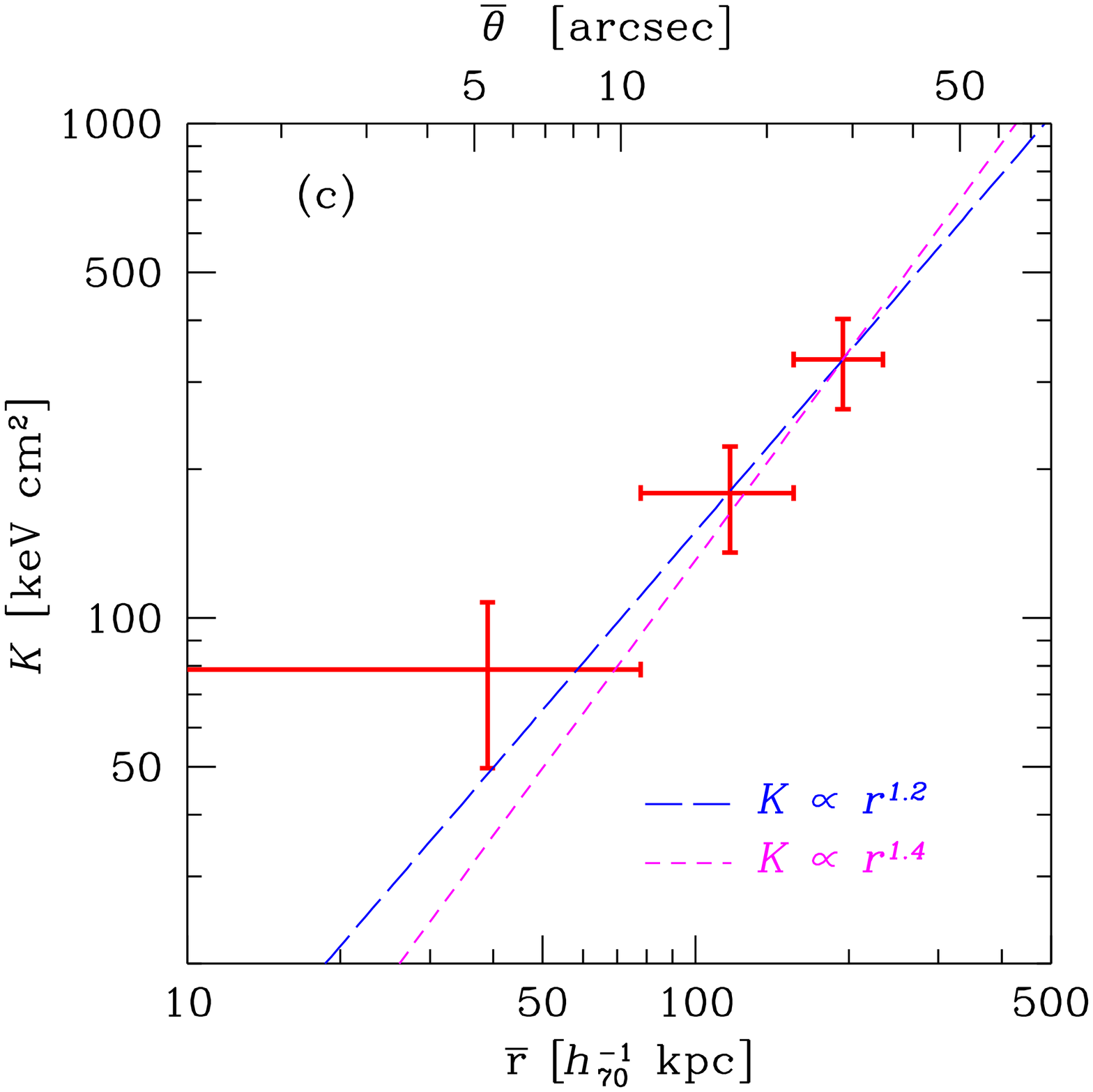}
  \includegraphics[width=8.4cm]{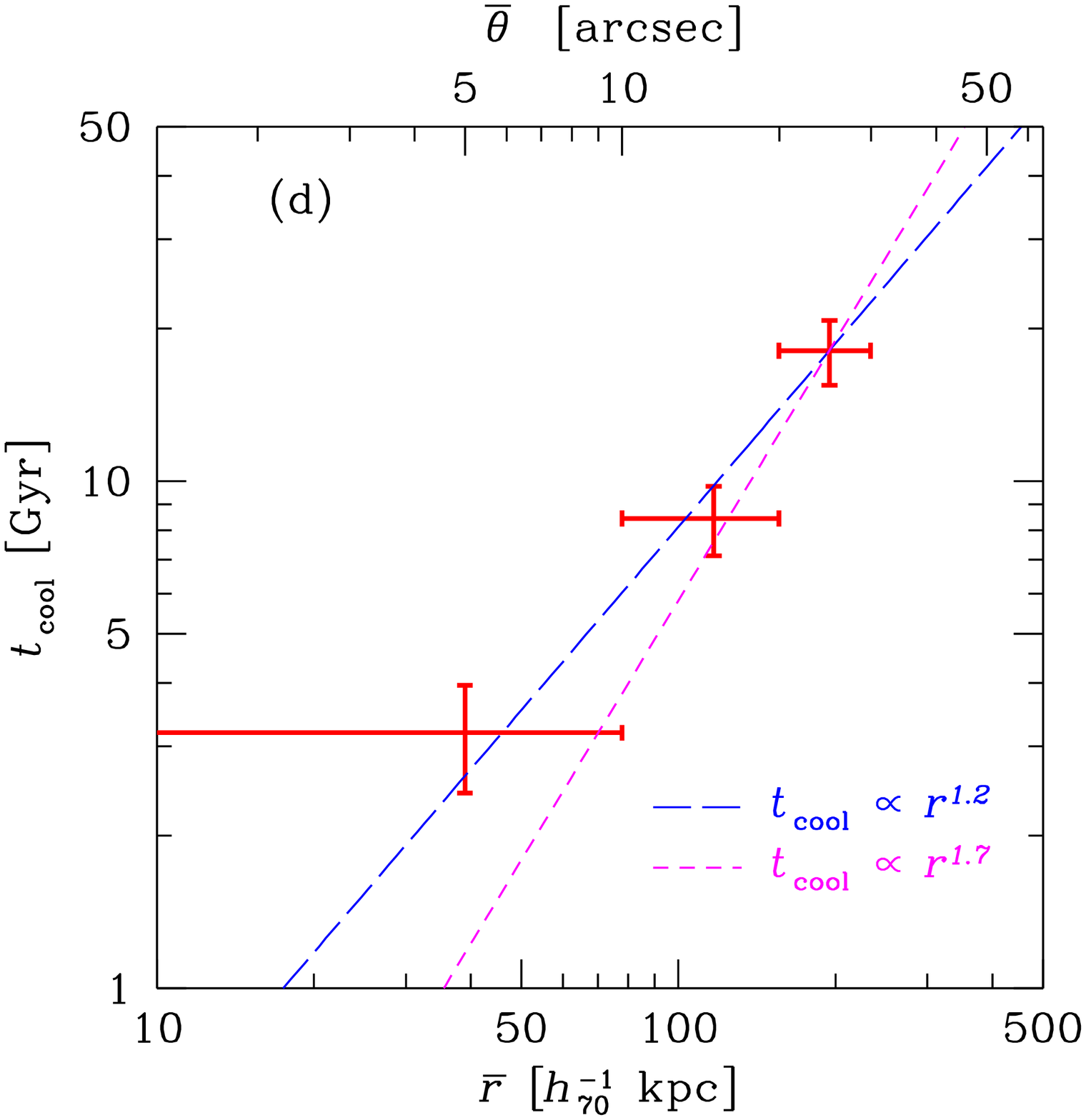}     
 \end{center}
 \caption{Deprojected quantities of \targeta\ from SZE and X-ray images as a function of the mean radius $\bar{r}$, or the corresponding angular size $\bar{\theta}=\bar{r}/d_{\rm A}$.  
 (a) Electron temperature. Lines indicate the mean profiles of
 non-cool-core (short-dashed) and cool-core (long-dashed) clusters at
 $0.6<z<1.2$ by \citet{McDonald14b}, both normalized to match the data
 point in the outer-most bin.  Shaded region show the $1\sigma$ range
 inferred from the X-ray spectrum within $ 45''$ around the center
 of \targeta.  (b) Electron density. The dashed line indicates the
 isothermal $\beta$-model (equation \ref{eq-nbeta}) inferred from the X-ray
 data alone.  (c) Entropy.  Lines indicate the model profiles $K\propto
 r^{1.2}$ (long-dashed; \cite{Voit05}) and $K\propto r^{1.4}$
 (short-dashed; \cite{Voit11}), both normalized to match the data point
 in the outer-most bin. (d) The radiative cooling
 time. Lines indicate the relations $t_{\rm cool}
 \propto r^{1.2}$ (long-dashed) and $t_{\rm cool} \propto r^{1.7}$
 (short-dashed; \cite{Kitayama20}), both normalized to match the data
 point in the outer-most bin.  For definiteness,
 $h_{70}=1$ and $\eta=1$ are assumed; the data points vary approximately
 as $T_{\rm e} \propto \left(\frac{h_{70}}{\eta}\right)^{0.70}$, $n_{\rm
 e} \propto \left(\frac{h_{70}}{\eta}\right)^{0.32}$, $ K \propto
 \left(\frac{h_{70}}{\eta}\right)^{0.49}$, and $t_{\rm cool} \propto
 \left(\frac{h_{70}}{\eta}\right)^{0.09}$ (see text).}
 \label{fig-deproj}
\end{figure*}

The quantities obtained by a joint SZE and X-ray
analysis have different dependencies on distance from those by the X-ray
spectral analysis. The predicted SZE and X-ray intensities vary
approximately as $I_{\rm SZ} \propto n_{\rm e} T_{\rm e}^{\alpha_{\rm
SZ}} h_{70}^{-1} \eta $ and $I_{\rm X} \propto n_{\rm e}^2 T_{\rm
e}^{\alpha_{\rm X}} h_{70}^{-1} \eta$, respectively, where $\alpha_{\rm
SZ} \simeq 0.97$ at 92 GHz and $\alpha_{\rm X} \simeq 0.52$ at $E_{\rm
obs}=0.4-7.0$ keV ($z=0.90$), over the range $3 < kT_{\rm e} < 10$ keV.
Jointly fitting $I_{\rm SZ}$ and $I_{\rm X}$ gives
\begin{eqnarray}
T_{\rm e} &\propto&
\left(\frac{h_{70}}{\eta}\right)^{\frac{1}{2\alpha_{\rm SZ}-\alpha_{\rm X}}} \simeq
\left(\frac{h_{70}}{\eta}\right)^{0.70}, \\
n_{\rm e} &\propto& 
\left(\frac{h_{70}}{\eta}\right)^{\frac{\alpha_{\rm SZ}-\alpha_{\rm X}}{2\alpha_{\rm
SZ}-\alpha_{\rm X}}} \simeq
\left(\frac{h_{70}}{\eta}\right)^{0.32}, \\ 
K &\propto&
\left(\frac{h_{70}}{\eta}\right)^{\frac{3 - 2 \alpha_{\rm SZ} +
2\alpha_{\rm X}}{3(2\alpha_{\rm SZ}  -\alpha_{\rm X})}} \simeq
\left(\frac{h_{70}}{\eta}\right)^{0.49}, \\
t_{\rm cool} &\propto& 
 \left(\frac{h_{70}}{\eta}\right)^{\frac{1 + \alpha_{\rm X} -
 \alpha_{\rm SZ} - 
\alpha_{\rm
 bol}}{2\alpha_{\rm SZ}-\alpha_{\rm X}}} \simeq
 \left(\frac{h_{70}}{\eta}\right)^{0.09},
 \label{eq-tc}
 \end{eqnarray}
 where we have used $\Lambda_{\rm bol} \propto T_{\rm e}^{\alpha_{\rm
 bol}}$ with $\alpha_{\rm bol} \simeq 0.42$ for $3 < kT_{\rm e} < 10$
 keV in equation (\ref{eq-tc}).  We present the results with $h_{70}/\eta=1$
 in figure~\ref{fig-deproj}.

 \subsubsection{Deprojected profiles}

Figure \ref{fig-deproj}(a) shows the deprojected electron temperature
profile of \targeta.  Also plotted for reference are the average
profiles of cool-core and non-cool-core clusters at $0.6<z<1.2$
\citep{McDonald14b}. The results are consistent with the mean X-ray
spectroscopic temperature of $kT_{\rm e}=5.90^{+0.79}_{-0.62} $ keV
(section \ref{sec-simmodel}).  The measured temperatures tend to decrease
moderately towards the center, in agreement with the average
profile of high-redshift cool-core clusters.  This is also consistent
with the fact that the observed $y$-parameter profile of \targeta\ is
better reproduced by a steeper pressure profile than the average of
non-cool-core clusters at $0.6<z<1.2$ (figure \ref{fig-yprof} and table
\ref{tab-yfit}).

The associated electron density profile is plotted in figure~\ref{fig-deproj}(b). It agrees well with the isothermal $\beta$ model
obtained solely from the X-ray data in section \ref{sec-simmodel}, which
provides a useful consistency test of the present
technique. The electron density profile of \targeta\
with $n_{\rm e0} \simeq 1.4 \times 10^{-2}$ cm$^{-3}$ and $r_{\rm
c}\simeq 100$ kpc is shallower than the average of cool-core clusters at
$0.2<z<1.9$ characterized by $n_{\rm e0} > 1.5 \times 10^{-2}$ cm$^{-3}$
and $r_{\rm c} \simeq 20-30$ kpc \citep{McDonald17}.

Figure \ref{fig-deproj}(c) illustrates that the entropy
decreases moderately to $\sim 80$ keV cm$^2$ within the core
($\theta_{\rm c} \simeq 13''$ or 100 kpc). Also plotted are a model
$K\propto r^{1.2}$ which tends to give a lower limit to non-radiative
clusters at $r \ltsim 0.5 r_{500}$ \citep{Voit05} and a prediction
$K\propto r^{1.4}$ for the steady-state cooling flow \citep{Voit11};
both profiles are normalized to match the data point in the outer-most
bin. The observed entropy profile shows a better agreement with
$K\propto r^{1.2}$ than $K\propto r^{1.4}$ indicating that radiative
cooling, if any, is modest in \targeta.

Figure \ref{fig-deproj}(d) further shows that the
radiative cooling time in the core is about half the age of the Universe
($\sim 6$ Gyr) at $z=0.9$. The radial profile of the cooling time is
represented approximately by a power-law of $t_{\rm cool} \propto
r^{1.2}$.  This is shallower than $t_{\rm cool} \propto r^{1.7}$
obtained from a similar analysis for the Phoenix cluster at $z=0.60$
\citep{Kitayama20}, in which efficient radiative cooling is suggested.

\subsection{Origin of the offset of the ACT SZE centroid of \targeta}
\label{sec-act}

The centroid position of \targeta\ in the ACT SZE map is offset to
the north-west direction by $\sim 30''$ (230 $h_{70}^{-1}$kpc) from the
centers of the ALMA SZE and the \Chandra\ X-ray images (figure
\ref{fig-xszJ2319}). We examined if this offset is due to the compact
millimeter sources (table \ref{tab-source1}) unresolved in the ACT map,
by means of cluster injection simulations similar to those done in
\citet{Hilton21}. 

We injected a model cluster and model sources to the
real ACT map, re-ran filtering and cluster-detection procedures, and
assessed the accuracy of the recovered cluster position. The compact
sources listed in table~\ref{tab-source1} were found to be far too weak
to produce the observed offset. The result is consistent with the fact that, 
according to figure 5 of \citet{Dicker21}, these sources would have changed the peak Compton $y$-parameter of \targeta\ in the ACT map ($\tilde{y}_0=(6.1 \pm 1.2)\times 10^{-5}$ in \cite{Hilton21})   
by at most $4\%$ for a spectral index of $-0.7$.  Figure 4 of \citet{Hilton21} further
indicates that the accuracy of position recovery in the ACT SZE maps is
sensitive to S/N and governed by the fluctuating cosmic microwave
background (CMB). At the S/N of 5.2 for \targeta, 22\% of the
injected clusters in the simulations are recovered at $> 30''$ away from
the position at which the cluster model was inserted. We thus conclude
that the observed offset in the ACT SZE map of \targeta\ is
statistically insignificant and likely due to the primary CMB
fluctuations.

We note that \targetb\ was detected by ACT with a higher S/N of 13.2
and exhibits agreement between the ACT and ALMA centroids
(figure \ref{fig-szJ0947}).  This also supports the above
interpretation that the position accuracy of the ACT SZE maps correlates
with S/N.

\section{Conclusions}

We have presented ALMA Band 3 measurements of the thermal SZE toward two
galaxy clusters at $z\sim 1$, \targeta\ and \targetb, and performed
joint analyses with available \Chandra\ X-ray data, optical data taken
by Subaru/HSC, and wider-field SZE data by ACT. 
Taking into account departures from spherical symmetry, we have
reconstructed the profiles of thermodynamic quantities
non-parametrically. This is one of the first such measurements for an
individual cluster at $z\simgt 0.9$.

In both clusters, the SZE is imaged at $5''$ resolution (corresponding
to the physical scale of $\sim 40 h_{70}^{-1}$kpc) within 300
$h_{70}^{-1}$kpc from the cluster center with the peak S/N exceeding
$7$.  The overall morphology of the SZE signal is well described by an
elliptical gas profile with axis ratio 
$\simlt 0.7$, clearly differing from unity. 
The elongated morphology is probably associated with highly asymmetrical galaxy
distributions found in these clusters. The inner pressure profile is
consistent with the average of clusters at $0.6<z <1.2$ by
\citet{McDonald14b} and is much shallower than that of local cool-core
clusters by \citet{Arnaud10}.

Of the two clusters studied in this paper, high quality X-ray data are
available only for \targeta. Both the centroid position and overall
morphology of the ALMA SZE map are in agreement with those of the
\Chandra\ X-ray brightness image.  We thus performed a non-parametric
deprojection of electron temperature, density, entropy, and the
radiative cooling time, combining SZE and X-ray images for \targeta. Our
results consistently indicate that \targeta\ hosts a weak cool core,
where radiative cooling is less significant than in local cool cores.
There are central radio sources as well as signs of subcluster
mergers.
We also suggest that the offset of the ACT SZE centroid
position seen in this cluster \citep{Hilton21} is likely due to primary
CMB fluctuations.

On the other hand, \targetb\ exhibits an even shallower pressure profile 
than \targeta\ and is more likely a non-cool-core cluster.  The
SZE centroid position is offset from the peaks of galaxy concentration
by more than 140 $h_{70}^{-1}$ kpc, suggesting stronger
impacts of mergers in this cluster than in \targeta. No
radio galaxies are found within 100 $h_{70}^{-1}$ kpc from either  
the SZE centroid or the peaks of galaxy concentration.
Additional X-ray observations will be quite useful for exploring the nature of this cluster further.

We conclude that both of these distant clusters
are at a very early stage of developing the cool cores typically found in
clusters at lower redshifts. Our results also imply that high angular resolution SZE
  observations provide a unique probe of thermodynamic structures of 
  such clusters. We have developed an image domain analysis that allows for a non-parametric reconstruction of physical quantities from the ALMA data. An alternative and complementary approach is to perform model 
  fitting in the visibility ($uv$) domain (e.g., \cite{Basu16,Luca19AA}), which we plan to investigate in our future publication. We expect that spectral and spatial ranges covered
  in future studies will be greatly enhanced by new facilities including
  the Band 1 receiver on ALMA \citep{Huang16}, the Large Submillimeter
  Telescope (LST; \cite{Kawabe16}), and the Atacama Large Aperture
  Submillimeter/millimeter Telescope (AtLAST; \cite{Klaassen19}). 

\begin{ack}
 We thank Grace Chesmore, Arthur Kosowsky, and Bruce Partridge for helpful comments. We also thank the anonymous referee for careful reading of the manuscript and insightful comments. 
 We are grateful to Daniel Espada,
 Atsushi Miyazaki, and Kazuya Saigo for their support on the ALMA data
 analysis.  This work was supported by JSPS KAKENHI Grant Numbers
 JP17H06130 (K.K.), JP18K03704 (T.K.), JP20H00181 (M.O.), JP20H01932
 (H.M.), JP20H05856 (M.O.), JP20K04012 (N.O.), JP21H00048 (S.T.),  
 JP21H01135 (T.A.), JP21H04495 (S.T.), and
 JP22H01260 (M.O.). S.U. acknowledges the support from Ministry of Science
 and Technology of Taiwan (MOST 111-2811-M-007-008 and 111-2112-M-001-026-MY3). J.P.H. acknowledges
 funding for SZE cluster studies from the NSF AAG program (grant number
 AST-1615657).

 This paper makes use of the following ALMA data:
 ADS/JAO.ALMA\#2018.1.00680.S and 2019.1.00673.S. ALMA is a partnership
 of ESO (representing its member states), NSF (USA) and NINS (Japan),
 together with NRC (Canada), MOST and ASIAA (Taiwan), and KASI (Republic
 of Korea), in cooperation with the Republic of Chile. The Joint ALMA
 Observatory is operated by ESO, AUI/NRAO and NAOJ. The National Radio
 Astronomy Observatory is a facility of the National Science Foundation
 operated under cooperative agreement by Associated Universities, Inc.
 
 This research has made use of data obtained from the \Chandra\ Data
 Archive (ObsID 5750, 7172, 7173, and 7174) and software provided by
 the Chandra X-ray Center (CXC) in the application packages CIAO and
 Sherpa.
 
 The Hyper Suprime-Cam (HSC) collaboration includes the astronomical
communities of Japan and Taiwan, and Princeton University. The HSC
instrumentation and software were developed by the National Astronomical
Observatory of Japan (NAOJ), the Kavli Institute for the Physics and
Mathematics of the Universe (Kavli IPMU), the University of Tokyo, the
High Energy Accelerator Research Organization (KEK), the Academia Sinica
Institute for Astronomy and Astrophysics in Taiwan (ASIAA), and
Princeton University. Funding was contributed by the FIRST program from
the Japanese Cabinet Office, the Ministry of Education, Culture, Sports,
Science and Technology (MEXT), the Japan Society for the Promotion of
Science (JSPS), Japan Science and Technology Agency (JST), the Toray
Science Foundation, NAOJ, Kavli IPMU, KEK, ASIAA, and Princeton
University.

This paper makes use of software developed for Vera C. Rubin Observatory. We thank the Rubin Observatory for making their code available as free software at http://pipelines.lsst.io/.

This paper is based on data collected at the Subaru Telescope and retrieved from the HSC data archive system, which is operated by the Subaru Telescope and Astronomy Data Center (ADC) at NAOJ. Data analysis was in part carried out with the cooperation of Center for Computational Astrophysics (CfCA), NAOJ. We are honored and grateful for the opportunity of observing the Universe from Maunakea, which has the cultural, historical and natural significance in Hawaii. 
 
\end{ack}

\appendix

\section{Line emissions from compact sources}
\label{sec-lines}

\begin{figure*}[tp]

 \begin{center}
  \includegraphics[width=7.2cm]{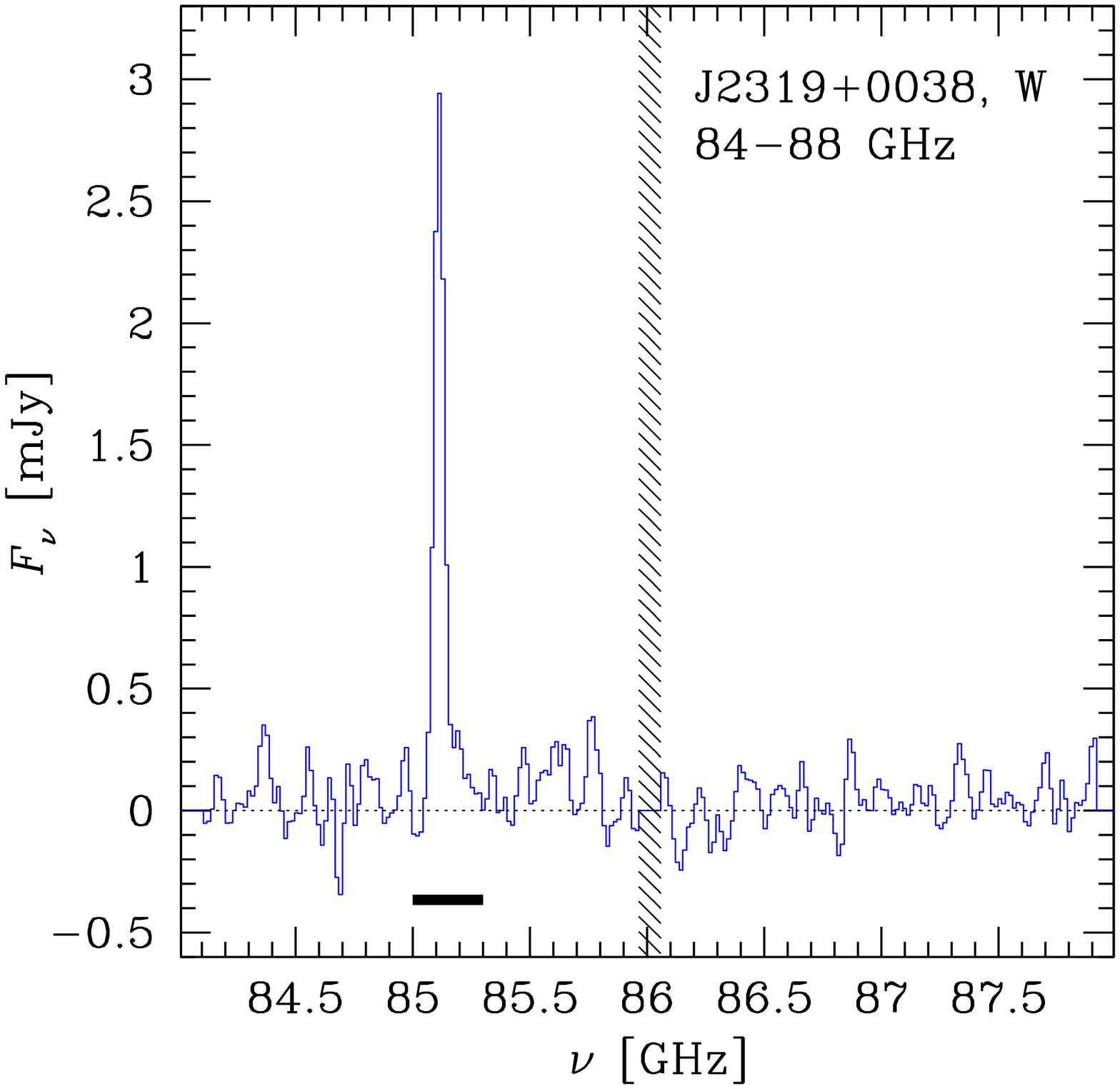}
  \includegraphics[width=7.2cm]{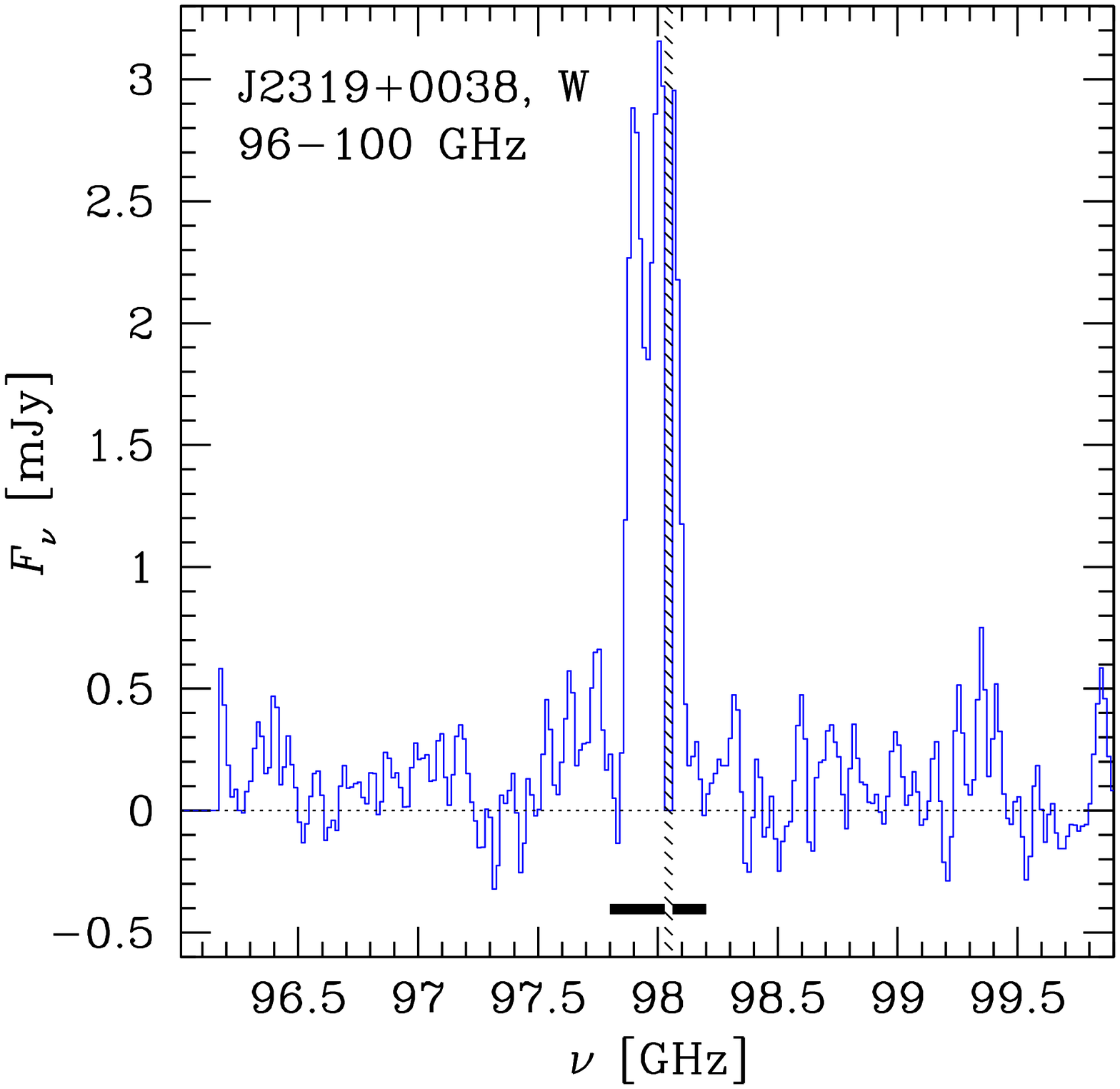}
  \includegraphics[width=7.2cm]{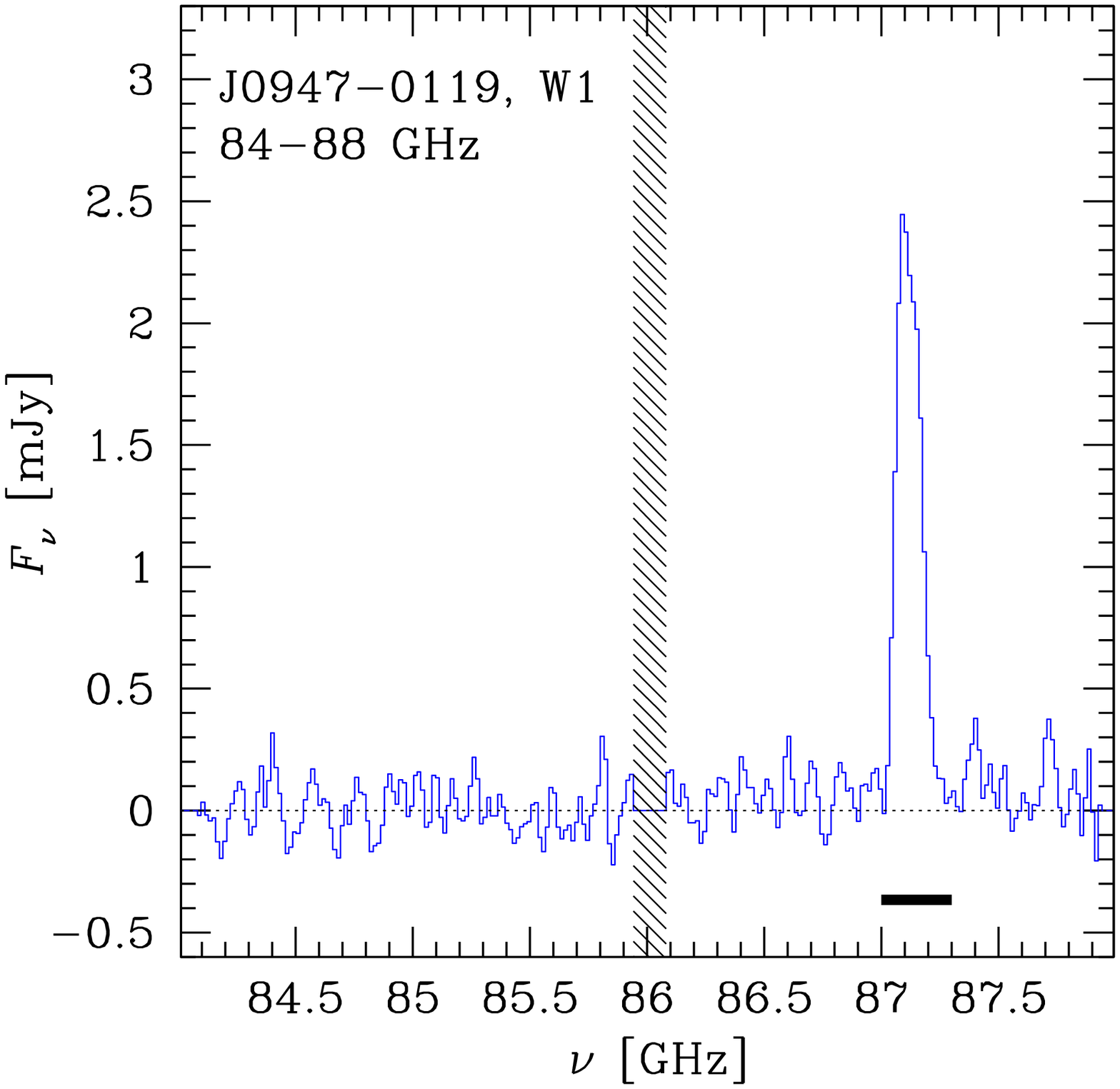}
  \includegraphics[width=7.2cm]{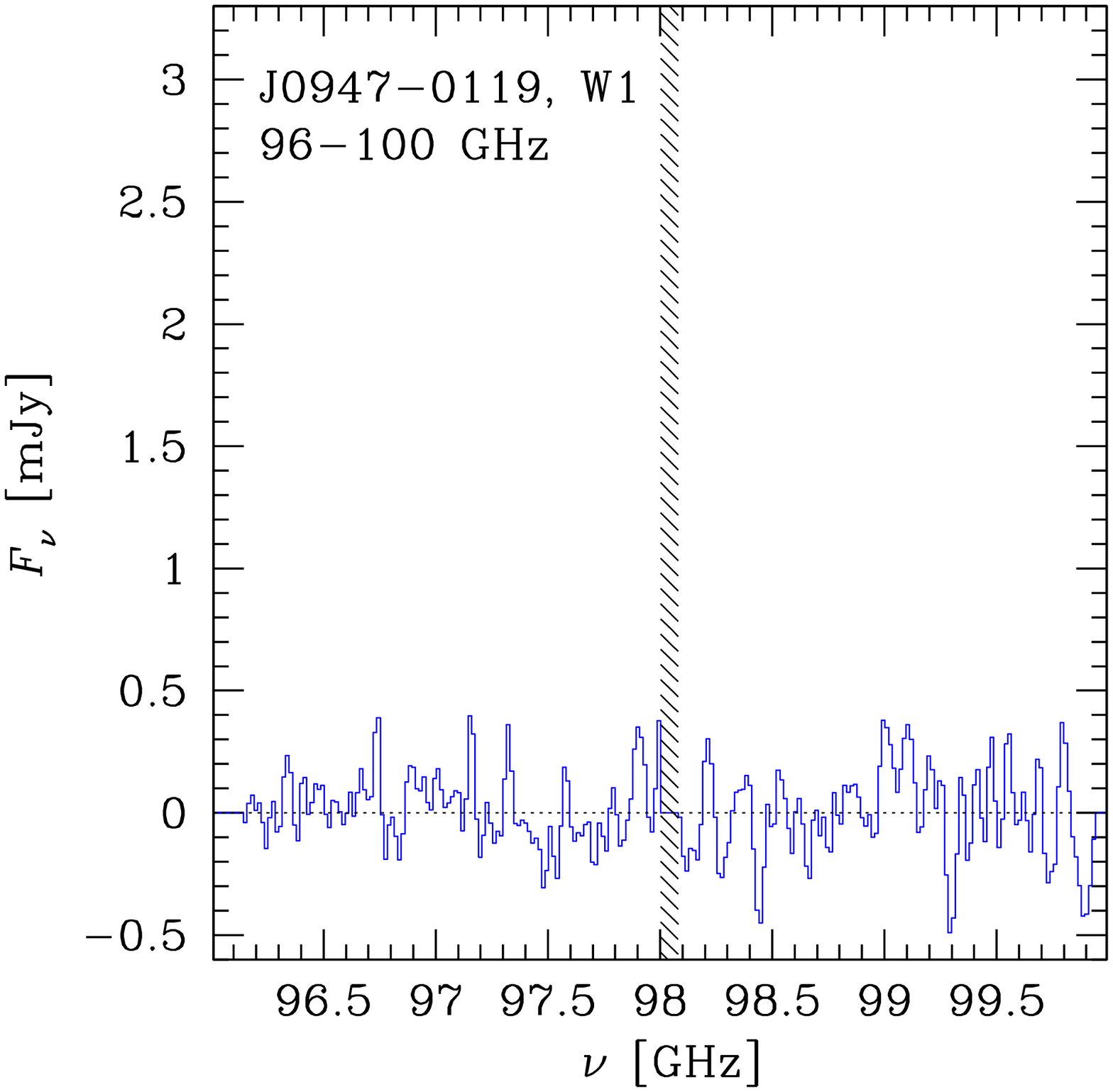}  
 \end{center}
 \caption{Spectra of source W toward \targeta\ (top) and source W1
 toward \targetb\ (bottom) against the observed frequencies at $84-88$
 (left) and $96-100$ (right) GHz. Thick horizontal bars mark the
 frequency ranges over which the integrated flux density in table
 \ref{tab-lines} is computed. Shaded regions show the frequency gaps
 over which the ALMA data are unavailable. Note that the data are not
 available at $88-96$ GHz, either. } \label{fig-lines}
\end{figure*}

 The ALMA Band 3 data presented in the present paper have a spectral
 resolution of $15.6$ MHz and can be used for probing line emissions
 from galaxies in the field-of-views. We find that sources W and W1
 toward \targeta\ and \targetb, respectively, host line emissions as
 listed in table \ref{tab-lines}. The position and the flux density are
 determined by fitting the visibility at $>15 $ k$\lambda$ over the frequency range shown in
 table \ref{tab-lines} using the CASA task {\it uvmodelfit}. We also plot in figure \ref{fig-lines} the spectra within a diameter of $7''$
 around the source position. Note that the continuum flux density of sources W and W1 in tables
 \ref{tab-source1} and \ref{tab-source2}, respectively, are obtained
 excluding the frequency ranges listed in table \ref{tab-lines}. The short baselines ($<15$ k$\lambda$) are excluded in the fit to eliminate possible contamination from the SZE signal.  The line and continuum components so determined are subtracted from the visibility
 to construct the SZE maps shown in this paper. We have checked that there is no
 other line emission in the field-of-views that would alter the detected SZE signal of
 \targeta\ and \targetb.

 The position of the line(s) at $97.8-98.2$ GHz (W-b) agrees with that
 of source W in table \ref{tab-source1} within $0.1''$.  On the other
 hand, the line at $85.0-85.3$ GHz (W-a) lies at $\sim 2''$ south-west
 of W-b and may be due to a different object from W and W-b.  The line at
 $87.0-87.3$ GHz (W1-a) lies at $\sim 1''$ west of source W1.

 It is difficult to determine the redshifts of the sources by a single emission line. For example,  their ranges are $z = 1.3 \sim 1.7$ and $2.5 \sim  3.1$ if the line is from CO(2-1) at 230.54 GHz and CO(3-2) at 345.80 GHz,
 respectively. 
 Detailed nature of these sources is beyond the scope of this paper and will be investigated elsewhere.

  \begin{table*}[h]

   \caption{Line emissions in the fields of \targeta\
   and \targetb. The integrated flux density is for the sum of line and continuum components over the frequency range shown.}
  \begin{center}
    \begin{tabular}{cccccc}
     \hline
     Field & ID & $\nu$ [GHz]
     &RA (J2000) & Dec (J2000) & Integrated flux density
     [Jy km/s]\\ \hline 
     \targeta & W-a & $85.0-85.3$&\timeform{23h19m49.58s} &\timeform{+00D37'55.40''}
		     & $0.76 \pm 0.03$ \\
        & W-b &  $97.8-98.2$&\timeform{23h19m49.70s} &\timeform{+00D37'56.47''}
		     & $^*1.64 \pm 0.03$ \\
     \targetb & W1-a & $87.0-87.3$ &\timeform{9h47m55.39s} &\timeform{-1D20'26.66''}
		 & $1.06 \pm 0.02$ \\
     \hline
    \end{tabular}
      \end{center}
    \begin{tabnote}
    $^*$ The flux density over the frequency gap at 98.028--98.060 GHz (figure \ref{fig-lines} top-right) is excluded. It is not relevant to the analysis of this paper.
    \end{tabnote}
 \label{tab-lines}
  \end{table*}

\bibliographystyle{apj}
\bibliography{hscrefs,refs}
\end{document}